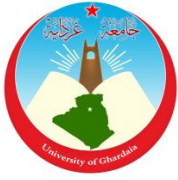 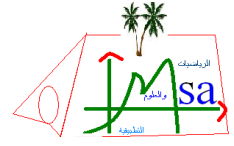

الجمهورية الجزائرية الديمقراطية الشعبية
République Algérienne Démocratique et Populaire
وزارة التعليم العالي والبحث العلمي
Ministère de l'Enseignement Supérieur et de la Recherche Scientifique

جـــامعة غـرداية

N° d'enregistrement

Université de Ghardaïa

كلية علوم الطبيعة والحياة وعلوم الأرض
Faculté des Sciences de la Nature et de la Vie et Sciences de la Terre

قسم البيولوجيا
Département de Biologie

# THÈSE

En vue de l'obtention du diplôme de Doctorat 3ème Cycle LMD
**Domaine :** Sciences de la Nature et de la Vie
**Filière :** Sciences biologiques
**Spécialité :** Ecologie saharienne

## Etude comparative des activités biologiques des extraits aqueux de deux plantes spontanées récoltées au Sahara Algérien

Soutenue publiquement le : 16/01/2020

Par
**CHERIF Rekia**
Devant le jury composé de :

| | | | |
|---|---|---|---|
| **Mr. KHENE Bachir** | MCA | Univ. Ghardaïa | **Président** |
| **Mr. OULD EL HADJ M. Didi** | Pr. | Univ. K.M. Ouargla | **Directeur de thèse** |
| **Mr. KEMASSI Abdellah** | MCA | Univ. Ghardaïa | **Co- directeur de thèse** |
| **Mme. OULD EL HADJ-KHELIL Aminata** | Pr. | Univ. K.M. Ouargla | **Examinatrice** |
| **Mr. BOUALALA Mohammed** | MCA | Univ. Adrar | **Examinateur** |

**Année universitaire : 2019 /2020**

# Remerciement

Avant tout, je remercie le Dieu, tout puissant, pour m'avoir donné la force et la patience de réaliser ce travail.

Je suis très honoré de remercier Monsieur OULD EL HADJ Mohamed Didi, Professeur à l'U.K.M. -Ouargla ; pour l'encadrement de ce modeste travail. Je vous remercie profondément pour m'avoir accordé votre temps, votre patience, votre assistance, votre expérience et vos conseils fructueux et pertinents dans toutes les étapes de ce travail. J'ai l'honneur d'exprimer mes très profondes reconnaissances et mes sentiments les plus sincères et mon admiration.

Je tiens à remercier vivement Monsieur KEMASSI Abdellah, Maitre de conférence A, à l'Université de Ghardaïa, pour la confiance que vous m'avez témoignée en acceptant la direction scientifique de mes travaux. Je vous suis reconnaissante de m'avoir fait bénéficier tout au long de ce travail de votre grande compétence, de votre rigueur intellectuelle, de votre dynamisme, et de votre efficacité certaine que je n'oublierai jamais, soyez assuré de mon attachement et de ma profonde gratitude.

J'exprime mes profondes remerciements à Monsieur KHENE Bachir, Maitre de conférence A, des sciences agronomiques de la faculté des sciences de la Nature et de la Vie de l'université Ghardaïa, pour l'honneur que vous me faites par votre participation à mon jury de thèse en qualité de président de jury, pour le temps consacré à la lecture de cette thèse, et pour les suggestions et les remarques judicieuses que vous m'avez prodiguées.

Je tiens à remercier Madame OULD EL HADJ-KHELIL Aminata, Professeur à l'U.K.M.-Ouargla, qui a bien voulu juger et examiner une grande partie de ce travail. Je vous remercie pour le temps consacré à la lecture de ce travail ainsi que pour les commentaires m'ayant permis de l'améliorer.

Monsieur BOUALALA Mohammed, Maitre de conférences A à l'université Adrar, d'avoir accepté de faire partie du jury de cette thèse. Je vous remercie pour vos conseils scientifiques que vous m'apportez en qualité d'examinateur, en jugeant une partie de cette thèse, ainsi que pour votre immense aide pour mener à bien ces travaux.

# *Dédicace*

*A*

- ☙ *La mémoire de ma Mère (paix à son âme) ;*
- ☙ *Mon cher père, lumière de ma vie ;*
- ☙ *Mon fils Aness 'Abderrahmane' ;*
- ☙ *Mes chères sœurs et frères ;*
- ☙ *Mes adorables nièces et neveux ;*
- ☙ *Toute ma famille CHERIF*

*Ce modeste travail est dédié.*

*CHERIF Rekia*

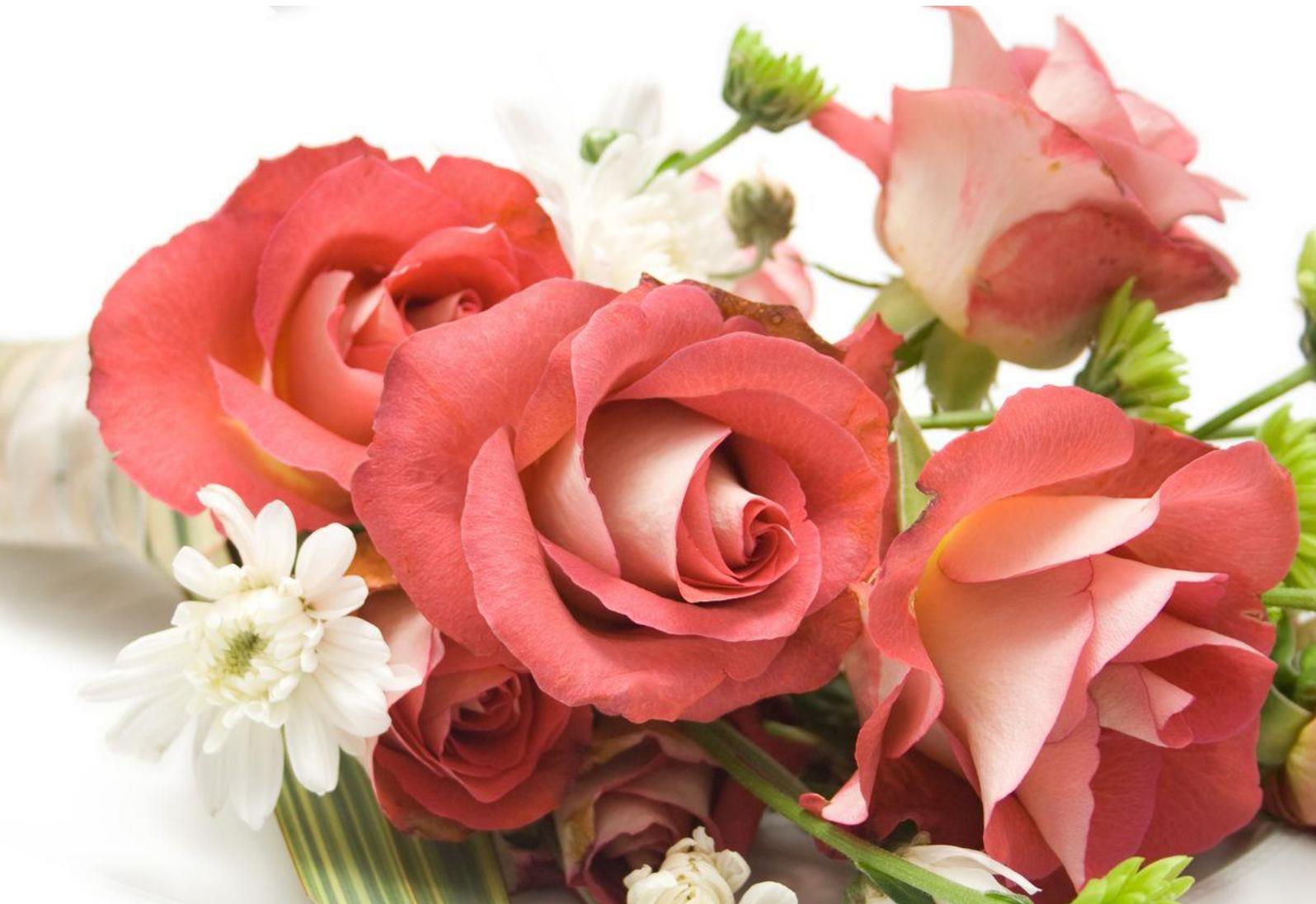

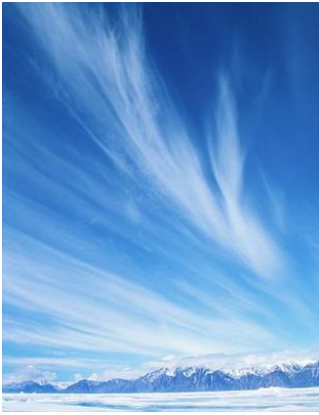

# Table des matières

# Table des matières

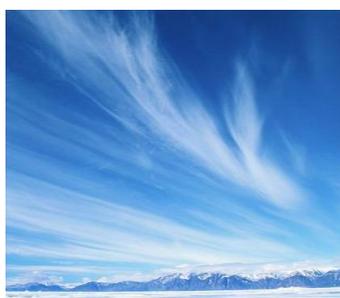

# Liste des figures

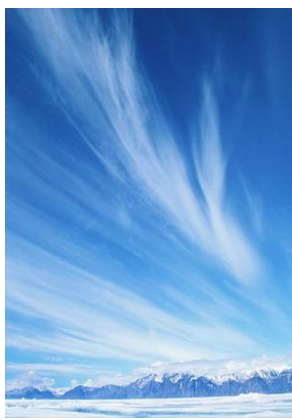

# Liste des Tableaux

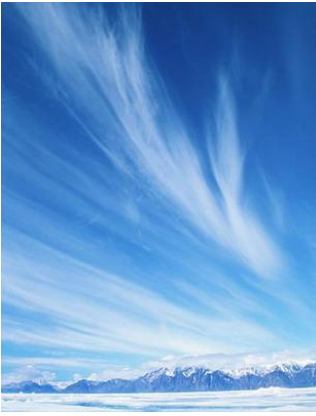

*Thème de thèse Doctorat 3$^{ème}$ cycle LMD*

*Etude comparative des activités biologiques des extraits aqueux de deux plantes spontanées récoltées au Sahara Algérien*

# Etude comparative des activités biologiques des extraits aqueux de deux plantes spontanées récoltées au Sahara Algérien

## Résumé

La présente étude porte sur les effets insecticides et herbicides des extraits foliaires de deux plantes spontanées récoltées au Sahara septentrional Est-Algérien. Il s'agit de *Cleome arabica* (*Capparaceae*) et *Pergularia tomentosa* (*Asclepiadaceae*).

L'évaluation de l'efficacité des extraits des plantes a été faite par la méthode d'extraction par reflux. Le Criblage phytochimique des extraits aqueux de *C. arabica* montre une richesse remarquable en principes actifs en comparaison avec l'extrait de *P. tomentosa*; dont des flavonoïdes, des saponosides, des glycosides, des terpènes, des stérols, des polyphénol et des désoxyoses et des alcaloïdes totaux.

Les imagos de *Tribolium confusum* traités par les extraits aqueux de *C. arabica* et de *P. tomentosa* aux doses 80% à 100% présentent respectivement des taux de mortalité de 73.33% à 96.67%, et de 36.67% à 86.67%.

L'estimation du temps létal 50 ($TL_{50\%}$) de l'extrait aqueux de *C. arabica* est de 6.41jour, et de 6.94jour pour l'extrait *P. tomentosa* pour les imagos de *T. confusum*. Les extraits de *P. tomentosa* semblent moins toxiques que les extraits de *C. arabica*.

Les potentiels allélopathiques de *C. arabica* et de *P. tomentosa* testés sur la germination des graines d'une adventice *Dactyloctenium aegyptium* (*Poaceae*) et deux espèces cultivées dont *Hordeum vulgare* et *Triticum durum* (*Poaceae*), montrent que l'effet inhibiteur des extraits de *C. arabica* est très hautement significatif. Il se manifeste sur la croissance de la partie aérienne et souterraine des plantules d'*H. vulgaire* et *T. durum*. Le taux d'inhibition est supérieur à 84.44% pour les graines chez *D. aegyptium* traitées par les différentes concentrations. Les taux d'inhibition oscillent entre 75.56% et 91.11% pour les graines de blé dur aspergées aux doses de 80% à 100%, mais ne sont que de 55.56% à 77.78% pour les graines d'orge traitées par les mêmes concentrations (80% à 100%). L'effet inhibiteur de l'extrait de *P. tomentosa* est modéré par rapport à l'extrait de *C. arabica*.

L'estimation des concentrations d'efficacités $CE_{50\%}$ et $CE_{90\%}$, pour l'extrait de *C. arabica*, il est enregistré 0,020mg/ml et 0.037mg/ml pour les graines d'orge, 0.012mg/ml et 0.028mg/ml pour les graines de blé dur, et 0.0001mg/ml et 0.0014mg/ml pour les graines dactylocténion. Pour l'extrait de *P. tomentosa*, elles sont de l'ordre de 0.026mg/ml et 0.041mg/ml pour les graines d'orge, 0.023mg/ml et 0.035mg/ml pour les graines de blé dur, et 0.0009mg/ml et 0.004mg/ml pour les graines dactylocténion. Les graines *D. aegyptium* sont plus sensibles à l'action de l'extrait foliaire de *C. arabica* et *P. tomentosa* comparativement aux graines d'orge et de blé dur.

**Mots clés :** *Cleome arabica*, *Pergularia tomentosa*, extraits aqueux, activités biologiques, étude comparative, Sahara.

# Comparative study of the biological activities of the aqueous extracts of two spontaneous plants harvested in the Algerian Sahara

**Abstract.**

The present study investigates the insecticidal and herbicidal effects of leaf extracts from two plants were harvested in the Northern Algerian Sahara. These are *Cleome arabica* (*Capparaceae*) and *Pergularia tomentosa* (*Asclepiadaceae*).

The efficacy of the extracts from the plants was evaluated by the reflux extraction method. The phytochemical screening of the aqueous extracts of *C. arabica* shows a remarkable richness in active principles in comparison with the extract of *P. tomentosa*; including flavonoids, saponosides, glycosides, terpenes, sterols, deoxyose, polyphenols and total alkaloids.

The imago of *Tribolium confusum* treated with aqueous extracts of *C. arabica* and *P. tomentosa* at doses of 80% to 100% respectively have mortality rates of 73.33% to 96.67%, and 36.67% to 86.67%.

The lethal time 50 ($TL_{50\%}$) of the aqueous extract of *C. arabica* was estimated about 6.41 days, and 6.94 days for the extract *P. tomentosa* for the imago of *T. confusum*. The extracts of *P. tomentosa* is less toxic than the extracts of *C. arabica*.

The allelopathic potentials of *C. arabica* and *P. tomentosa* tested on germination of the seeds of a weed *Dactyloctenium aegyptium* (*Poaceae*) and two cultivated species, including *Hordeum vulgare* and *Triticum durum* (*Poaceae*), show that the inhibitory effect of extracts of *C. arabica* is very highly significant. It manifests itself in the growth of the aerial and underground part of the *H. vulgar* and *T. durum*. The inhibition rate is more than 84.44% for *D. aegyptium* seeds treated with the different concentrations. The inhibition rates range from 75.56% to 91.11% for *T. durum* wheat irrigate at 80% to 100%, but are only 55.56% to 77.78% for barley seeds treated with the same concentrations (80% to 100%). The inhibitory effect of the extract of *P. tomentosa* was moderated by contribution to the extract of *C. arabica*.

The estimation of the $EC_{50\%}$ and $EC_{90\%}$ efficacy concentration, for the extract of *C. arabica*, was recorded 0.020mg/ml and 0.037mg/ml for the barley seeds, 0.012mg/ml and 0.028mg/ml for durum seeds, and 0.0001mg/ml and 0.0014mg/ml for dactyloctenion seeds. For the extract of *P. tomentosa*, they are of the order of 0.026 mg/ml and 0.041 mg/ml for barley seeds, 0.023mg/ml and 0.035 mg/ml for durum seeds, and 0.0009 mg/ml and 0.004mg/ml for the seeds dactyloctenion. The dactyloctenion seeds are more sensitive to the leaf extract of *C. arabica* and *P. tomentosa* compared to barley and durum seeds.

# دراسة مقارنة للنشاط البيولوجي لمستخلصات نباتات برية المقطوفة من صحراء الجزائر
## *Cleome arabica (Capparaceae)* و *Pergularia tomentosa (Asclepidaceae)*

## الملخص

تمت الدراسة حول أثر مستخلص أوراق النبتتين *C. arabica* و *P. tomentosa* المقطوفتين من شمال صحراء الجزائر، كمبيد للحشرات والأعشاب.

تم تقييم فعالية المستخلصات النباتية بواسطة طريقة الاستخلاص Reflux. بين التحليل الكيميائي لمكونات المستخلصات، وجود وفرة في العناصر الأساسية الفعالة عند مستخلص *C. arabica* مقارنة بمستخلص نبتة *P. tomentosa*، مثل: الفلافونويدات، السابونوزيدات، الغليكوزيدات، تربنويد، ستيرول، البوليفينول والألكالويدات.

من خلال دراستنا، تبين أن حشرات *T. confusum* المعالجة بمستخلص نبتتي *C. arabica* و *P. tomentosa* بالتراكيز 100% و 80% أعطت نسبة وفيات تتراوح من 73.33% إلى 96.67% لنبات *C. arabica* و 36.67% إلى 86.67% لنبات *P. tomentosa* على التوالي.

قدر الوقت اللازم بـ 6.41 يوم للقضاء على 50% من حشرة *T. confusum* لمستخلص نبات *C. arabica*. بينما في مستخلص نبات *P. tomentosa* قدر بـ 6.94 يوم. ما يُبين أن مستخلص *P. tomentosa* أقل سمية من *C. arabica*.

تم اختيار نبات *P. tomentosa* و *C. arabica* لاختبار إمكانية التثبيط على نمو بذور *Dactyloctenium aegyptium (Poaceae)* ونوعين مزروعين هما *Hordeum vulgare (Poaceae)* و *Triticum durum (Poaceae)*.

أظهرت النتائج نمو بذور الأعشاب *H. vulgare*، *T. durum* و *D. aegyptium* المعالجة بمستخلصين *C. arabica* و *P. tomentosa* أن تأثير المستخلص *C. arabica* كان بارزا جدا على تثبيط نمو البذور. ويتجلى ذلك أكثر على نمو الجزء الهوائي والترابي لنبتتي *T. durum* و *H. vulgare*. نسبة التثبيط كانت أكبر من 84.44% عند نمو بذور *D. aegyptium* المعالجة بمختلف التراكيز.

نسبة التثبيط تتراوح بين 75.56% و 91.11% عند بذور القمح الصلب المسقية بالتراكيز 80% إلى 100%. بينما تنخفض هذه النسب إلى 55.56% و 77.78% عند بذور الشعير المعالجة بنفس التراكيز (80% إلى 100%). التأثير التثبيطي لمستخلص *P. tomentosa* كان أقل جزئياً مقارنة بمستخلص *C. arabica*.

يظهر تقدير التراكيز الفعالة EC50% و EC90% للمستخلص *C. arabica* بـ 0.020ملغ/مل و 0.037ملغ/مل لبذور الشعير، 0.012ملغ/مل و0.028ملغ/مل لبذور القمح الصلب، 0.0001ملغ/مل و0.0014ملغ/مل لبذور *Dactyloctenium*. أما بالنسبة لمستخلص *P. tomentosa* قدرت بـ 0.026 ملغ/مل و0.041 ملغ/مل لبذور الشعير، 0.023 ملغ/مل و0.035 ملغ/مل لبذور القمح الصلب، 0.0009 ملغ/مل و 0.004 ملغ/مل لبذور *dactyloctenion*.

تعد بذور dactyloctenion الاكثر حساسية لمفعول مستخلص أوراق *C. arabica* و *P. tomentosa* مقارنة ببذور الشعير والقمح الصلب.

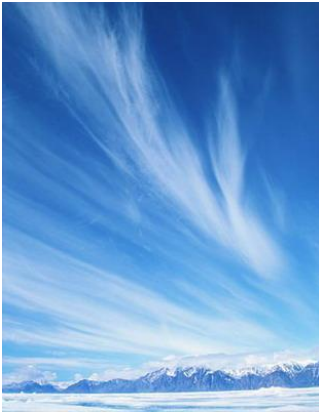

# Introduction

# Introduction

L'agriculture constitue un pilier de l'économie algérienne. Le secteur agricole occupe près de 12.3% du produit intérieur brut (PIB) avec 11% de la population active (BERKOUK, 2015).

La plupart des activités agricoles se situent dans les hauts plateaux en Algérie. La diversité agricole se caractérise par la dominance des grandes cultures telles que la céréaliculture qui occupe une place importante dans la production agricole, suivie par les légumineuses et les fourrages (GAUTHIER, 1980).

Les données du recensement de la production agricole durant la période de 2014 à 2017, montrent un taux de croissance de la production céréalière de près de 19%, suivie par la production de la pomme de terre avec un taux de croissance près de 9.4%. Ces données exhaustives restent incomparables avec les données de la production agricole de l'année 2000, où il est enregistré un taux de croissance de près de 83% entre du blé dur et du blé tendre et environ 12% pour la production de pomme de terre (APS, 2018). Cette déficience dans la production agricole est probablement liée aux différents facteurs biotiques en tenant compte de la concurrence des herbes envahissantes pour les cultures associées, les molécules allélochimiques emmagasinées dans le sol, etc. et des facteurs abiotiques ou anthropiques liés à des activités agricoles (épandage), l'introduction des produits phytosanitaires pour la protection des cultures, l'utilisation excessive et irrationnelle des pesticides lors de traitement, etc. (HE *et al.*, 2005).

L'utilisation des pesticides en agriculture a pour objectif de gérer les populations (ennemis, insectes, champignons et maladies) des cultures, ainsi que contre les mauvaises herbes qui les privent d'une partie des ressources naturelles dont l'eau, la matière organique et la lumière. Ceci permet donc une production agricole de haute qualité et de quantité (rendement), grâce à leurs avantages, mais elle suscite aussi de nombreuses inquiétudes liées notamment à leur toxicité, aux normes d'applications des doses, leurs impacts nocifs sur l'être humain et les écosystèmes terrestres (biosphère).

En revanche, un pesticide de synthèse est un mélange de substances, produites pour empêcher ou détruire l'activité d'un ravageur donné soit un vecteur de maladie, espèces nocives durant la période de production, transformation, entreposage ou stockage de denrées alimentaires. La majorité des formulations de pesticides contiennent des matières actives, divers adjuvants qui servent de solvants ou améliorant l'efficacité d'absorption (FAO, 1986).

Selon le parasite à contrôler, la classification des pesticides repose sur trois grandes catégories, dont les herbicides qui servent à l'élimination des adventices de culture, les insecticides pour détruire les insectes nuisibles et les fongicides permettent de lutter contre les maladies cryptogamiques (AYAD MOKHTARI, 2012).

Selon l'Organisation des Nations Unies pour l'alimentation et l'agriculture, la consommation Algérienne des pesticides en kilogramme par hectare est estimée à près de 9,64 en 2002 et de 22,32 en 2016. Cette consommation accrue des produits phytosanitaires au cours du temps, l'emploie abusif de ces produits, la réalisation de plusieurs applications chimiques durant les stades végétatifs et l'ignorance de respect des doses, normes, et le stade de traitement engendrent des fluctuations sur le plan écologique, sociale (diffusion de maladies), économique (pauvreté), et industriel.

Lors de traitement d'une culture, il existe une proportion non négligeable de produit disséminée dans le biotope. Certains pesticides (organochlorés) sont peu ou pas dégradés dans le sol et les milieux contaminés, de ce fait, ils vont s'accumuler dans les plantes voire dans les graisses animales et donc se concentrer tout au long de la chaine alimentaire ce qui provoque un déséquilibre de réseaux trophique. De plus, il est rare que les pesticides ont des effets sélectifs (cible pour une seule espèce), puis qu'ils interviennent sur les processus physiologiques et fondamentaux du métabolisme dont la photosynthèse, la division et élongation cellulaire, la reproduction, etc. (AUBERTOT *et al.*, 2005).

N'oublions pas les méfaits des pesticides sur la faune, ce sont des espèces au sommet de la chaine alimentaire. Les insectes tels que les abeilles sont les plus touchées. Ils sont considérés comme des pollinisateurs, environ de 80% des angiospermes sont pollinisés par les insectes et les animaux (SCDB, 2008).

De même, l'impact de ces matières actives internes qui les constituent n'est pas généralement pris en considération pour la santé humaine. Elles constituent une proportion élevée du pesticide commercial et peuvent avoir des effets nocifs et plus grave. C'est le cas des organophosphorés qui sont des insecticides appliquées par voie de digestion, contacte, et ou inhalation au niveau de l'insecte. Ce sont des molécules neurotoxiques qui bloquent l'activité enzymatique des acétylcholinestérases et empêchent ainsi la transmission de l'influx nerveux (OMS, 1991). De plus, ils contaminent les aliments que nous mangeons, ainsi que la diffusion des maladies pathologiques notamment les cancers, les stérilités, les malformations congénitales, les déficiences mentales, des troubles neurologiques et de reproduction (TANDIA *et al*., 2003).

Face à ces menaces, il est nécessaire de contrôler de manière rigoureuse l'usage des pesticides lors de traitement des cultures. A la lumière de ce constat, la présente étude se situe dans le cadre de la valorisation des potentialités de la flore saharienne et la recherche de molécules bioactives d'origine naturelle multi-usages soit pour la lutte contre les mauvaises herbes, mais aussi contre des insectes nuisibles, des maladies, etc.
Elle porte sur les activités biologiques des extraits de deux espèces végétales spontanées et communes au Sahara Algérien. Il s'agit de *Cleome arabica* (*Capparidaceae*) et *Pergularia tomentosa* (*Asclepiadaceae*).

Le travail est structuré en 2 chapitres. Le premier chapitre, porte sur la méthodologie de travail. Le second chapitre regroupe les résultats obtenus et leur discussion. Une conclusion générale achève ce travail.

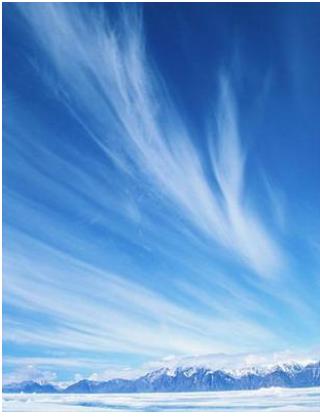

*Chapitre I*

# Méthodologie de travail

# Chapitre I- Méthodologie de travail

## I.1.- Principe adopté

La biodiversité au Sahara dispose d'une richesse floristique exceptionnelle, constituée d'environ 500 espèces végétales, ces plantes sont connues pour leurs multiples usages pratiqués par la population autochtone, tant sur le plan pharmaceutique, agronomique, économique, écologique mais aussi stratégique (KEMASSI *et al*., 2012).

Afin de mieux caractériser et valoriser les potentialités de la flore du Sahara Algérien, la présente étude recherche des activités biologiques des extraits à partir de deux plantes spontanées récoltées au Sahara septentrional est-algérien soient *Cleome arabica* (*Capparaceae*) et *Pergularia tomentosa* (*Asclepiadaceae*).

Le travail porte sur l'analyse phytochimique (criblage) des extraits aqueux des parties aériennes de *Cleome arabica* et *P. tomentosa*, ainsi que l'évaluation des activités insecticide et allélopathique vis-à-vis de certaines espèces testées.

## I.2.-Choix des plantes

Près de 6 377 espèces de plantes sont utilisées en Afrique, dont plus de 400 sont des plantes médicinales, constituant environ 90% de la médecine traditionnelle. Selon OMS, dans certaines pays en voie de développement, près de 75% de la population dépend de la médecine traditionnelle particulièrement chez les populations autochtones en milieu rural a fait le recoure aux plantes spontanées médicinales pour se soigner les maladies (ZEGGWAGH *et al*., 2013).

Dans le contexte socio-économique des pays en voie de développement, l'étude des plantes peut aboutir à l'obtention de réponses thérapeutiques adéquates et de faible prix, joignant à une efficacité scientifique prouvée, une acceptabilité culturelle optimale. La valorisation scientifique de la médecine traditionnelle doit conduire notamment à la mise au point de médicaments à base de plantes.

Il a été estimé que les principes actifs provenant des végétaux représentent 25% des médicaments prescrits soit un total de 120 composés d'origine naturelle provenant de 90 plantes différentes (POTTERAT et HOSTETTMANN, 1995). Actuellement, 205 drogues végétales entrent dans la composition de médicaments dits de phytothérapie et bénéficient d'un dossier allégé d'autorisation de mise sur le marché (ZEGGWAGH *et al*., 2013).

La famille Capparidacées est connue pour sa richesse en métabolites secondaires. Divers groupes de ces métabolites secondaires ont été identifiés dans la plante *Cleome arabica* soient les triterpènes, anthraquinones, flavonoïdes, saponines, stéroïdes, résines, glycosides, tannins et alcaloïdes (MADI, 2018).

Plusieurs recherches sont effectuées sur *P. tomentosa*. ALGHANEM et EL-AMIER (2017) notent que *P. tomentosa* présente une richesse remarquable de principes actifs à savoir les saponines, tanins, alcaloïdes, flavonoïdes, glycosides et stéroïdes.

A cet effet, l'étude porte sur la caractérisation qualitative des principes actifs présents dans la partie foliaire des espèces végétales investies, ainsi que, l'évaluation de pouvoir biocides d'extraits aqueux foliaires de ces deux espèces choisies.

## I.2.1.- Récolte

La collecte des plantes choisies est réalisée dans la commune de Metlili. La région est située à 40km au sud de la wilaya de Ghardaïa (Algérie). Metlili est à une latitude de 32° 29' Nord et de longitude de 3° 60' Est. Elle couvre une superficie de 7300 km².

Les collectes de *P. tomentosa* et *C. arabica* sont effectuées entre le mois de Mars et celui d'avril 2017, au stade de végétation dans leur biotope naturel. La situation géographique de la zone de collecte est située en amont l'Oued Metlili à El-Guemgouma à une altitude de 922 m à 32°17'46" Nord et de longitude de 3°34'10" Est de 526 m (photo 01).

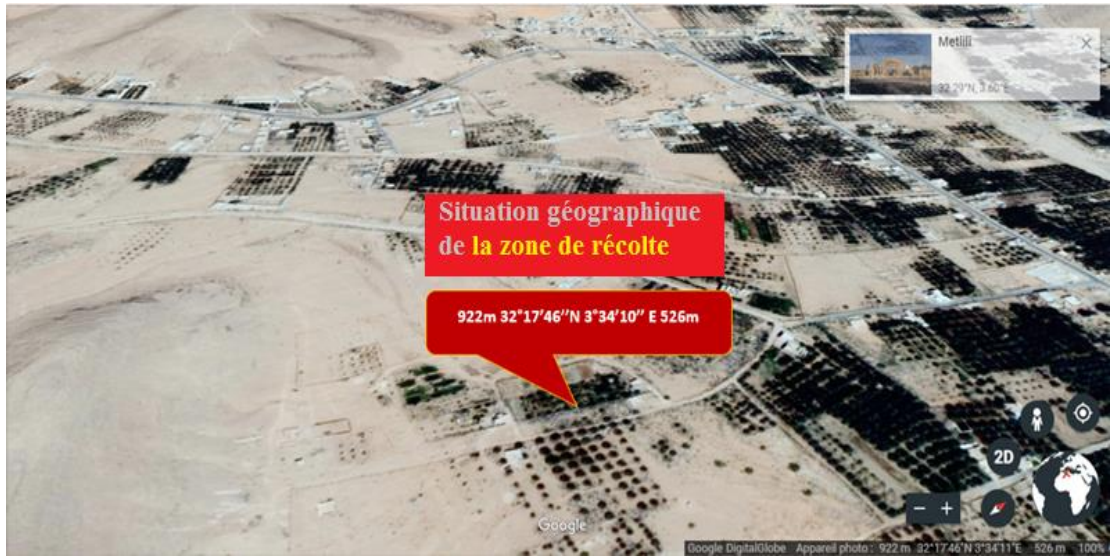

**Photo 01.-** Position géographique de la région de collecte en amont de l'Oued Metlili El-Guemgouma (Google-Earth ; version 2018).

### I.2.2.- Présentation de *Pergularia tomentosa* L. *(Asclepidaceae)*

La famille Asclepidaceae regroupe environ 355 genres et près de 3 700 espèces répandues essentiellement dans les régions tropicales et subtropicales, mais avec quelques représentants essentiellement herbacés dans les zones tempérés (GOOD, 1951).

### I.2.2.1.-Description botanique

*P. tomentosa* (*Asclepidaceae*) est un arbuste grimpant à partir d'un rhizome ligneux (photo. 2). Les jeunes tiges se tortillent autour des tiges âgées, contenant du latex blanc ; tiges et inflorescences densément et courtement pubescentes à poils raides. Les feuilles sont opposées, simples et entières, stipules absentes ; pétiole de 0,5 à 1,5 (-3) cm de long ; limbe légèrement charnu, largement ovale à presque orbiculaire, de 1-3(-4.5) cm x 1-3(-4.5) cm, base profondément cordée, apex obtus à aigu, à poils courts denses sur les deux faces, gris argenté (BAERTS et LEHMANN, 2012).

L'inflorescence se fait en grappe extra-axillaire, ombelliforme lorsque jeune, pédoncule de 1 à 3cm de long. Les fleurs sont bisexuées et régulières à 5 mètres parfumées avec un pédicelle de 1.5 à 2 cm de long, courtement poilu ; sépales ovales à oblongs de 2 à 3,5mm de long, apex aigu, à poils courts denses. Le corolle vert jaunâtre ou violet brunâtre, le tube cylindrique est de 2 à 4mm de long, lobes ovales à elliptique, de 4-(5-6(-8)mm x (2-)3-3.5(-4)mm, apex quasi aigu ou aigu, glabre à l'intérieur sauf la base et les bordes qui portent les longs poils ; couronne externe membraneuse à lobes ciliés, lobes de la couronne interne minces à trapus, de 3,5 à 5 (6) mm de long, la projection apicale s'étendant légèrement étalées ; ovaire supère, deux loculaires (ABE et YAMAUCHI, 2000). Les fruits sont habituellement une paire de follicules, réfléchis le long du pédicelle, lancéolés, de 4 à 5,5cm x 1 à 1,5 (2) cm, courbés en un bec atténué court ou long, lisses ou couverts de protubérances à poils courts et doux. Graines ovoïdes, aplaties, de 7 à 9mm x environ 9mm, bords pâles, à poils courts denses, munies d'une touffe de poils à une extrémité d'environ 3cm de long (SCHMELZER et GURIB, 2013).

Au Sahara, *P. tomentosa* fleurit toute l'année, mais vers les lisières du Nord et Est, il fleurit au printemps. Elle se multiplie par graines. Le poids moyen de 1000 graines est de 16g. Le genre *Pergularia*, tout comme *Asclepias*, *Calotropis* et *Gomphocarpus*, appartient à la subtribu des *Asclepiadinae*. Le genre comprend deux espèces polymorphes ; l'autre espèce, *Pergularia daemia*, qui a un port plus petit, se rencontre dans les déserts et les régions sèches de l'Afrique australe jusqu'en Somalie, et jusqu'en Asie du Sud en passant par la péninsule Arabique. Les formes intermédiaires entre les deux espèces sont présentes à Socotra (Yémen) et elles viennent d'être incluses dans *P. tomentosa* (ABE et YAMAUCHI, 2000).

*P. tomentosa* pousse bien dans les déserts où les précipitations ne dépassent pas les 100mm par an, dans les lits des oueds et sur les plateaux, sur les sols argileux à sablonneux, graveleux et pierreux. Il est présent depuis le niveau de la mer jusqu'à 1000m d'altitude. Le long de la mer rouge, il se trouve à la fois avec *Aerva lanata* (L) Juss., *Capparis decidua* (Forssk) Edgew et *Salsola spinescens* Moq. Au sein des communautés de plantes qui dominent les plaines sablonneuses (BAERTS et LEHMANN, 2012).

La plante *P. tomentosa* était largement répandue en Libye, en Palestine, au Pakistan et en Arabie saoudite (GOHAR *et al*., 2000 ; AL-FARRAJ et AL-WABEL, 2007), en Égypte dans le désert, sur la côte de la mer Rouge et à Sinaï (BOULOS, 2000), de même elle est très répandue au Sahara Algérien (ABEGAZ et DEMISSEW, 1998).

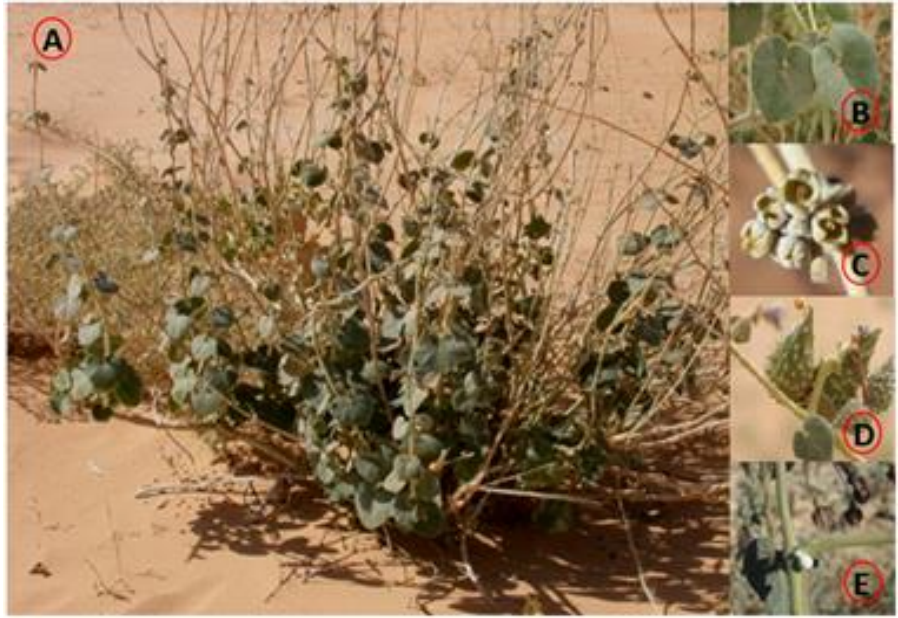

**Photo 02-** *P. tomentosa* en pied isolé

(Oued Metlili, Wilaya de Ghardaïa ; Avril 2017).

**(A)** : Plant entier, **(B)**; Feuille **(C);**Fleurs en grappe; **(D)** Fruits; **(E)** : Latex blanc.

## I.2.2.2.- Position systématique

La classification de *P. tomentosa* présentée dans le tableau 1 est faite selon MAMAN (2003) et in KEMASSI *et al*., (2008).

**Tableau 01 :** Position systématique de l'espèce *P. tomentosa*

| | |
|---|---|
| Règne : | Plantes |
| Embranchement : | Spermaphytes |
| Sous-embranchement : | Angiospermes |
| Classe : | Magnoliopsidae |
| Sous-classe : | Rosidae |
| Ordre : | Gentianales |
| Famille : | Asclepiadaceae |
| Genre : | Pergularia |
| Espèce : | *Pergularia tomentosa* L. |

## I.2.2.3.- Intérêt socio-économique

Au Sahara central, une décoction de pâte de racine avec la viande de chèvre se prend pour traiter la bronchite et la tuberculose. Un morceau de la racine fraîche s'applique dans le rectum pour traiter les hémorroïdes. En Côte d'Ivoire, la plante broyée, parfois avec l'ajout de pigments, se boit ou se donne en lavement contre la dysenterie et comme vermifuge. Le jus de feuille, appliqué en collyre, est considéré comme un remède excellent contre les maux de tête, alors qu'au Niger, il est administré la pâte de l'écorce de racine réduite en poudre dans le nez à cet effet. Au Niger et dans le Nord du Nigeria, l'infection par le ver de Guinée se traite par l'application de quelques gouttes de latex dans des incisions sur la coque. Au Niger, la racine fait partie d'un médicament contre la fatigue généralisée, la plante est un ingrédient d'un poison de flèche à base de Strophanthus, et elle aurait également des propriétés médicales. En Afrique du Nord, la plante sert à provoquer l'avortement (ABIOLA et *al.*, 1993).

Dans le Nord du Nigeria, dans les monts du Hoggar (Sud Algérien) et en Egypte, la plante est réduite en pâte ou sa décoction est frottée sur les peaux à apprêter. La pâte est laissée pendant une nuit après quoi les poils peuvent être ôtés de la peau. Il est utilisé le latex de la même façon. Les peaux peuvent aussi être trempées dans une décoction de la plante ou dans un bain de cendres de bois ; dans lequel la plante a pu fermenter pendant quelques jours afin d'améliorer l'absorption des tanins. Au Maroc et dans d'autre pays, le latex s'utilise comme dépilatoire cosmétique. Au Soudan oriental, les berges frottent le latex contre la mamelle des vaches pour augmenter la production laitière (ABIOLA *et al.*, 1993).

## I.2.2.4.-Caractéristiques générales de l'espèce végétale

Le latex de *P. tomentosa* est corrosif et peut sérieusement endommager la peau. Toutes les parties de la plante sont des sources riches en hétérosides cardénolides, avec l'uzariénine. La coroglaucigénine et la pergularine comme aglycones, et le glucose et le digitoxose comme sucres. Les racines contiennent des hétérosides (Ghalakinosides, Calactine) et des dérivés ; les feuilles et le latex contiennent Ghalakinosides, de la Pergularine, de la 16α-acétoxycalotropine de la calactine et de la coroglauigénine. En dehors des cardénolides la plante contient également des alcaloïdes, des poly-phénols, des terpènoïdes, des flavonoïdes, des coumarines, des anthraquinones, et les tanins (ABICH et REICHSTEIN, 1962).

Plusieurs des Cardénolides sont caractérisés par le double lien entre la fraction sucre et l'aglycone, ce qui leur donne une structure dioxanoides. Les hétérosides sont très cardiotoxiques lorsque administrés par voie intraveineuse. Ils sont peu absorbés lorsqu'intégrés par voie buccale, ce qui pourrait expliquer l'information contradictoire concernant l'utilisation de la plante comme fourrage. Certaines de ces cardénolides issues de la racine et des parties aériennes ont montré une activité cytotoxique contre plusieurs lignées de cellules cancéreuses humaines in vitro (SCHMELZER et GURIB, 2013).

## I.2.3.- Présentation de *Cleome arabica* (*Capparaceae*)

Sur le plan taxonomique, la famille Capparidacées est numériquement importante. Elle ne compte pas moins de quarante-cinq genres et environ sept cents espèces distribuées essentiellement dans les zones tropicales et subtropicales (DUVIGNEAUD et VAN BOCKSTAL, 1976 et HEYWOOD, 1979). Pour OZENDZ (1991), les Capparidaceae sont modestement représentées au Sahara septentrional et central. Au Sahara méridional, environs 20 espèces sont recensées, regroupées dans les genres Capparis, Cleome et Maerua. Du point de vue des caractères botaniques marquants on notera que les Capparidacées sont généralement des herbes, arbustes ou lianes, à feuilles alternes, simples ou transformées en épines les fruits sont des capsules, des baies ou des siliques (BOGNOUNOU, 1994).

### I.2.3.1.-Description botanique

Plante vivace de 30 cm de hauteur, à tiges dressées et ramifiées, feuilles petites poilues, trifoliées à folioles lancéolées, les fleurs ont des pétales dont la couleur va du jaune au pourpre-foncé et violet foncé à l'apex, le fruit est une gousse velue de 2 à 5 cm de longueur située à la base du pétiole (GUBB, 1913 et OZENDA, 1991). C'est une plante à odeur fétide, toxique et présente des effets hallucinogènes (GUBB, 1913) (photo 3).

*C. arabica*, fréquent dans les savanes désertiques de l'étage tropical, monte dans l'étage méditerranée inférieur sur les pentes pierreuses et dans les ravins sablonneux jusque vers 2300 m d'altitude (MAIRE, 1933). C'est une espèce commune dans tout le Sahara septentrional, en Egypte et en Afrique tropicale (KEMASSI *et al*., 2012). Selon OZENDA (1991) dans la région saharienne, *C. arabica* se trouve sur des rocailles, du sable et des graviers.

Cleome est le plus grand genre de Cleomoideae comprenant 180 à 200 espèces herbacées annuelles ou vivaces et arbustes largement distribués dans les régions tropicales et subtropicales (RAGHAVAN, 1993). La fleur, calice à 4 sépales, 4 pétales, 6 étamines, 4 ovaires à 1 loge, portés par un pied court ou nul, fruit est capsule stipitée, siliquiforme, à 2 valves se séparant des placentas (QUEZEL et SANTA, 1962).

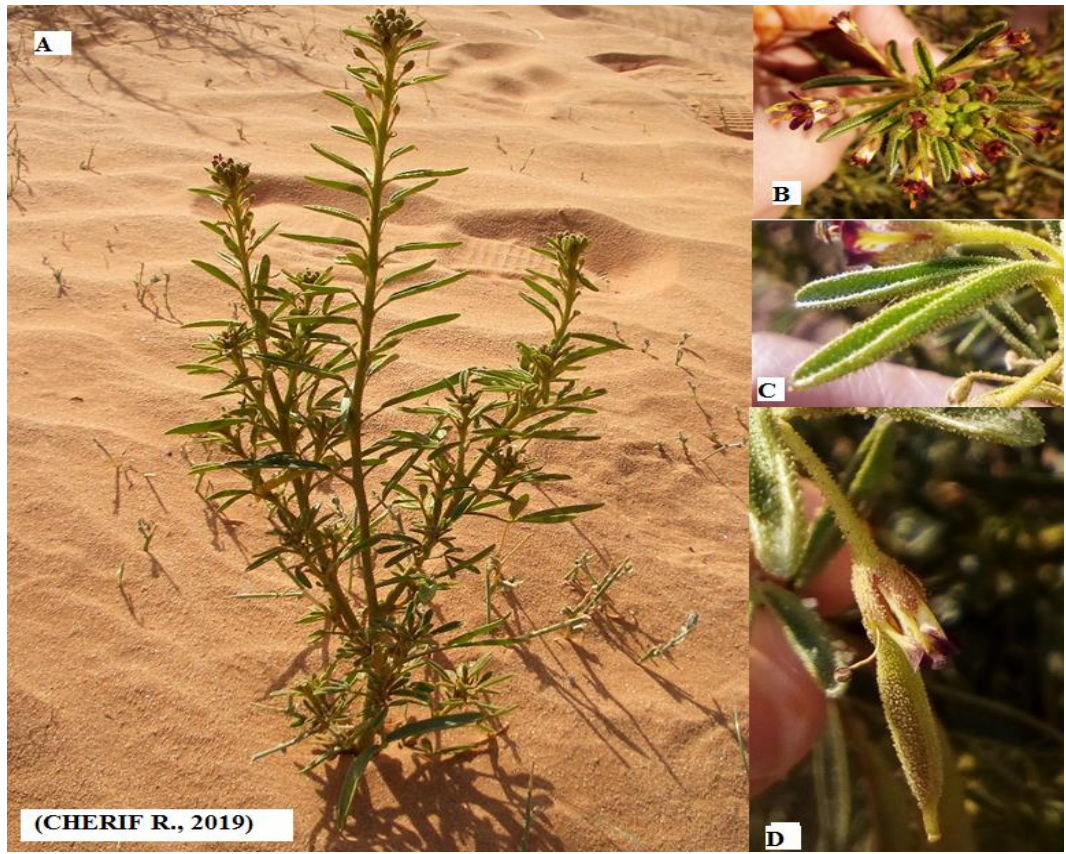

**Photo 3-** *C. arabica* en pied isolé (Oued Metlili, Wilaya de Ghardaïa; Janvier 2019). (A: Plante entière; B: Fleurs en grappe; C: Feuilles palmes; D: Silique.

### I.2.3.2.-Position systématique

MAIRE (1965); OZENDA (1991); classent l'espèce *C. arabica* selon le tableau 2.

**Tableau 02 :** Position systématique de la plante spontané *C. arabica*

| | |
|---|---|
| Règne | Plantes |
| Embranchement : | Spermaphytes |
| Sous-embranchement : | Angiospermes |
| Classe : | Dicotylédones |
| Sous-classe : | Dilleniidae |
| Ordre : | Capparales |
| Famille : | Capparidaceae |
| Genre : | Cleome |
| Espèce : | *Cleome arabica* (OZENDA, 1991) |

### I.2.3.3.-Intérêt socio-économique

En Mauritanie, les feuilles grillées sont cuites en aliments qui se prennent en cas d'infections des reins et du dos et comme aphrodisiaque. Au Niger, les feuilles séchées et réduites en poudre sont ajoutées aux aliments pour leurs propriétés diurétiques, pour provoquer la transpiration ou pour traiter les rhumatismes. En Arabie Saoudite, la plante entière s'utilise pour soigner la gale, la fièvre rhumatismale et l'inflammation (SCHMELZER et GURIB, 2013).

Les chameaux, les chèvres et les moutons n'en mangent que très peu. Les autochtones l'utilisent comme diurétique et contre les rhumatismes (MAIRE, 1933 et SCHMELZER et GURIB, 2013). Elle est utilisée en médecine traditionnelle par les nomades du Sahara comme analgésique des douleurs névralgiques (SHARAF *et al.*, 1992).

### I.2.3.4.-Caractéristique générales de l'espèce végétale

Les composés isolés des parties aériennes comprennent le stigma 4-3 one, le lupéol et le taraxastérol ainsi que les tripterpènes de type dammarane suivants : amblyone, $15\alpha$-acétoxycléoamblnol A, cléoamblynol A et cléomblynol B, cléocarpanol, et cabraléahydroxy lactone. Le lutéoline 3-méthyl éther (une flavone) et son 7-glycoside ont aussi été isolés. Les extraits à éthanol des parties aériennes ont montré une activité antifongique variable contre une série de champignons pathogènes (SCHMELZER et GURIB, 2013).

## I.3.-Préparation des extraits

Après la récolte, les parties aériennes des plantes subissent un séchage à l'air libre à température ambiante afin d'avoir une drogue homogène. Il est indispensable de suivre cette technique de séchage (NADEAU et PUIGGALI, 1995; TOM, 2015).

### I.3.1.-Séchage

Les parties aériennes sont rincées à l'eau de robinet pour éliminer la poussière et toute matière collée sur les feuilles. Ensuite, elles sont étalées sur une plaque mince, dans une chambre aérée à température ambiante. La durée de séchage diffère d'une plante à une autre, mais elle ne dépasse guère les trente jours. Lorsque les parties sont complètement sèches, elles sont coupées en petits morceaux afin de faciliter leur broyage et conservées dans des bocaux en verre pour à la fin être transformée en poudre (KEMASSI *et al.*, 2015).

## I.3.2.- Broyage

A l'aide d'un broyeur électrique, les feuilles subissent un broyage afin d'obtenir une poudre fine et homogène, ensuite la poudre est passée à travers un tamis à mailles inférieur à 0,5mm. Les broyats sont stockés dans des boites en verre et conservés à l'abri de la lumière et de l'humidité. Chaque boite porte le nom de l'espèce, la partie utilisée, la date et le lieu de récolte. Le broyat des feuilles constitue le matériel végétal final utilisé pour la préparation des extraits aqueux.

## I.3.3.-Extraction par reflux

Après le broyage, la poudre végétale subit une extraction par reflux dans une solution (2/3) méthanol et (1/3) d'eau distillée.

L'extrait aqueux est obtenu par solubilisation des fractions actives dans de l'eau distillée et du méthanol. 100 grammes de la poudre végétale sont mis dans un ballon de 1000 ml de capacité avec suffisamment de solution aqueuse de méthanol (2:1) (2/3 de méthanol et 1/3 d'eau distillée). Le ballon est surmonté par un réfrigérant permettant la condensation des fractions volatiles organiques lors de l'extraction. Le mélange est porté à ébullition à 50°C pendant 6 heures. L'homogénat est refroidi et filtré à l'aide d'un papier filtre standard de type FIORONI R0122A00010.

Pour éliminer le méthanol, le filtrat est soumis à une évaporation sous vide à l'aide d'un rotor-vapor pendant deux à trois heures. Le produit obtenu est un extrait aqueux conservé dans un bocal hermétiquement fermé et couvert par du papier aluminium.

## I.3.4.- Choix des concentrations

L'étude de la toxicité des extraits aqueux de *C. arabica* et *P. tomentosa*, récoltée au Sahara septentrional est-algérien vis-à-vis des espèces tests, nécessite de chercher la concentration minimale inhibitrice. De ce fait, dix concentrations en extrait végétal sont choisies respectivement : 100%, 90%, 80%,70%, 60%, 50%, 40%, 30%, 20% et 10% pour les extraits préparés des deux espèces végétales choisies.

## I.4.-Criblage phytochimique

Les différentes réactions chimiques ont pour objectif de caractériser et de rechercher les principaux groupes chimiques dont les alcaloïdes, flavonoïdes, triterpénoïdes, saponosides, glycosides, tanins et les quinones dans les extraits aqueux bruts de *C. arabica* et *P. tomentosa*. Le criblage est fait par une analyse phytochimique qualitative à partir des tests de coloration et/ou de précipitation (HARBONE, 1976).

Les résultats qualitatifs sont classés en :
- o Réaction très positive +++ : présence confirmée ;
- o Réaction positive ++ : présence modérée ;
- o Réaction plus au moins positive + : trace ;
- o Réaction négative - : absence.

### I.4.1.- Test des alcaloïdes

Le test des alcaloïdes est basé sur le phénomène de précipitation avec un réactif spécifique « Mayer» qui est considéré comme révélateur des alcaloïdes.

- **Réactif de MAYER :** soit 10g de iodure de potassium (KI) et 2,7g de chlorure mercurique ($HgCl_2$) dissous dans un 20 ml d'eau distillé.

    **Méthode 1 :** les alcaloïdes sont identifiés par le réactif de Mayer préalablement préparé. L'ajout de quelques gouttes de réactif de Mayer à 2ml d'extrait végétal entraine la formation d'un précipité blanc-jaune indiquant la présence d'alcaloïdes (DEBRAY, 1970).

- **Réactif de BOUCHARDAT :** soit de 2g d'iode bi-sublimé et 2g d'Iodure de potassium (KI) dans 100ml d'eau distillé.

    **Méthode 2 :** le réactif de BOUCHARDAT est également utilisé pour mettre en évidence la présence des alcaloïdes dans les extraits. Il se forme un précipité brun-noir, brun-terre, ou jaune brun avec les alcaloïdes.

### I.4.2.- Test des flavonoïdes

Dans un tube à essai, il est ajouté quelques millilitres d'extrait végétal, quelques gouttes d'une solution de soude (NaOH) au 1/10. L'apparition d'une coloration jaune-orange indique la présence des flavonoïdes (HARBONE, 1976).

### I.4.3.-Test des Saponosides

La détection des saponines est réalisée en ajoutant un peu d'eau à un millilitre de l'extrait végétal. Par la suite, cette solution est fortement agitée. Après 15 min de repos. La détection des saponines se traduit par la persistance d'une mousse d'au moins un centimètre de hauteur après les 15 minutes (HARBONE, 1976).

### I.4.4.- Test des tanins

Dans un tube à essai, 2ml de chaque extrait est ajouté au trichlorure de fer ($FeCl_3$) à 2%. Ce test permet de détecter la présence ou l'absence des tannins. L'apparition d'une couleur brun-noir confirme la présence des tannins galliques (tannins hydrolysables) et la couleur bleu verdâtre pour la présence des tannins catéchiques (ou tannins condensés).

### I.4.5.-Test des stérols et triterpènes

Dans un tube à essais contenant quelques millilitres d'extrait végétal, il est ajouté un millilitre de $H_2SO_4$ (36N.) en inclinant le tube de 45°. L'apparition d'un anneau rouge au niveau de l'interface (la zone de contact entre les deux liquides) indique la présence de stérols insaturés, tandis que l'apparition d'un anneau rouge ou violet indique la présence des triterpènes. Les deux phénomènes peuvent être observés simultanément (BOTOSOA, 2010).

### I.4.6.- Test des Glycosides

Dans un tube à essais contenant 2ml d'extrait végétal, il est ajouté du Liqueur de Fehling A et B. Ensuite, il est chauffé pendant 3 à 5 minutes. La réduction de la liqueur de Fehling montre la présence des glycosides, la couleur devient plus intense confirmant la présence des glycosides.

## I.5.-Tests biologiques

Pour réaliser le présent travail, douze (12) lots sont constitués, dont deux lots témoins positif et négatif et dix lots pour les traitements. Chaque lot constitué est caractérisé par une concentration en extrait végétal de *C. arabica* et *P. tomentosa* soit l'extrait à 100%, 90%, 80%,70%, 60%, 50%, 40%, 30%, 20% et 10% pour chaque lot, trois répétions sont réalisées.

### I.5.1.- Effet insecticide

Les céréales et les légumineuses constituent la principale source de l'alimentation chez les populations humaines. A l'échelle mondiale, les pertes de ces produits agricoles occasionnées par les ravageurs des denrées stockées sont estimées à près de 10% selon les récentes statistiques de la FAO. Ils peuvent causer des pertes importantes en réduisant la qualité et la quantité des produits stockés (GOERGEN, 2005).

L'utilisation des insecticides de synthèse constitue la technique la plus pratiquée pour lutter contre les insectes ravageurs. Cependant, l'emploi abusif de ces insecticides occasionne des intoxications des utilisateurs, des désordres écologiques et l'apparition des espèces résistantes (GUEYE, 2011). Le recours aux produits d'origine botanique apparait comme la meilleure alternative de lutte propre contre ces ravageurs (ABBASSI *et al.,* 2005, SENTHIL-NATHAN *et al*., 2006 ; JBILOU *et al*., 2008).

L'étude de l'effet insecticide est consacré pour l'évaluation des extraits aqueux du *Cleome arabica* L. et *Pergularia tomentosa* L. vis-à-vis des imagos du *Tribolium confusum* Duval (*Coleoptera-Tenebrionidae*).

## I.5.1.1.-Insecte test *Tribolium confusum* J. Duval   (*Coleoptera-Tenebrionidae*)

Les Tenebrionidae sont des coléoptères de taille comprise entre 2mm et 80mm. Cette famille regroupe environ 20000 espèces réparties sur 190 genres. Comme tous les Coléoptères, les espèces de cette famille se caractérisent par des téguments rigides, épais, noir mat ou luisant, de teinte sombre, coloré par interférence, avec des yeux généralement grands, ovales ou ronds chez certaines sous-familles. Par des antennes de 10 à 11 articles, aptères ou ailées (BALACHOWSKY, 1962). Un certain nombre de *Tenebrionidae* ont été signalés comme nuisibles sur les plantes cultivées et d'autres s'attaquent aux denrées alimentaires stockées. Parmi, ces dernières le genre *Tribolium* qui comprend deux espèces principales cosmopolites et nuisibles soit *Tribolium confusum* J. Duval et *Tribolium castaneum* Herbst (SMITH et WHITMAN, 1992).

Les insectes nuisibles aux denrées stockées à long terme ont un cycle de vie classique. L'œuf, pondu par une femelle préalablement fécondée, donne la vie à une larve d'allure vermiforme, qui, après une croissance par étape à l'éclosion de mues successives se métamorphose (stade nymphal), pour donner finalement un adulte qui se reproduira peu de temps après l'émergence. C'est le cas des coléoptères et les Lépidoptères (CLOUTIER, 1992).

L'adulte de *Tribolium confusum* présente une taille moyenne de 3,4 mm et une largeur thoracique 1,02mm. Leur poids varie en fonction du sexe ; le mâle (1,48mg) et la femelle attient 1,78mg (BRINDELY, 1930 ; CROFT, 1990; THOMAS, 2002)

Le régime alimentaire de *T. confusum* est psichophage où le développement est impossible sur les grains entiers, il est très ralenti sur grain simplement éclatés. Leur développement optimal est situé entre 24 et 28°C pour 70 à 75% d'humidité relative. La durée de la phase de l'œuf à l'adulte sur la farine de blé ou de seigle est 39 jours à 30,5°C à 31,5°C.

99 jours à 19°C à 20°C. La période pré-reproductrice dure une ou deux semaines à 25°C à 30°C. La fécondité varie de 239 à 400 œufs à 25 à 26°C. De 236 à526 œufs à 28°C ; la période de reproductrice peut durer plusieurs années. La longévité est très élevée jusqu'à 4ans (CROFT, 1990).

Le cycle de vie de *Tribolium confusum* comprend quatre stades distincts : œuf, larve, pupe et adulte (BENOIT *et al*., 1998).

Une femelle adulte de la farine peut pondre de 300 à 500 œufs dans une vie, et généralement pond 2 à 3 œufs par jour (YOUNG, 1970). À 30°C, un œuf à reformuler 5 à 7 jours peut éclore. Les larves passent généralement 6 à 11stades, avant qu'elles soient suffisamment grandes pour se nymphoser. Chaque phase larvaire entre le délestage des exosquelettes est appelée instar (stade). Les exosquelettes du *Tribolium* peuvent être trouvés dans le milieu. Les coléoptères restent généralement dans la phase larvaire pendant 2 à 3 semaines**.** Les pupes (nymphes) ne se déplacent pas à moins qu'elles ne soient dérangées et ne se nourrissent pas. Le stade nymphal est d'environ 8 jours. Une fois que la pupe devient adulte, elle passe trois à quatre jours dans un stade adulte immature et non reproducteur qui est appelé le stade adulte lâche. Après avoir atteint la maturité, le *Tribolium* peut vivre pendant un à deux ans (RYAN *et al*., 1970).

## I.5.1.2.-Elevage de masse

Dans le souci d'obtenir une population homogène et un nombre suffisant d'individus, un élevage de masse était maintenu dans le laboratoire de la faculté des sciences de la nature et de la vie de l'université de Ghardaïa (Algérie) avec des adultes de *Tribolium confusum Duval*. Les imagos de *Tribolium confusum* proviennent d'un stock de semoule de maison situé dans la région de Ghardaïa. L'élevage de masse est réalisé dans des bocaux en verre (18cm de hauteur et 11cm de diamètre) contenant 500g de semoule commerciale. Les bocaux ainsi préparés sont mis dans une chambre à l'obscurité à une température oscillant entre 32 et 35°C et une humidité relative de 70% conformément à la méthode utilisée par TAPONDJOU *et al*. (2003). Les boites sont laissées environ un mois et plus pour assurer un grand nombre d'individus.

## I.5.1.3.- Application de traitement

L'étude de la toxicité concernant les extraits aqueux de *Cleome arabica* et *P. tomentosa* récoltés au Sahara septentrional est-algérien vis-à-vis des imagos *Tribolium confusum*. Dix concentrations différentes sont préalablement préparées (100%, 90%, 80%, 70%, 60%, 50%, 40%, 30%, 20% et 10%) d'extrait végétal. Pour les lots de témoins, il est utilisé un témoin positif qui correspond à un insecticide de synthèse (Appellation commerciale : HILAC) dilué et un témoin négatif correspondant à de l'eau distillée.

L'insecticide HILAC est un insecticide à large spectre d'action. Il appartient au groupe des pyréthrinoides de synthèse. Le produit HILAC, est présenté sous la forme EC (Émulsion concentrée). Il contient de la Cyperméthrine (50g/l), tétraméthrine (10g/l) photostable pyréthrinoides, utilisés à titre d'agent tuant et photolabile pyréthrinoides (knock-down). HILAC agit particulièrement par contact et aussi par ingestion (RICHOU-BAC et VENANT, 1985; DHMPE, 2017). HILAC est destinée à être diluée dans l'eau soit un litre de HILAC dilué dans 19 litres d'eau.

Un lot constitué de dix (10) imagos de *T. confusum* fraîchement prélevés de leur milieu d'élevage a été introduit dans chaque boîte de Pétri contenant une quantité 5g de milieu de culture (semoule). Ensuite traité par 4ml d'extrait végétal et 1ml pour le témoin positif. Les boites sont ensuite immédiatement refermées. Trois répétitions ont été effectuées pour chaque dose et les individus morts sont comptés (et maintenus dans les boîtes) toutes les 24 heures pendant dix jours (fig. 1).

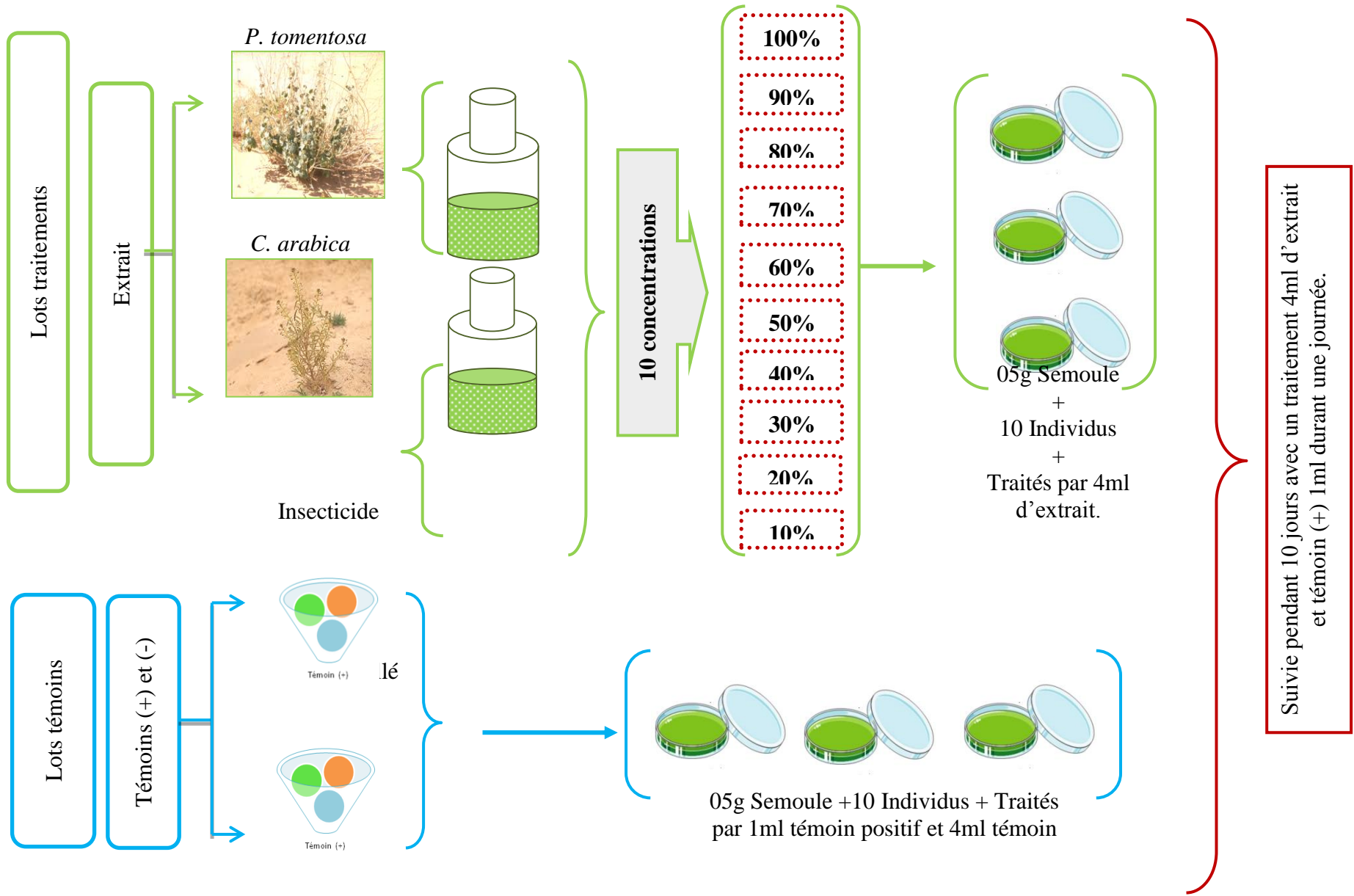

**Figure 1 :** Protocol expérimental.

### I.5.1.4.- Paramètres étudiées

Pour la présente étude, quatre paramètres sont étudiées soient : le taux de mortalité, cinétique de mortalité, dose létale $DL_{50\%}$ et $DL_{90\%}$ et les temps létaux $TL_{50\%}$ et $TL_{90\%}$.

### I.5.1.4.1.- Taux de Mortalité

La mortalité est le premier critère de jugement de l'efficacité d'un traitement chimique ou biologique (KEMASSI *et al*., 2015). Le pourcentage de la mortalité observée chez les imagos de *Tribolium confusum* est estimé en appliquant la formule suivante :

Mortalité observée = (Nombre des individus morts/ Nombre total des individus) ×100%

OULD EL HADJ *et al*., (2006), KEMASSI *et al*., (2010) (2012) (2015)

### I.6.2.4.2.- Dose létale ($DL_{50\%}$, $DL_{90\%}$)

La $DL_{50\%}$ est la quantité d'une matière administrée en une seule fois, qui cause la mort de 50% d'un groupe d'insectes d'essais. La $DL_{50\%}$ est une façon de mesurer le potentiel toxique à court terme (toxicité aigüe) d'une matière. La méthode de FINNEY (1971) basée sur la régression des Probit de mortalités en fonction des logarithmes des doses a permis de déterminer la $DL_{50\%}$ et $DL_{90\%}$.

### I.6.2.4.3.-Temps létaux ($TL_{50\%}$ et $TL_{90\%}$)

Le temps létal correspond au temps nécessaire pour que 50% des individus d'une population meurent suite à un traitement par une substance quelconque. Il est calculé à partir de la droite de régression des Probit correspondants au pourcentage de la mortalité corrigée en fonction des logarithmes du temps de traitement. Il est utilisé la formule de SCHNEIDER (TEDONKENG PAMO *et al*., 2002) et la table des Probit.

Formule de SCHNEIDER (in KEMASSI *et al*. 2014 ; 2015) :

Mortalité corrigée $=\left(\frac{M_o - M_t}{100 - M_t}\right) \times 100\%$

$M_o$ : mortalités cumulées dans les boîtes traitées.

$M_t$ : mortalités cumulées observées dans les boîtes témoins.

### I.6.2.4.4.-Analyse statistique (analyse de la variance "ANOVA")

Les résultats sont comparés par ANOVA (Analysis of variance) en utilisant le logiciel «MINITAB version 15 English.Ink. Selon BERK et STEAGALL (1995), le test d'analyse de variance est un test de comparaison de deux variances ; la variance intergroupe en fonction de la variance intragroupe. Si la variation intergroupe est plus élevée que la variation intragroupe, les deux groupes sont significativement différents. Une probabilité inférieure à 0,01 correspond à un effet hautement significatif. Pour une probabilité de 0,05, l'effet est significatif. Si celle-ci est supérieure à 0,05, l'effet est considéré comme non significatif.

**I.5.2.- Effet herbicide**

Dans le cadre de la valorisation des ressources phyto-génétiques et la diversité de la flore du Sahara septentrional est-algérien, une partie de l'étude vise une recherche des effets allélopathiques des extraits aqueux préalablement préparés de *C. arabica* et de *P. tomentosa* sur la germination des graines des espèces tests. Il sera entamé la présentation des espèces tests et le protocole expérimental adopté pour le test biologique.

**I.5.2.1.- Espèces tests**

L'essai consiste à évaluer l'effet inhibiteur des extraits aqueux de *C. arabica* et *P. tomentosa* vis-à-vis des graines de mauvaises herbes pour la culture céréalière *Dactyloctenium aegyptium* (*Poaceae*) et deux espèces cultivées dans les champs des céréales soient *Hordeum vulgare* et *Triticum durum* (*Poaceae*). Les espèces sont monocotylédones de la famille des poacées. Les semences sont achetées dans le commerce d'un même lot dans la région de Ghardaïa. Elles ne doivent pas être traitées aux fongicides ni aux insecticides et elles doivent être homogènes en taille. Les semences sont conservées pour une durée maximale d'un an jusqu' à expérimentation.

**I.5.2.2.- Tests biologiques**

La réalisation d'un test de toxicité nécessite des conditions environnementales optimales au cours de la période d'exposition, lesquelles sont adaptées spécifiquement à l'organisme biologique utilisé afin d'éviter de fausses réponses positives.

Tous les tests de germination sont réalisés dans les pots en plastique à 10cm de diamètre et d'une hauteur de 10cm. Chaque pot est étiqueté avec une étiquette portant la dose de traitement et le nom de l'espèce test, la date de début traitement (référence du lot).

Quatre lots sont constitués, dont le premier est le témoin négatif (eau distillé) et témoin positif (herbicide Ghyphon) (deuxième lot). Le troisième lot est pour l'extrait de *C. arabica* et le quatrième pour l'extrait de *P. tomentosa*. Ainsi, dix concentrations sont maintenues pour chaque extrait soient 100%, 90%, 80%, 70%, 60%, 50%, 40%, 30%, 20% et 10% et trois pots pour chaque concentration.

L'herbicide choisi le (Glyphon), est un herbicide systémique non sélectif, utilisé dans la lutte contre les adventices annuels, biannuel et pérennes des cultures céréalières. Le Glyphon est utilisé par pulvérisation foliaire, qui lui permet de pénétrer à l'intérieur de la mauvaise herbe et diffuser jusqu'à atteindre les racines. L'herbicide utilisé contenant 480g/l de Glyphosat, pour une dose d'application normalisée de l'ordre de 1,5 litre de Glyphon par hectare.

Pour préparer un milieu favorable à l'ensemencement, des quantités de sable de dune ont été échantillonnées de la zone sableuse de Brizina (Oued Metlili région de Ghardaïa). Les échantillons du sol ont été tamisés par un tamis de 4mm (pour éliminer les grosses particules).

Au cours de la préparation des pots pour l'ensemencement, une quantité de 400g du sable de dune est déposée par pot, cette quantité correspond à une épaisseur de sable de 4cm.

Après la détermination de la quantité de sable et de l'eau convenable pour les tests, et après la préparation des pots, et à l'aide d'une pince, la semence des espèces végétales tests est insérée à une profondeur de 1cm. Chaque pot doit contenir 15 graines déposées de façon homogène en respectant un certain espacement.

Après la préparation des lots (pots ensemencés), à l'aide d'une pipette graduée, 4ml d'extrait végétal sont additionnés à 100ml d'eau. Cette solution est utilisée pour l'irrigation des pots uniquement le premier jour du traitement. Alors que pour les pots du témoin négatif, il est irrigué par 104ml d'eau. Pour le témoin positif, une quantité de 3ml d'herbicide est ajoutée à 100ml d'eau (cette dose est estimée en fonction des normes d'application pratique de cet herbicide dans les champs de céréales). Afin d'éviter le dessèchement du sable dans tous les lots, une quantité d'eau de 4ml est additionnée quotidiennement.

Vu que l'objectif se limite essentiellement à l'effet de ces traitements sur la germination des graines, la durée de suivi expérimental est limitée à dix jours, en notant chaque jour le nombre des graines germées et toute sorte d'anomalie de croissance post-germination (fig. 2).

Une foi la période de suivi expérimental (10 jours) achevée, des mesures morpho-métriques sont effectuées soit la longueur de la racine et de la partie aérienne, le poids frais des différents organes, l'observation des anomalies de croissance (flétrissement, jaunissement ou noircissement des extrémités du feuillage et le dessèchement) (Annexe I).

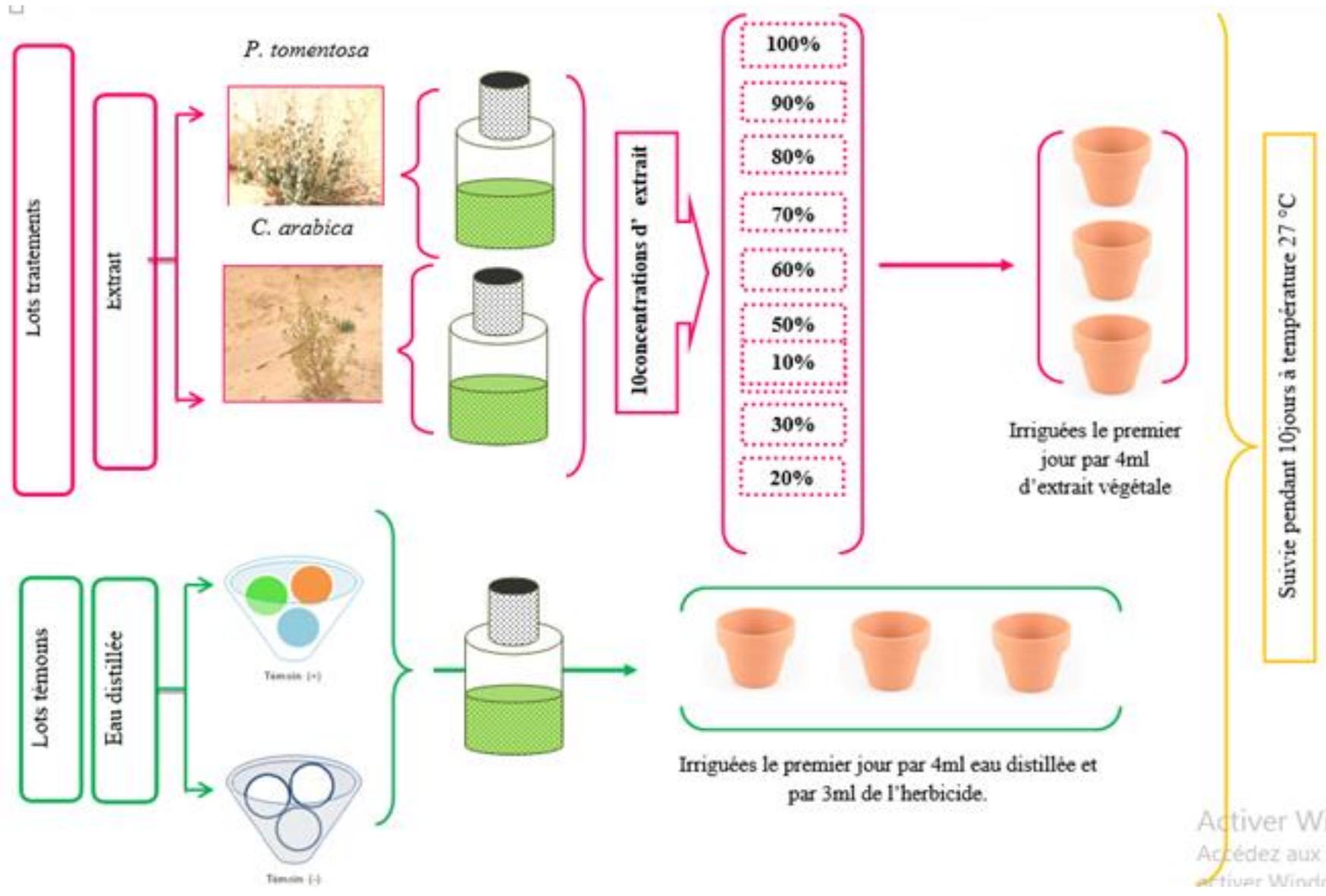

Figure 02- Dispositif expérimental.

### I.5.2.3.- Traitement des résultats

Afin de mettre en exergue les effets des traitements effectués sur la germination des graines des espèces végétales maintenues et leur croissance, de nombreux paramètres sont estimés.

### I.5.2.3.1.- Taux d'inhibition de la germination

La capacité d'un extrait ou d'une substance à inhiber la germination des graines est exprimée par la relation suivante :

$$\text{Taux d'inhibition} = \left[\frac{\text{Nombre des graines semées} - \text{Nombre des germées}}{\text{Nombre des graines semées}}\right] \times 100\%$$

(CÔME, 1970)

### I.5.2.3.2.- Cinétique de germination

La cinétique de germination représente la variation du taux de germination des graines de la plante testée au fil du temps. Graphiquement, il est présenté par une courbe de pourcentage de germination en fonction du temps. Il permet une vision précise de l'évolution de la germination d'un lot de semences placé dans des conditions spécifiques (CÔME, 1970; ENNACERIE *et al*., 2018).

### I.5.2.3.3.- Concentration d'efficacité (CE)

La concentration d'efficacité à 50% ($C.E._{50\%}$) est la quantité d'une matière pouvant induire un pourcentage de succès de 50% de la population traitée. En d'autre terme, c'est celle qui provoque une inhibition de la moitié (50%) des graines traitées. La concentration d'efficacité est une façon de mesure du potentiel toxique à court terme (toxicité aiguë) d'une matière donnée. Les concentrations d'efficacité sont estimées selon la méthode des Probit. Plusieurs auteurs ont utilisés cette même méthode pour l'évaluation quantitative de la toxicité de leurs extraits végétaux dont DJEUGAP *et al*. (2011) et KEMASSI *et al*. (2014; 2015).

### I.5.2.3.4.- Longueur et poids des parties aériennes et souterraines

Pour les mesures de la croissance en longueur et en poids, aucune valeur n'est attribuée aux graines n'ayant pas germé. Les longueurs moyennes de la partie aérienne et souterraine des graines germées dans les lots témoins et traitements sont mesurés à l'aide d'une règle graduée et/ou d'un papier millimétré. La détermination du poids des parties aériennes et souterraines est effectuée par une balance de précision à 0,0001g (MOUKRAD *et al*., 2014; ENNACERIE *et al*., 2018).

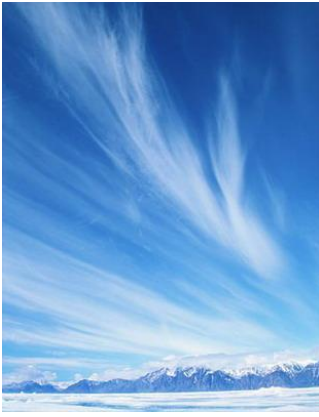

*Chapitre II*

# Résultats et Discussion

# Chapitre II- Résultats et discussion

## II.1.- Criblage phytochimique

Les tests de caractérisation phytochimique des extraits aqueux obtenus de *Cleome arabica* et de *Pergularia tomentosa,* permettent d'obtenir une vision qualitative de la présence ou l'absence des molécules bioactives retrouvées au niveau des deux extraits. Il est noté que les extraits sont riches en métabolites secondaires dont les flavonoïdes, Saponosides, glycosides et les terpènes et une quantité moyenne des stérols, désoxyose, polyphénols et des alcaloïdes totaux. Pour les deux plantes choisies, il y a absence de quinones, de tanins galliques et de polyphénols. L'extrait foliaire de *C. arabica* est plus riche en métabolites secondaires par rapport aux extraits foliaires de *P. tomentosa* (tab. 3).

**Tableau 3.-** Criblage phytochimique de deux plantes spontanés
(- : absence, + : faible présence, ++ : présence moyenne, +++ : forte présence)

| Plantes | Groupes chimiques | | | | | | | | | | |
|---|---|---|---|---|---|---|---|---|---|---|---|
| | | | | Tanins | | | | | | | |
| | Stérols | Poly-phénols | Flavonoïdes | Catéchiques | Galiques | Saponosides | Triterpènes | Glycoside | Quinones | Désoxyoses | Alcaloïdes totaux |
| *Cleome arabica* | +++ | + | +++ | ++ | - | +++ | +++ | +++ | - | ++ | +++ |
| *Pergularia tomentosa* | ++ | + | ++ | + | - | ++ | ++ | ++ | - | ++ | ++ |

Il se distingue deux groupes de métabolites: les métabolites primaires qui sont des molécules organiques, se trouvent dans toutes les cellules de l'organisme d'une plante pour y assurer sa survie (glucides, protéines, lipides et acides nucléiques) et les métabolismes secondaires qui n'exercent aucune fonction directe au niveau des activités fondamentales de l'organisme végétale (croissance, développement, reproduction…etc.) mais peuvent jouer différentes rôles pour la survie du végétal lui-même, rôle de défense, rôle de résistance (GUIGNARD, 2000).

Les résultats du criblage phytochimique des extraits bruts qui se base sur des réactions de coloration et/ou de précipitation, montrent que l'extrait aqueux de *C. arabica* est plus riche en métabolites secondaires dont les alcaloïdes, flavonoïdes, terpènes, stérols et tanins par rapport à l'extrait du *P. tomentosa*.

Les métabolites secondaires sont classés en trois grands groupes soit les alcaloïdes, polyphénols (flavonoïdes, tanins, acides phénoliques…etc.) et les terpènoïdes et leurs dérivés (BRUNETON, 1999). Les alcaloïdes notamment constituent un vaste groupe de substances secondaires, près de 12000 des alcaloïdes sont inventoriés. En raison de leurs propriétés d'origine végétale, ils sont toujours présentés dans les produits pharmaceutiques (MERGHEM, 2009).

La teneur des alcaloïdes dépend de divers facteurs, elle est souvent due au caractère variétale, n'étant pas la même dans divers espèce du même genre. Elles sont susceptibles de fluctuations liées aux conditions de croissance par exemple l'exposition au soleil ou à l'ombre. Elle est également fonction de l'âge de la plante (MERGHEM, 2009). Les flavonoïdes représentent une autre classe de métabolites secondaires largement répandue dans le règne végétal. Ce sont des pigments quasiment universels des végétaux qui sont en partie responsables de la coloration des fleurs, des fruits et parfois les feuilles. Le terme flavonoïde regroupe une très large gamme de composés naturels polyphénoliques (STÖCKIGT *et al*., 2002).

Le genre *Cleome* est représenté par environ 200 espèces qui présentent des utilisations thérapeutiques, alimentaires ou artisanales. Plusieurs études menées sur une quinzaine d'espèces du genre *Cleome* ont montré la présence de différentes classes de métabolites secondaires dont des polyphénols, des terpènoïdes, des alcaloïdes et des métabolites primaires (PHAM, 1999).

MADI *et al*. (2017); ont isolés au niveau des feuilles, des graines et des racines de *C. arabica* de grands groupes chimiques, qui permet de ressortir que l'extrait foliaire est plus riche en métabolite secondaire par rapport aux extraits issus des graines et des racines. L'extrait des feuilles révèle la présence d'alcaloïdes, de flavonoïdes, de saponines, de triterpènes, de tannins, de stéroïdes et de coumarines.

DJERIDAN *et al.* (2010) confirment la présence de stéroïdes et de flavonoïdes dans l'extrait foliaire de *C. arabica.* TAKHI *et al.* (2011) signalent la présence des alcaloïdes dans l'extrait de la plante. HARRAZ *et al.* (1995) notent que *C. amblyocarpa*, une sous-espèce de *C. arabica*, présente des terpènoïdes dans sa composition. SHARAF *et al.* (1997) confirment la présence de flavonoïdes dans la même espèce.

Les résultats de criblage phytochimique montent la présence des saponosides, des tanins, des flavonoïdes, des alcaloïdes, des terpènoïdes, des quinones. BOULENOUAR (2011) et TLILI *et al*. (2015) signalent la présence de flavonoïdes, de tannins, de coumarines, et d'alcaloïdes dans la partie foliaire de *Pergularia tomentosa* (Asclepiadaceae).

Le screening phytochimique montrent une richesse en métabolites secondaires soit des flavonoïdes, des saponosides, des glycosides, des terpènes, des stérols et des alcaloïdes totaux pour *C. arabica* et *P. tomentosa* (tab. 3).

## II.2.- Effet insecticide

Les bio-essais ont été effectués au laboratoire dans des boîtes de Pétri de 9cm de diamètre à une température moyenne de 24±1°C et une humidité relative de 70±5%%, pendant dix jours. Il est noté quotidiennement le nombre des individus morts et toutes anomalies ou mals formations observées chez les populations traitées ou non traitées.

Il faut signaler que les teneurs en matière sèche des solutions mères des deux extraits appliqués sur les insectes test sont différentes. Elles sont de l'ordre de 3,15% pour *Cleome arabica* et de 1,51% pour l'extrait de *Pergularia tomentosa*. De ce fait, la concentration définie par le pourcentage ne correspond guère et ne permet pas une réelle comparaison de la toxicité entre les deux extraits, seule l'estimation de la concentration d'efficacité permet cette comparaison.

### II.2.1.- Effet sur le taux de mortalité cumulée

La figure 3 illustre l'évolution des pourcentages des mortalités cumulées au niveau des lots témoins et traités par les extraits foliaires de *C. arabica* et *P. tomentosa*. Au vu des résultats illustrés sur la figure 3, il est noté une variation du taux de mortalité en fonction de la dose de l'extrait appliqué. Pour les lots traités par l'extrait foliaires de *C. arabica*, les plus fortes doses soient 100% et 90%, engendrent une mortalité remarquable. Elle est de 96,67±0,58% et 90±1,00% respectivement. Pour les autres concentrations appliquées dont 80%, à 70%, 60% et 50%, les taux de mortalité cumulée rapportés sont respectivement de l'ordre de 73,33±0,58%, 46,67±0,58%, 43,33±0,58% et 33,33±0,58%. Toutefois, pour les faibles concentrations soit 30%, 20% et 10%, les pourcentages de mortalités cumulées oscillent entre 10±0,00 et 20±0,00%.

L'analyse de la variance à un critère de classification montre une différence très hautement significative au niveau des lots traités par les concentrations dont 100%, 90%, 80%, 70%, 60%, 50%,

40% et 30% de *Cleome arabica* par à apport aux lots témoins. Les valeurs de facteur F sont notés pour une probabilité inferieure (P<0,001) (Annexe 2).

Pour le lot traité par la concentration 20% de l'extrait *C. arabica,* une différence hautement significative est rapporté (F=25.00 ; P= 0,007) et elle est non significative pour le lot traité par l'extrait à 10% de concentration.

Bien que pour les individus traités par l'extrait foliaires de *P. tomentosa*, il est enregistré des taux de mortalités inférieurs à ceux rapportés chez les individus traités par l'extrait de *C. arabica*; ils sont de l'ordre de 86,67±0,58%, 63,33±0,58%, 36,67±0,58%, 33.33±0,58%, 16.67±0,58%, 13,33±0,58%, 13,33±0,58%, 10,0±0,00%, 6,67±0,58% et 3,33±0,58% pour les concentrations en extrait foliaire de *P. tomentosa* à 100%, 90%, 80%, 70%. 60%, 50%,40%, 30%, 20% et 10% respectivement.

De même pour les lots traités par l'extrait aqueux de *P. tomentosa,* l'analyse de la variance des pourcentages de mortalité enregistrés chez les imagos de *T. confusum* traités par l'extrait aux concentrations 100%, 90%, 80% et 70% par rapport aux valeurs témoin négatif, une différence très hautement significative est notée, les valeurs du facteur F enregistrées sont pour une probabilité inferieure à (P<0,001) (Annexe 2).

Elle est hautement significative (F= 25,.00 ; P= 0,007) pour les individus traités par la concentration 60%.Tandis que, chez les imagos traités par les concentrations dont (50%, 40%, 30%), une différence significative (P≤0,05) est notée. Alors que pour les autres concentrations soient 20% et 10%, elle est non significative (F=4,00 ; P= 0,116) pour la première et de (F=1,00 ; P=0,374) pour la seconde (Annexe 2).

En revanche, aucune mortalité n'est observée chez les individus du lot témoin négatif (eau distillée), bien que les individus traités par l'insecticide HILAC$_{(EC)}$ (témoin positif), le taux de mortalité est de 100%.

Suite aux résultats obtenus, il est constaté que les deux extraits végétaux appliqués montrent des possibilités d'utilisation comme un bio-insecticide exceptionnel. Les extraits appliqués aux fortes concentrations (100%, 90% et 80%) engendrent des taux de mortalités notables, peuvent être comparables à ceux obtenus par le traitement insecticides utilisé comme témoin positif. De même, au cours de la période de suivi expérimental, des symptômes d'intoxications sont observés dont la réduction de l'activité motrice et des mouvements incohérents. Pour les lots traités par les extraits à 100%, 90%, 80%, 70%, 60%, 50%, 40%, 30% et 20%, les individus survivant meurent au bout 10jours (mortalité retardé). Par contre pour certains individus exposés aux extraits à la plus faible concentration (10%), des accouplements, des pontes et même des larves néonates sont observés, cela témoigne probablement de l'effet faible ou limité des extraits à cette dose sur les imagos de *Tribolium confusum*.

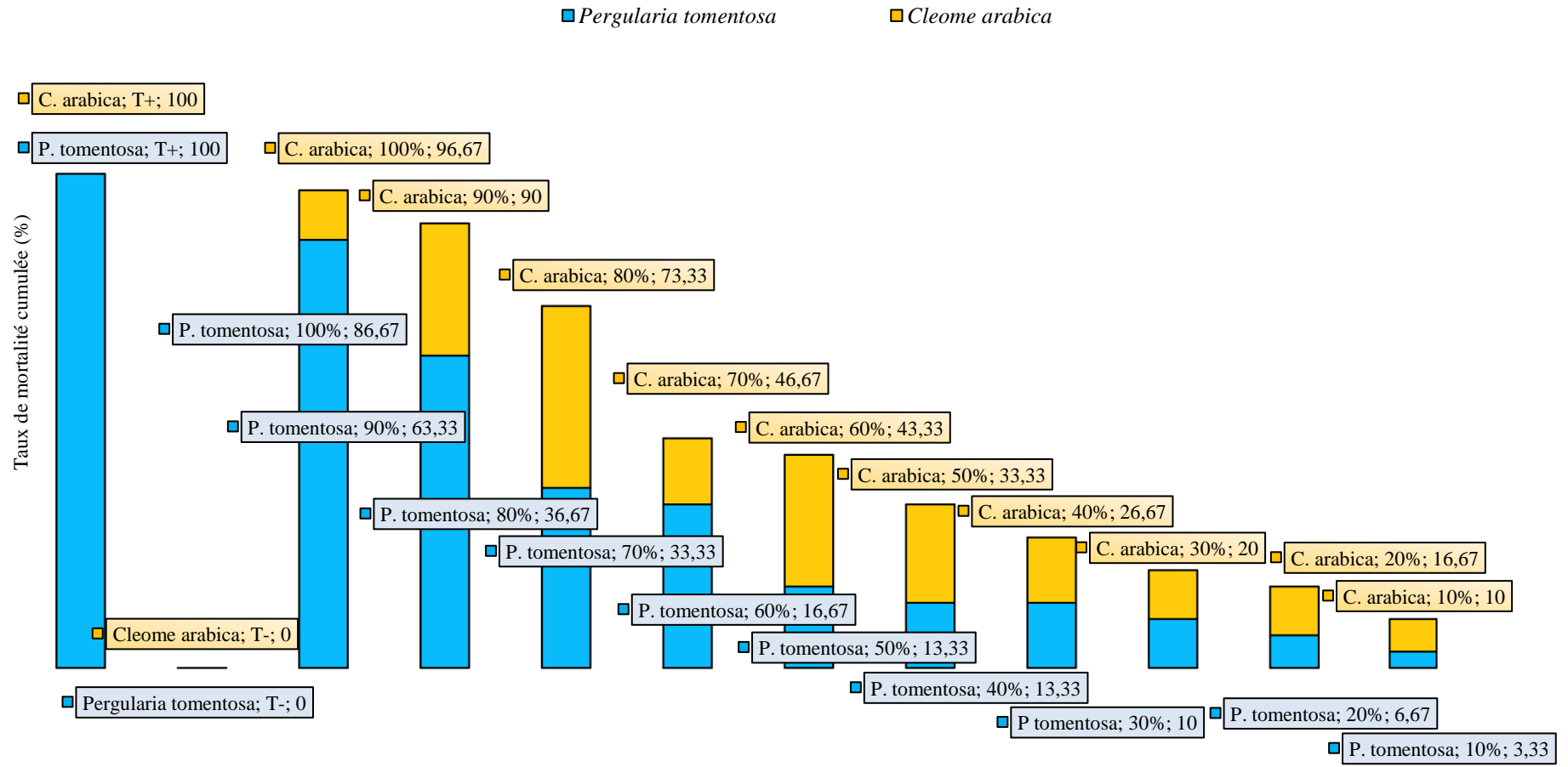

**Figure 3 :** Variation des pourcentages des mortalités cumulée des adultes de *T. confusum* en fonction de la dose d'extrait des feuilles de *C. arabica* et *P. tomentosa*.

*Tribolium confusum* est un insecte qui se déplace rapidement lorsqu'il est dérangé. Son corps de forme allongée est recouvert par une cuticule qui lui donne un aspect rigide et le protège contre les agressions anthropiques et/ou naturelles. Ces caractéristiques biologiques lui permettent de s'adapter aux conditions rudes. Pendant la période d'exposition des insectes aux extraits végétaux, les individus traités sont actifs, ce qui témoigne que le traitement appliqué ne présente guère d'effet par contact. De ce fait, la présence de cette cuticule rigide ''la sclérite'' qui couvre les pattes, appendices et l'abdomen de l'insecte étudié, permet de résister au traitement appliqué. La présence de ce revêtement rigide est un obstacle structural vis-à-vis des produits appliqués selon le mode du traitement adopté (GENEVIEVE, 2006).

*Pergularia tomentosa* (*Asclepiadaceae*) contient un latex corrosif blanc laiteux sécrété par des cellules spécialisées, lors des lésions tissulaires et ne joue aucun rôle dans le métabolisme primaire (GOYDER, 2006 ; AGRAWAL et KONNO, 2009). Le latex contient une variété de produits chimiques et de protéines, tels que divers terpènoïdes, alcaloïdes et cardénolides, ainsi que diverses protéines et enzymes, telles que les protéases, les chitinases et les glucosidases (KONNO, 2011). MILADI *et al.* (2018) confirment que le latex de *P. tomentosa* présente un effet insecticide contre *Locusta migratoria* (*Orthoptera-Acrididae*). Le traitement des larves de $4^{ème}$ stade nouvellement émergées a entraîné une mortalité significative qui atteint 96,49 ± 6,07%.

NGAMO et HANCE (2007), déclarent que la plante *Boscia senegalensis* (*Capparidaceae*) est parmi les espèces de plantes tropicales testées pour leurs activités insecticides sur les ravageurs des céréales *Sitophilus zeamais* et *Tribolium castaneum*. KEMASSI *et al.*, (2012), enregistrent un taux de mortalité de 33,33% chez les imagos de *Schistocerca gregaria* (Forskål, 1775) (*Orthoptera-Acrididae*) traités par l'extrait foliaire brut de *Cleome arabica* (*Capparidaceae*).

BOUNECHADA et ARAB (2011) montrent que la pulvérisation directe de la poudre de *Peganum harmala* L. (*Zygophyllaceae*) à la dose 30% engendre une mortalité de 100% chez les imagos de *Tribolium castaneum* Herbst (*Coleoptera- Tenebrionidae*).

Dans ce même contexte, HASSAINE (2014), note qu'une quantité de 0,6g de la poudre des feuilles de *Tetraclinis articulata* L. (*Cupressaceae*) mélangée avec 10g de semoule de blé provoque 70% de mortalité des imagos de *Tribolium castaneum* Herbst. (*Coleoptera-Tenebrionidae*).YAHAOUI (2005) dans son travail sur l'efficacité insecticide par contact des huiles essentielles de la menthe verte *Mentha spicata* (*Lamiaceae*) sur le *Tribolium confusum* Duval (*Coleoptera-Tenebrionidae*) et *Rhyzopert hadominica* L. *(Coleoptera-Bostrichidae)* à la dose de

3,12% rapporte que cet extrait végétal est très toxique vis-à-vis des deux espèces d'insecte; une mortalité de 100% est rapportée.

## II.2.2.-Effet sur la cinétique de la mortalité cumulée

Les taux journaliers de la mortalité cumulée des individus *T. confusum* témoins et traités par les extraits foliaires de *C. arabica* et *P. tomentosa* sont illustrés sur la figure 4 $_{(A,B)}$.

Chez les imagos traités par l'extrait *C. arabica* (figure 4$_A$). Les résultats montrent un début de mortalité, après 48 heures de l'application de traitement pour les doses 100%, 90% et 80%. Pour les doses testées 70%, 60%, 50%, 40%, 30%, 20%, l'effet de la mortalité commence 96 heures après l'application des extraits sur les imagos de *T. confusum*, à l'exception de la dose 10% où la mortalité enregistrés que après de 168 heures de traitement.

Pour les doses d'extraits végétaux de *Cleome arabica* de 60%, 50%, 40%, 30%, 20%, le taux de mortalité est de 3,33% au bout du 5$^{ème}$ jour (144heures). Pour les fortes doses soient 100%, 90% et 80%, des pourcentages de moralités de 10%, 6,67%, 6,67% sont rapportées après 48 heures de traitement. Chez Certains individus traités à faible dose (10%), un retard de 8 jours dans l'apparition des premières formes de mortalité est signalé.

Pour les individus traités par l'extrait foliaires de *P. tomentosa*, les résultats illustrent un retard de mortalité en comparaison avec les individus traités par les différentes concentrations de *C. arabica*. Au vu des résultats présentés sur la figure 4$_B$, la mortalité des individus de *T. confusum* a commencé après le 4$^{ème}$ jour (96 heures) du traitement pour la majorité des doses appliquées, à l'exception des faibles doses (30%, 20% et 10%) où la mortalité débute à partir du 8$^{ème}$jour d'exposition.

Le taux de mortalité attient 3,33% au bout 5$^{ème}$ jour (144heures) pour les doses 80%, 70%, 60%, 50%, 40%. Pour les fortes doses soient 100% et 90%, des taux de mortalité de 16,67 et 10% sont observés respectivement après 96 heures. Chez les individus exposés aux faibles doses dont 30%, 20% et 10%, un retard de mortalité est observé, les premiers cas de mortalité n'apparaissent qu'après 9 jours d'exposition.

En revanche, aucune mortalité n'est enregistrée chez les individus traités par l'eau distillée (témoin négatif), pendant la période de suivi expérimentale. Les individus traités par l'insecticide HILAC$_{(EC)}$ (témoin positif), présentent un taux de mortalité atteint 100% au bout 48 heures. Il est utile de signaler que l'effet insecticide des extraits aqueux de *C. arabica* et *P. tomentosa* semble notable pour les fortes concentrations (100%, 90% et 80%). La sensibilité des imagos de *T. confusum* vis-à-vis des extraits de deux plantes testées est bien marquée.

L'extrait de *C. arabica* et de *P. tomentosa* possède des possibilités insecticides à l'égard de *T. confusum*. Son effet toxique se traduit par la mortalité ou par des signes d'intoxication chez les individus traités. La toxicité de l'extrait est d'autant plus élevée que les doses sont importantes.

En effet, l'étude de l'action de l'extrait aqueux de *C. arabica* sur les individus de *T. confusum*, montre que son action est rapide aux fortes concentrations soient 100%, 90% et 80% avec un taux de mortalité oscillent entre 73,33% et 96,67%. BOUZOUITA *et al*., (2008) confirment que les tests réalisés avec les huiles essentielles à une concentration 0,1% de *Juniperus phoenicea* L. (*Cupressaceae*), ont donné 90% de mortalité par ingestion des adultes *Tribolium confusum*. NDOMO *et al*., (2009) signalent un taux de mortalité de 97,5% chez les bruches *Acanthoscelides obtectus* Say (*Coleoptera-Bruchidae*) au bout du quatrième jour après un traitement par la forte dose 0,40µl/g des feuilles de *Calliste monviminalis* (*Myrtaceae*). HASSAINE (2014), preuve que le traitement des adultes *Tribolium castaneum* Herbst par la poudre des feuilles de *Tetraclinis articulata* (*Cupressaceae*) et *Pistacia lentiscus* (*Anacardiaceae*) à 0,6g provoque 70% de mortalité au bout de sixième jours respectivement.

KEMASSI *et al*., (2012) dans son étude sur la toxicité des extraits de trois plantes *Peganum harmala* (*Zygophyllaceae*), *Citrillus colocynthis* (*Cucurbitaceae*) et *Cleome arabica* (*Capparidaceae*) vis-à-vis des larves du $5^{ème}$ stade et des adultes de *Schistocerca gregaria* (*Orthoptera-Acrididae*) révèle une mortalité de l'ordre de 16,66% pour *P. harmala*, 16,66% pour *C. colocynthis* et de 33,33% pour *C. arabica* respectivement au bout du $14^{ème}$ jours pour les larves $L_5$ et pour les adultes. KEMASSI et *al*. (2013) signalent un taux de mortalité de 100 % chez les larves de $5^{ème}$ stade de *Schistocerca gragaria* à partir de 14 jours après un traitement par l'extrait d'*Euphorbia guyoniana* (*Euphorbiaceae*) à fortes doses, récoltée au Sahara septentrional, est-Algérien.

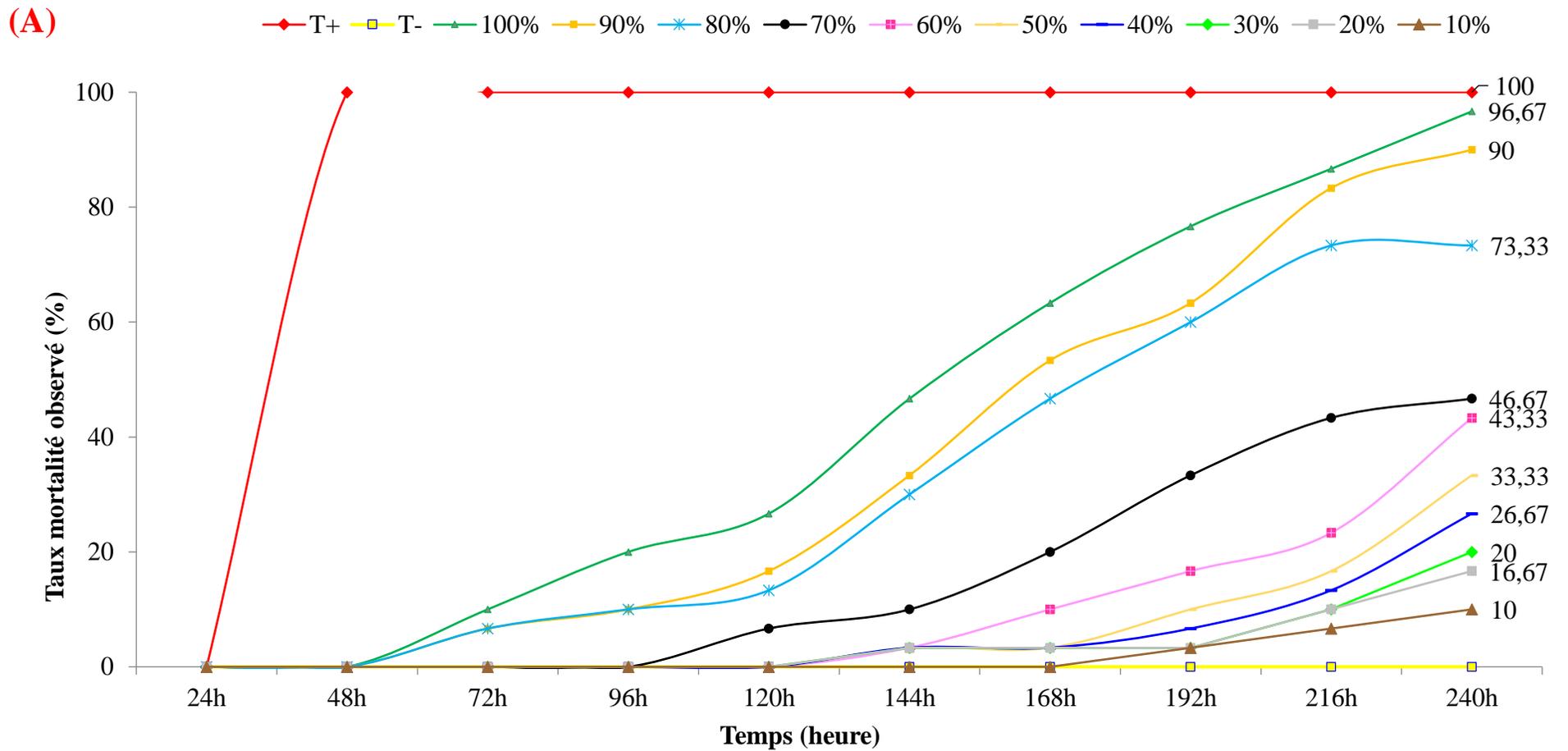

**Figure 4 (A) :** Cinétique de mortalité cumulée observée chez les adultes *T. confusum* témoins et traités par les extraits végétaux de deux plantes récoltées au Sahara Algérien **A :** Extrait foliaire aqueux de *C. arabica*

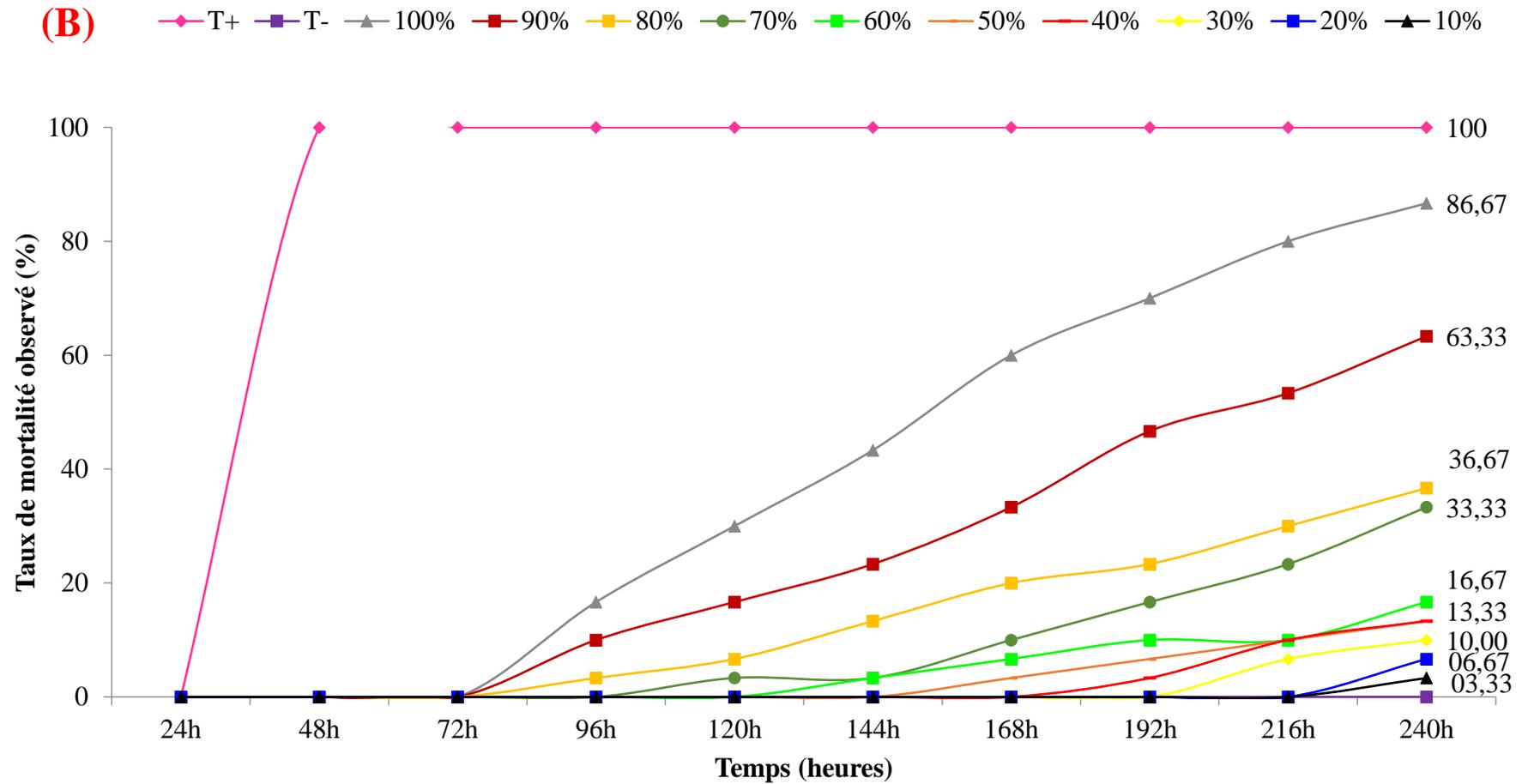

**Figure 4(B) :** Cinétique de mortalité cumulée observée chez les adultes *T. confusum* témoins et traités par les extraits végétaux de deux plantes récoltées au Sahara Algérien. **B :** Extrait foliaire aqueux de *P. tomentosa*

Selon OULD EL HADJ *et al*. (2007), l'étude de la toxicité des extraits de *Melia azedarach* (*Meliaceae*), d'*Azadirachta indica* (Meliaceae) et d'*Eucalyptus globulus* (Myrtaceae) vis-à-vis des larves du $5^e$ stade et des adultes de *S. gregaria* révèle une mortalité de 100 % au bout de 10 jours pour les $L_5$ et de 13 jours pour les adultes traités avec *A. indica*. Pour *M. azerdarach*, elle est de 11 jours pour les $L_5$ et 14 jours pour les adultes. Les individus traités à l'*Eucalyptus* meurent quelques jours plus tard. Les larves du $5^e$ stade s'avèrent être plus sensibles que les imagos à ces extraits

Environ 200.000 espèces appartiennent à la famille des *Asclepiadaceae* réparties dans les régions tropicales et tempérées du monde (ARRIBERE *et al*., 1998). La famille est réputée pour ses plantes contenant des cardénolides, telles que les *Pergularia tomentosa* L., *Solenostemma argel* et *Calotropis procera* Ait (NENAAH, 2013). En effet, dans une étude réalisée par ACHEUK *et al*., (2014), une forte activité insecticide de l'extrait éthanolique brut de *Solenostemma argel* (*Asclepiadaceae*) vis-à-vis les larves $5^{ème}$ stade de *Locusta migratoria*, un taux de mortalité atteint 100% au bout de 14 jours pour la faible dose ($300\mu g/L_5$), tandis que la forte dose ($3000\mu g/L_5$) provoque 100% de mortalité après 24 heures. Pour la dose moyenne ($1000\mu g/L_5$), la mortalité est attient de 100% au $11^e$ jour.

## II.2.3.-Evaluation de la dose létale ($DL_{50\%}$, $DL_{90\%}$)

Les doses létales $DL_{50\%}$ et $DL_{90\%}$ pour les deux extraits bruts de *C. arabica* et de *P. tomentosa* sont estimées à l'aide des équations de régressions (Fig. $5_{A, B}$)

D'après les résultats obtenus, il apparait que les valeurs de $DL_{50\%}$ et $DL_{90\%}$ obtenues pour l'extrait brut de *C. arabica* sont successivement d'ordre 0.16mg/ml et de 0,051mg/ml, bien qu'elles soient de l'ordre de 0,14mg/ml et 0,048mg/ml, pour l'extrait brut de *P. tomentosa* respectivement (tabl. 5).

La partie aérienne de *Pergularia tomentosa* (feuilles) s'avère toxique pour les imagos de *Tribolium confusum*. De ce fait, les composantes chimiques de cet extrait expliquent son action insecticide vis-à-vis l'espèce test. En effet, les deux extraits aqueux de *C. arabica* et *P. tomentosa* sont riches en métabolites secondaires dont les alcaloïdes, flavonoïdes, triterpénoides, tanins, glucosides et les saponosides. Les effets insecticides de ces constituants sont mentionnés dans la première partie (criblage phytochimique).

Selon SCHMELZER et GURIB-FAKIM (2013), toutes les parties de la plante *P. tomentosa* sont des sources en hétérosides cardénolides, avec l'uzariénine. Les racines sont riches en hétérosides et dérivés. Les feuilles et le latex contiennent des ghlakinosides et de la pergularine.

La plante *P. tomentosa* contient également des alcaloïdes, des polyphénols, des terpènoïdes, des flavonoïdes, des coumarines, des anthraquinones et des tanins.

BOUNECHADA et ARAB (2011), signalent que l'activité insecticide de *Melia azedarach* (*Meliaceae*) vis à vis des adultes *Tribolium castaneum* est due à la présence de Triterpénoides qui ont un effet anti-nutrionnel et inhibent la prise alimentaire des insectes phytophages. Aussi, ils provoquent la mort et des malformations chez les descendants (CARPINELLA *et al*., 2003). Les tanins possèdent des propriétés insecticides, larvicides et répulsives, ils influencent la croissance, le développement et la fécondité de plusieurs insectes phytophages (ACHEUK *et al*., 2014).

BOUCHELTA *et al*. (2005), notant que les effets des alcaloïdes sont plus actifs que les saponines et les flavonoïdes de l'extrait *Capsicum frutescens* (*Solanaceae*) sur *Bemisia tabaci* (Gennadius) (*Homoptera-Aleyrodidae*). Les alcaloïdes se sont avérés plus efficaces et peuvent avoir des effets toxiques par contact et par ingestion chez les adultes de *B. tabaci*. Les flavonoïdes se sont montrés sans ou à faibles effets sur l'éclosion des œufs et la survie des adultes de *B. tabaci*. Aussi l'effet des composés saponines semble être toxique ; ainsi appliqués contre *Callosobruchus chinensis* (*Coleoptera-Bruchidae*), les saponines inhibent le développement de l'insecte (APPLEBAUM *et al*., 1969).

Les résultats trouvés par OUEDRAOGO *et al*. (2016) sur l'évaluation de la toxicité des huiles essentielles de *Cymbopogon nardus* (L) et *Ocimum gratissimum* (L) contre *Sitophilus zeamais* Motsch (*Coleoptera-Curculionidae*) et *Rhyzopert hadominica* F. (*Coleopetra-Bostrichidae*) indiquent que l'huile essentielle extraite de citronnelle *Cymbopogon nardus* (*Poaceae*), a une $DL_{50\%}$ de 57,1 ml/l et une $DL_{90\%}$ de 207,1 ml/l. Quant à l'huile essentielle d'*Ocimum gratissimum* (*Lamiaceae*), les $DL_{50\%}$ et $DL_{90\%}$ sont respectivement 43,18 ml/L et 78,83 ml/L vis-à-vis des adultes *Sitophilus zeamais*. Pour l'huile essentielle de *C. nardus*, la concentration capable de causer la mort de 50% ($DL_{50\%}$) des adultes de *R. dominica* est de 17,29 ml/l. Quant à la $DL_{90\%}$, elle est de 33,03 ml/l. Pour l'huile essentielle d'*Ocimum gratissimum*, les $DL_{50\%}$ et $DL_{90\%}$ sont respectivement de 8,03 ml/l et 28,74 ml/l.

**Tableau 4 :** Taux maximal de mortalité et Probit correspondants en fonction de la concentration d'extraits aqueux

| *Cleome arabica* | | | | | *Pergularia tomentosa* | | | | |
|---|---|---|---|---|---|---|---|---|---|
| **Doses** | | | **Mortalité corrigée** | | **Doses** | | | **Mortalité corrigée** | |
| (%) | (mg/ml) | log [mg/ml] | (%) | Probits | (%) | (mg/ml) | log [mg/ml] | (%) | Probits |
| **100** | 0,0315 | -1,50168945 | 96,67 | 6,85 | **100** | 0,0151 | -1,82102305 | 86,67 | 6,13 |
| **90** | 0,0284 | -1,54668166 | 90,00 | 6,28 | **90** | 0,0135 | -1,86966623 | 63,33 | 5,33 |
| **80** | 0,0252 | -1,59859946 | 73,33 | 5,01 | **80** | 0,012 | -1,92081875 | 36,67 | 4,67 |
| **70** | 0,022 | -1,65757732 | 46,67 | 4,92 | **70** | 0,0105 | -1,9788107 | 33,33 | 4,56 |
| **60** | 0,0189 | -1,7235382 | 43,33 | 4,82 | **60** | 0,009 | -2,04575749 | 16,67 | 4,05 |
| **50** | 0,0157 | -1,80410035 | 33,33 | 4,56 | **50** | 0,0075 | -2,12493874 | 13,33 | 3,87 |
| **40** | 0,0126 | -1,89962945 | 26,67 | 4,39 | **40** | 0,006 | -2,22184875 | 13,33 | 3,87 |
| **30** | 0,00945 | -2,02456819 | 20,00 | 4,16 | **30** | 0,0045 | -2,34678749 | 10 | 3,72 |
| **20** | 0,0063 | -2,20065945 | 16,67 | 4,05 | **20** | 0,003 | -2,52287875 | 6,67 | 3,52 |
| **10** | 0,00315 | -2,50168945 | 10,00 | 3,72 | **10** | 0,00151 | -2,82102305 | 3,33 | 3,25 |

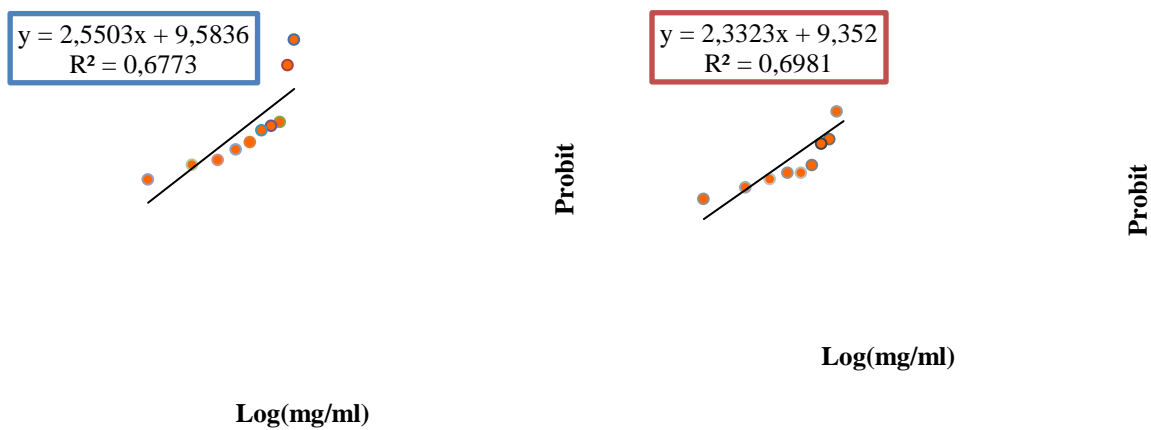

| | |
|---|---|
| **A-** Probits de taux de mortalité corrigée en fonction des logarithmes des doses en extraits aqueux de *C. arabica* appliquées | **B-** Probits de taux de mortalité corrigée en fonction des logarithmes des doses en extraits aqueux de *P. tomentosa* appliquées |

**Figure 05$_{A,B}$-** Droites de régressions des Probit en fonction des logarithmes des doses d'extraits aqueux de *C. arabica* et *P. tomentosa* vis-à-vis des imagos de *T. confusum*.

**Tableau 05-** Équation de droite de régression, coefficient de régression et valeurs de DL$_{50\%}$ et DL$_{90\%}$ pour l'extrait testé (*C. arabica* et *P. tomentosa*).

| Extrait aqueux | Équations | Coefficient de régression | DL$_{50\%}$(mg/ml) | DL$_{90\%}$(mg/ml) |
|---|---|---|---|---|
| *C. arabica* | y = 2,5503x+ 9,5836 | R² = 0,6773 | 0,016 | 0,051 |
| *P. tomentosa* | y = 2,3323+ 9,352 | R² = 0,696 | 0,014 | 0,048 |

## II.2.4.- Temps létaux (TL$_{50\%}$, TL$_{90\%}$)

L'estimation des temps létaux 50% et 90% ont été effectués en dressant la droite de régression des Probit correspondants aux pourcentages des mortalités corrigées en fonction des logarithmes des temps de traitement.

A vu des résultats regroupés dans le tableau 6 et 7, il ressort que l'extrait de *Cleome arabica* a donné un bon résultat relatif à la toxicité sur les individus de *Tribolium confusum*, cette efficacité est confirmée par la mort des imagos de *T. confusum*. Le temps létal le plus court est 6,41 jours et 6,73 jours enregistrés pour les deux fortes concentrations soit 100% et 90%. Le TL$_{50\%}$ oscille entre 11,28 jours à 20,89 jours pour les doses 70%, 60%, 50%,40%. Pour les doses assez faibles 30%, 20% et 10%, le temps létal est plus de 23,82 jours.

Les individus exposés aux les fortes doses 100%, 90% et 80% de *P. tomentosa* représentent un temps létal d'ordre 6,94 jours, 8,67 jours et le 10,94 jours respectivement. Pour les doses 70%, 60%, 50% et 40%, les temps létaux oscillent entre 14,04 à 59,15 jours. Pour les doses faibles soient les extraits à 30%, 20% et 10%, le temps létal est plus 213,81 jours.

Selon les résultats obtenus, les fortes doses 100%, 90% et 80% de l'extrait aqueux de *C. arabica* semblent toxiques et présent une rapidité d'action bien marquée ; 50% de mortalité est observée au bout de 6$^{ème}$ jours.

En revanche, les valeurs des temps létaux 90% de estimées pour les imagos de *T. confusum* traités par l'extrait *Cleome arabica* à 100%, 90% et 80% est respectivement de 9,15 jours, 10,44 jours et 11,76 jours. Pour les doses 70%, 60%, 50%, 40%, 30%, 20% et 10%, les temps létaux sont supérieurs à 18,20 jours. Ces résultats restent incomparables avec les résultats de temps létal 90% des individus traités par les doses 100%, 90% et 80 % de l'extrait *Pergularia tomentosa* où le TL$_{90}$% est 10,32 jours, 13,69 jours et 18,29 jours respectivement. Pour les doses 70%, 60%, 50%, 40%,30%, 20% et 10%, les TL$_{90}$% estimés dépassent 23,91 jours.

**Tableau 6-** Equation des droites de régression, et les valeurs de temps TL$_{50\%}$ et TL$_{90\%}$ enregistrés pour l'extrait de *Cleome arabica*.

| (%) | Equation | Coefficient | TL$_{50\%}$ (heure) | TL$_{50\%}$ (jour) | TL$_{90\%}$ (heure) | TL$_{90\%}$ (jour) |
|---|---|---|---|---|---|---|
| **100%** | y = 7,085x - 10,31 | R² = 0,921 | 153,86 | 6,41 | 219,56 | 9.15 |
| **90%** | y = 6,703x - 9,800 | R² = 0,924 | 161,42 | 6.73 | 250,56 | 10,44 |
| **80%** | y = 6,244x - 9,022 | R² = 0,911 | 176,06 | 7,33 | 282,27 | 11,76 |
| **70%** | y = 6,158x - 9,980 | R² = 0,764 | 270,77 | 11,28 | 436,98 | 18,20 |
| **60%** | y = 5,455x - 9,124 | R² = 0,664 | 388,31 | 16,17 | 666,54 | 27,77 |
| **50%** | y = 5,035x - 8,420 | R² = 0,662 | 462,74 | 19,28 | 666,54 | 27,77 |
| **40%** | y = 4,853x - 8,104 | R² = 0,660 | 501,40 | 20,89 | 920,32 | 38,34 |
| **30%** | y = 4,588x - 7,650 | R² = 0,654 | 571,73 | 23,82 | 920,32 | 38,34 |
| **20%** | y = 4,539x - 7,563 | R² = 0,654 | 606,10 | 25,25 | 1121,47 | 46,72 |
| **10%** | y = 3,193x - 5,525 | R² = 0,416 | 1978,21 | 82,42 | 4979,09 | 207,46 |

**Tableau 7-** Equation des droites de régression, et les valeurs de temps $TL_{50\%}$ et $TL_{90\%}$ évalué pour l'extrait de *Pergularia tomentosa*.

| (%) | Equation | Coefficient | $TL_{50\%}$ (heure) | $TL_{50\%}$ (jour) | $TL_{90\%}$ (heure) | $TL_{90\%}$ (jour) |
|---|---|---|---|---|---|---|
| **100 %** | y = 7,435x - 11,52 | R² = 0,849 | 166,69 | 6,94 | 247,78 | 10,32 |
| **90 %** | y = 6,454x - 9,963 | R² = 0,839 | 208,16 | 8,67 | 328,65 | 13,69 |
| **80%** | y = 5,734x - 8,872 | R² = 0,845 | 262,57 | 10,94 | 439,01 | 18,29 |
| **70 %** | y = 5,536x - 8,993 | R² = 0,767 | 337,00 | 14,04 | 573,91 | 23,91 |
| **60 %** | y = 4,699x - 7,809 | R² = 0,645 | 531,98 | 22,16 | 996,08 | 41,50 |
| **50%** | y = 4,276x - 7,285 | R² = 0,543 | 746,46 | 31,10 | 1487,15 | 61,96 |
| **40 %** | y = 3,521x - 6,099 | R² = 0,417 | 1419,80 | 59,15 | 3279,15 | 136,63 |
| **30%** | y = 2,554x - 4,476 | R² = 0,283 | 5131,66 | 213,81 | 16271,70 | 677,98 |
| **20%** | y = 1,327x - 2,351 | R² = 0,144 | 346388,06 | 14432,83 | 3192601,31 | 133025,05 |
| **10%** | y = 1,176x - 2,084 | R² = 0,144 | 1056354,10 | 44014,75 | 12949258,4 | 539552,43 |

Les résultats obtenus permettent de mettre en évidence l'effet insecticide de *Cleome arabica* et *Pergularia tomentosa* sur les imagos *Tribolium confusum*. Cet effet est plus marqué aux fortes concentrations de l'extrait utilisé. L'expérimentation menée montre que la mortalité des imagos de *Tribolium confusum* augmente proportionnellement avec le temps et la dose appliquée. Les résultats obtenus sont plus proches ceux de VIJAYKUMAR *et al.* (2015) qui a confirmé en général que la mortalité des adultes de *T. castaneum* augmente avec le temps d'exposition. Quel que soit l'extrait utilisé, les temps létaux oscillent entre 6,41 jours à 10,94 jours pour les fortes doses. Les individus traités par les faibles doses, montrent un temps létal supérieur de 14,04 jours, cela est probablement lié à l'effet faible ou limité de l'extrait appliqué aux faibles doses.

BOUNECHADA et ARAB (2011) notent que le $TL_{50\%}$ des larves et des adultes *Tribolium castaneum* Herbst (*Coleoptera-Tenebrionidae*) est de 6,8 jours et 7,4 jours respectivement avec la poudre de *Peganum harmala* (*Zygophyllaceae*). Pour les individus traités par la poudre de *Melia azedarach* (*Meliaceae*), le $TL_{50\%}$, est plus court, il est de 3,9 jours chez les larves et de l'ordre de 4,3 jours chez les adultes de *Tribolium castaneum*.

BOUGHDAD *et al.* (2011) l'activité biologique des huiles essentielles de *Mentha* (*Mentha pulegium* L. *Lamiaceae)* sur un ravageur des graines du pois chiche entreposées au Maroc *Callosobruchus maculatus* (*Coleoptera-Bruchidae*). Les résultats de test par fumigation montent que $TL_{50\%}$ est d'ordre 3.89jours pour les mâles et 3,59 jours pour les femelles du *C. maculatus*

Pendant la période expérimentale, des accouplements, des exuvies et des larves néonates sont observés. Au bout du deuxième jour après le traitement, les imagos de *T. confusum* se regroupent sous forme d'agrégats dans les boites de Pétri. De plus, des anomalies de comportement sont observées, telles que réduction de l'activité motrice et difficulté de déplacement. S'ajoute à ces manifestations, le renversement des individus sur leur face dorsale. Ces manifestations sont probablement liées à l'influence des extraits testés sur le métabolisme hormonal de l'insecte dont les ecdystéroïdes et l'hormone juvénile qui contrôlent la mue et la métamorphose (TRUMAN et RIDDIFORD, 1999).

BRINDLEY (1930) a signalé, le cycle de vie de *Tribolium confusum* dans des conditions d'humidité et de température soigneusement contrôlées. Il a soumis les coléoptères à une température de 29,7 °C et une humidité relative de 73%. Ces données mettent en évidence, le fait que toutes les étapes larvaires se transforment assez rapidement, sauf pour les périodes d'œuf et de nymphe. Ce dernier stade larvaire affiche également la plus grande variation de temps de sa durée. Il est intéressant de noter que cela est lié à l'observation de CHAPMAN (1918) qui souligne que le dernier stade larvaire est plus influencé par les changements écologiques, par rapport à sa durée, que n'importe lequel des autres stades de développement. La durée moyenne requise pour l'achèvement d'un cycle de vie de *T. confusum* dans les conditions de température et d'humidité de BRINDELY et les aliments est de 30 jours. Les données de BRINDELY démontrent que, pour un ensemble particulier de conditions, et un stock présument élever, la variation de la durée du cycle est de petite taille pour une observation basée sur 100 individualités différentes. Le coléoptère confus a complété son développement en environ 25 jours (œufs en quatre jours ; Larve en 16 jours ; imagos en six jours) (HOWE, 1960).

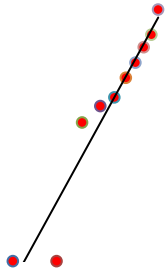 A  y = 7,085x - 10,317  R² = 0,9218  **100%**

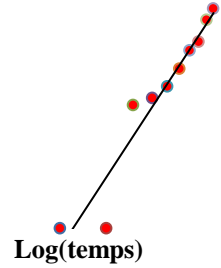 B  y = 6,7033x - 9,8001  R² = 0,9245  90%

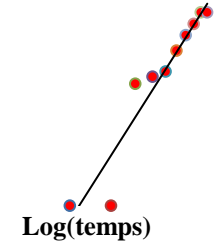 C  y = 6,2442x - 9,0224  R² = 0,9117  80%

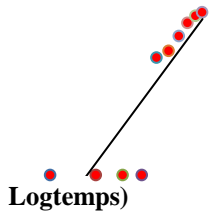 D  y = 6,1588x - 9,9805  R² = 0,7644  70%

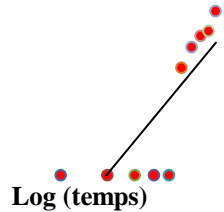 E  y = 5,4558x - 9,124  R² = 0,6641  60%

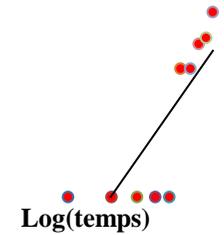 F  y = 5,0355x - 8,4202  R² = 0,662  50%

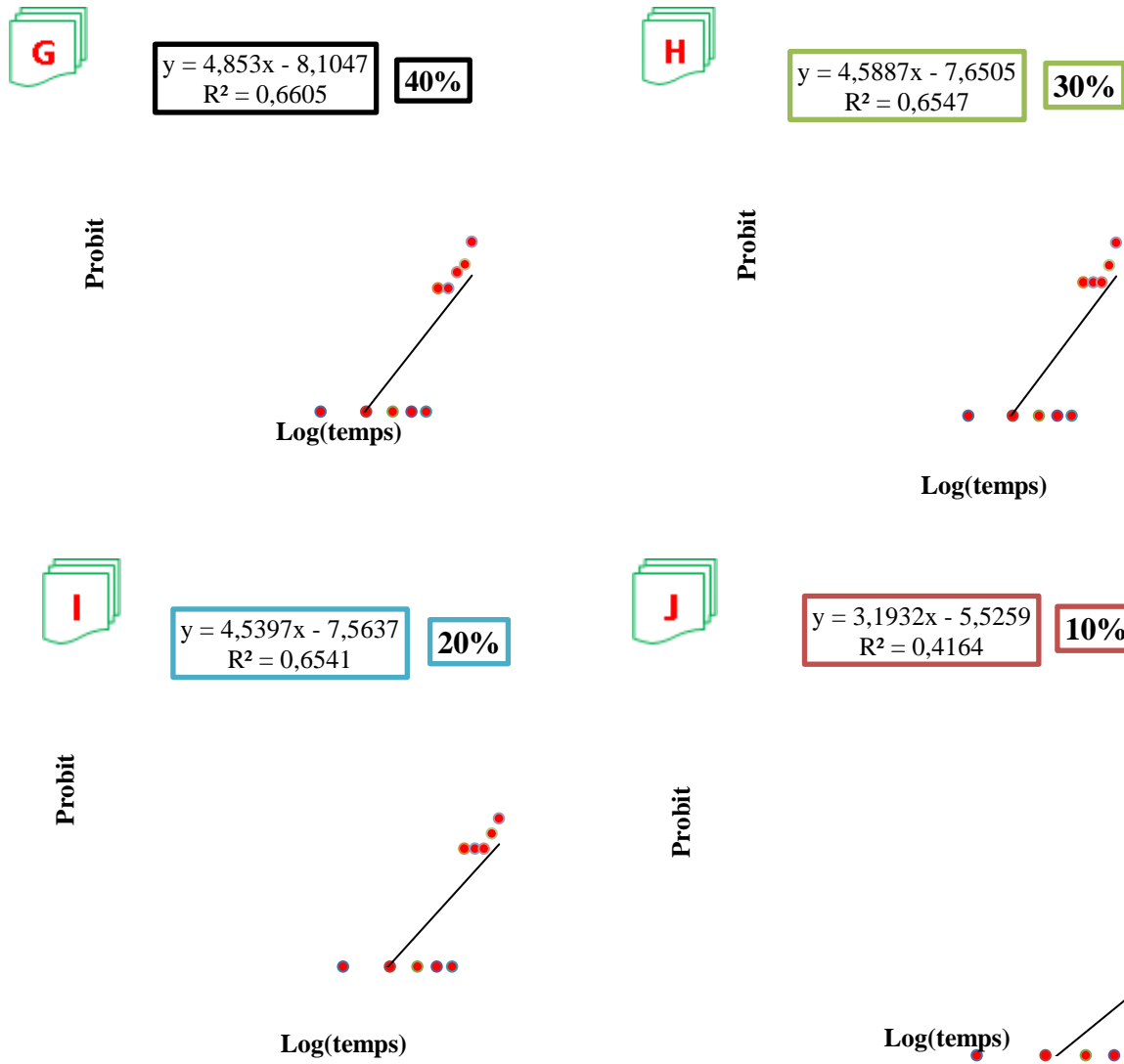

**Figure 6** A B C D E F G H I J**-** Droite de régression des Probit en fonction de Log (temps) *Cleome arabica*.

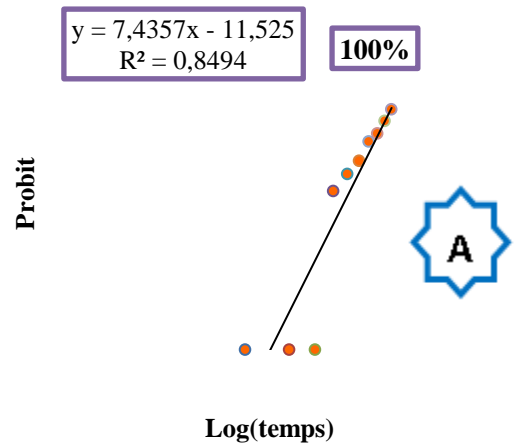
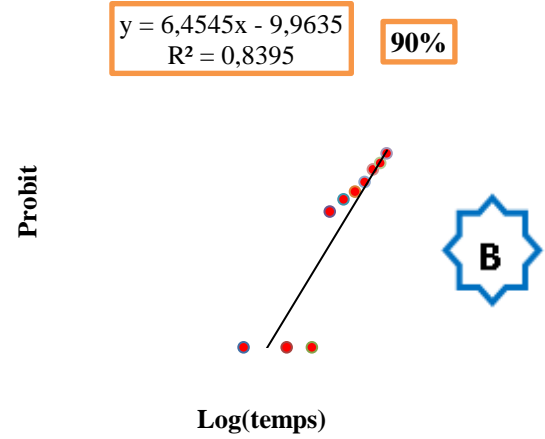
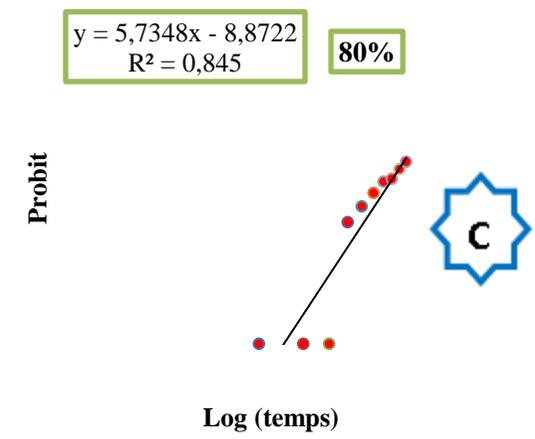
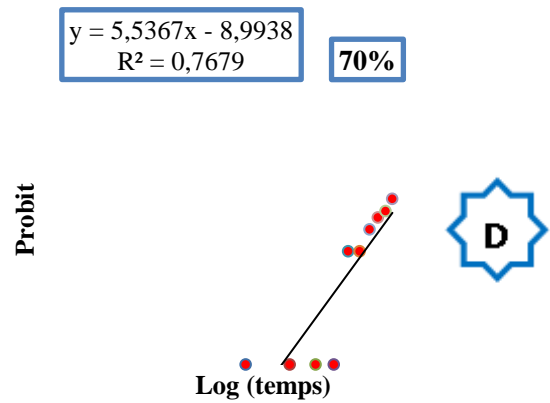
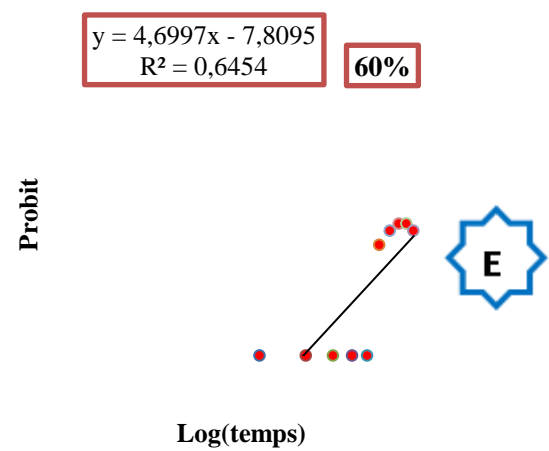
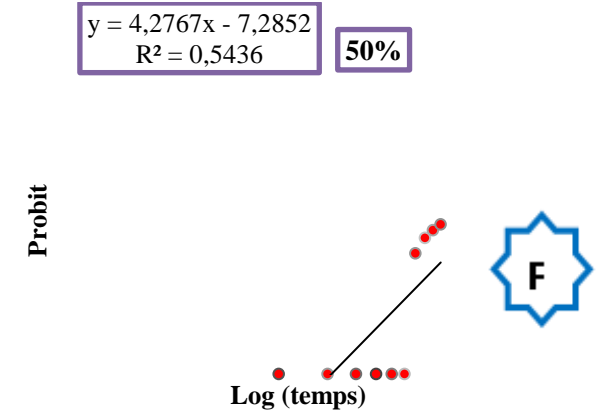

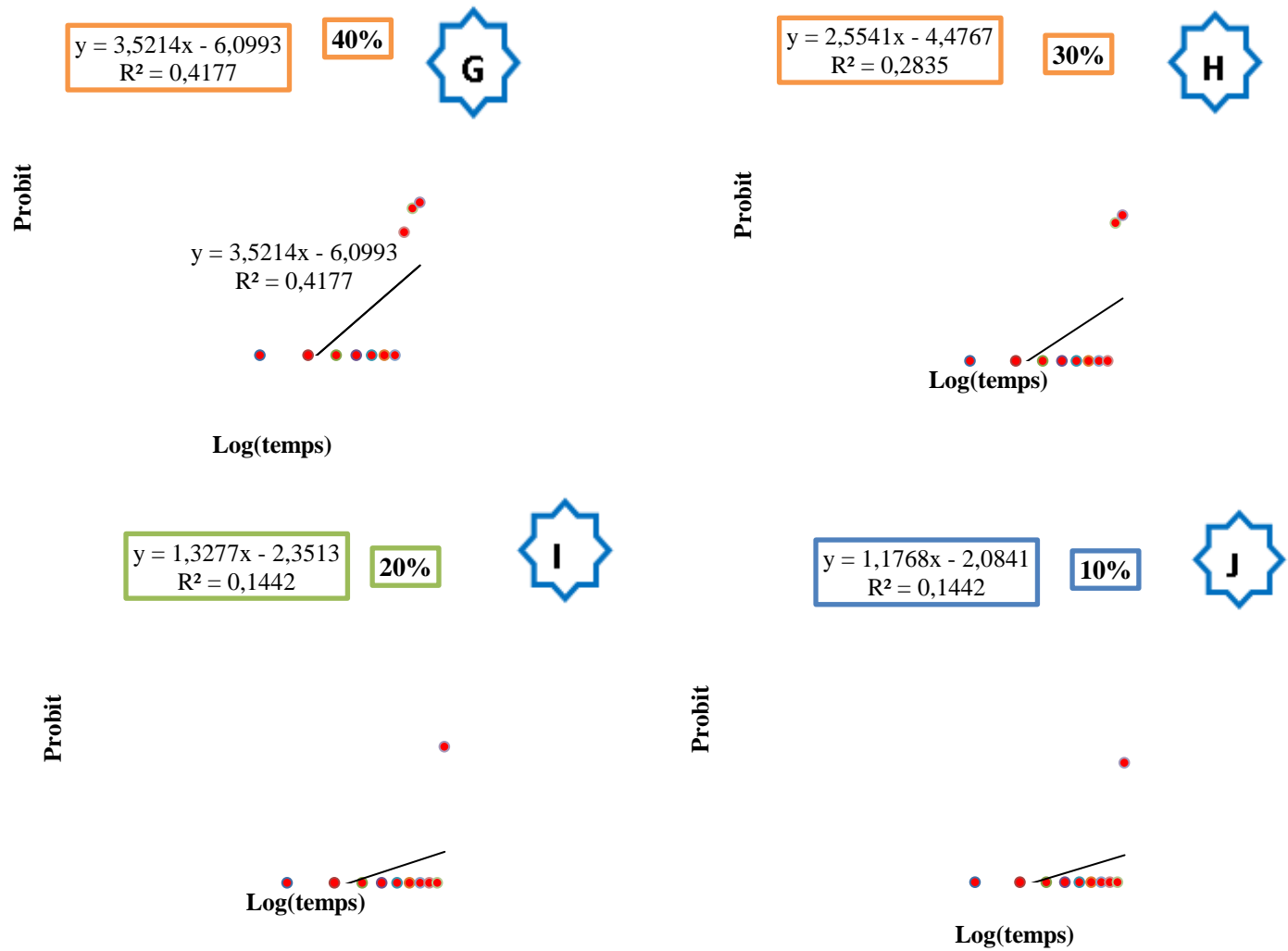

**Figure 7**(A B C D E F G H I J)**-** Droite de régression des Probit en fonction de Log (temps) *Pergularia tomentosa*.

## II.3.- Effet herbicide

L'activité allélopathique porte sur l'évaluation des effets toxiques des extraits aqueux des feuilles de *C. arabica* et *P. tomentosa,* sur la germination des graines d'*H. vulgaire*, *T. durum* et *D. aegyptium*, durant dix jours, en notant quotidiennement le nombre de graines germées et toute anomalie de croissance observée. Les critères d'appréciations sont non seulement le taux d'inhibition de germination mais aussi l'évolution dans le temps de la germination, les concentrations d'efficacité et la biométrie de la partie aérienne et souterraine.

## II.3.1.-Taux d'inhibition

Les variations des taux d'inhibition en fonction de la dose des extraits appliquées de *C. arabica* (fig. 8) et *C. arabica* (fig. 9), laissent remarquer les effets inhibiteurs des extraits testés et leurs influences sur la germination des graines d'*H. vulgaire*, *T. durum* et *D. aegyptium*.

Pour l'extrait de *C. arabica,* les graines de *D. aegyptium* exposées à des concentrations 100%, 90%, 80%, 70%, 60%, 50%, 40%, 30% et 20%, aucune germination des graines n'est observée. L'inhibition est notable pour tous les traitements appliqués. Le pourcentage d'inhibition de la germination est de 100% pour les concentrations 100%, 90%, 80%, 70%, 60%, 50%, 40%, 30% et 20%, alors qu'il est de 84,44% pour la concentration la plus faible soit l'extrait à 10% (fig. 8).

Les graines de blé dur traitées par l'extrait de *C. arabica* pour des concentrations de 100%, 90%, 80%, 70%, 60% et 50%, les taux d'inhibition rapportés sont de l'ordre de 91,11%, 84,44%, 75,56%, 68,67% et 60% respectivement, tandis qu'ils oscillent entre 22,22% et 46,67% pour les autres concentrations appliquées soient de 40%, 30%, 20% et 10% (fig. 8). De même, il apparait que l'orge est plus résistant aux effets inhibiteurs de la germination de l'extrait appliqué. Les doses 100%, 90% et 80% en extrait de *C. arabica* engendre des taux d'inhibition de la germination de 77,78%, 71,11% et 55,56% successivement. Pour les concentrations d'extrait de 70%, 60%, 50%, 40%, 30% et 20%, les taux d'inhibition enregistrés sont successivement de l'ordre de 46,44%, 35,56%, 35,56%, 28,89%, 13,33% et 8,89%. Pour la concentration 10%, le taux d'inhibition de la germination des graines d'orge est de 0%; à cette concentration, l'extrait de *C. arabica* ne présente aucun effet inhibiteur sur la germination des graines d'orge (fig. 8).

L'effet inhibiteur de l'extrait de *C. arabica* sur l'inhibition de la germination des trois espèces testées au seuil de P≤0,001, révèle que les traitements présentent des différences très hautement significatives.

Il est noté un taux d'inhibition maximal chez les graines de *D. aegyptium* aspergées par la majorité des concentrations appliquées. L'analyse de la variance montre que l'effet inhibiteur de l'extrait *C. arabica* aux doses 100%, 90%, 80%, 70%, 60%, 50%, 40%, 30%, 20% et 10%, induit une différence d'inhibition très hautement significative (P≤0,001) (Annexe 2).

La comparaison des moyennes des taux d'inhibition des graines de blé dur et le témoin négatif, montre que l'extrait de *C. arabica* aux concentrations 100%, 90%, 80%, 70%, 60% et 50% engendre une inhibition très hautement significative (P≤0,001) (Annexe 2).

Pour le lot traité à 40%, une différence hautement significative est notée (F= 49,00 ; P= 0,002). Pour les concentrations 30%, 20% et 10%, il est noté une inhibition très hautement significative (P≤0,001) (Annexe 2).

L'analyse de la variance de l'effet inhibiteur de *C. arabica* sur l'inhibition de la germination des graines d'orge révèle que les traitements 100%, 90%, 80%, 70%, 60%, 50% et 40% provoquent une inhibition hautement significative voir très hautement significative; les valeurs de facteur F enregistrées sont pour une probabilité P≤0,01.

Pour la concentration 30%, elle est significative (F=12,00 ; P= 0,026), alors que pour les doses 20% et 10%, la germination des graines est maximale (100%).

Généralement, les résultats relatifs aux effets de l'extrait de *P. tomentosa* sur la germination de trois plantes testées montrent que cet extrait présente un effet inhibiteur de germination moins perceptible par rapport à l'extrait de *C. arabica*. Le taux d'inhibition attient le 100% au niveau des lots de graines de *D. aegyptium* traités par l'extrait de *P. tomentosa* à des concentrations 100%, 90%, 80%, 70%, 60%, 50%, 40% et 30%. Les graines de *D. aegyptium* traitées par le même extrait à des concentrations de 20% et de 10%, les taux d'inhibition notés sont de 91,11% et 75,56% respectivement (fig. 9).

Les graines de blé dur exposées à ce même extrait de *P. tomentosa* à différentes concentrations soient 100%, 90% et 80%, 70%, 60%, 50% et 40%, montrent des taux d'inhibition de 82,22%, 73,33% et 66,67%, 42,22%, 40%, 31,11% et 22,22% respectivement, bien qu'il est de 0% pour les autres doses testées dont l'extrait à 30%, 20% et 10%. Les résultats de l'effet inhibiteur de l'extrait de *P. tomentosa* aux concentrations 30%, 20% et 10% vis-à-vis de blé dur, sont similaires à ceux enregistrés pour ces mêmes concentrations appliquées sur l'orge, où un pourcentage de germination de 100% est noté. Cependant, l'extrait de *P. tomentosa* appliqué à des concentrations de 100%, 90%, 80%, 70%, 60%, 50% et 40%, les taux d'inhibition de la germination notés sont successivement de l'ordre de 64,44%, 55,56%, 53,33%, 37,78%, 33,33%, 13,33% et 8,89% (fig. 9).

Analysant les résultats obtenus, les tests réalisés montrent que l'effet inhibiteur de la germination varie en fonction de l'extrait végétal testé et l'espèce test; l'extrait de *C. arabica* a été plus toxique que celui de *P. tomentosa*. De même, *D. aegyptium* est l'espèce la plus vulnérable aux effets des deux extraits testés par rapport au blé dur et orge. Ce dernier semble plus résistant aux effets des extraits testés. L'effet de l'extrait *C. arabica* à différentes concentrations montre un pouvoir allélopathique notable. Il se manifeste par l'action inhibitrice de la germination très hautement significative ($P \leq 0,001$) dans la majorité des pots ensemencés.

FRIEDMAN (1995) déclare que le phénomène d'allélopathie ne se manifeste, que si la quantité des composés à effets allélochimiques atteint le seuil critique d'action qui varie en fonction de l'organisme cible. A partir de ce seuil critique, l'inhibition évolue en fonction de la dose. PARRY (1982) a signalé que les substances allélochimiques sont synthétisées par certaines plantes qui exercent des effets toxiques sur d'autres plantes avoisinantes.

L'analyse de la variance pour les valeurs du taux d'inhibition enregistrées chez les graines d'orge traitées par l'extrait de *C. arabica*, montrent une différence hautement significative. Les valeurs de facteur F sont mentionnées respectivement de ($F=44,26$ ; $P=0.003$), ($F=48,08$ ; $P=0,002$), ($F=48,00$ ; $P=0,002$) et ($F=48,08$ ; $P=0,002$) chez les graines d'orge aspergées par les concentrations 100%, 90%, 80% et 70% (Annexe 02).

Pour les lots traités à 60%, 50%, 30%, 20% et 10% de concentration, ils n'affectent pas significativement la germination des graines d'orge ($P>0,05$).

Une différence significative est rapportée ($F=16,00$ ; $P=0,016$) chez les graines d'orge traitées par la dose 40% de l'extrait de *C. arabica*.

En parallèle, une différence très hautement significative (P≤0.001) est notée dans les pots ensemencés de blé dur traités par les doses 100%, 90%, 60% et 40% de l'extrait de *P. tomentosa* en comparaison avec le témoin négatif (Annexe 2).

De même, pour les graines traitées par les concentrations 80% et 70% entrainent une inhibition hautement significative de la germination où le facteur F noté est de 23,08 ; P=0.009 et F=22,56 ; P=0,009 respectivement (Annexe 2).

Une différence significative est rapportée chez les graines de blé dur traitées par la dose 50%, le facteur F est de (F=28,00 ; P=0,006). Pour les autres doses soient 30%, 20% et 10%, n'affectent pas la germination des graines de blé dur (P>0,05) (Annexe 2).

En outre, au niveau les pots ensemencés par les graines de *D. aegyptium* et traitées par les concentrations 100%, 90%, 80%, 70%, 60%, 50%, 40% et 30%, elles engendrent une inhibition très hautement significative (F=*; P=*). Pour les doses 20% et 10%, une différence très hautement significative est rapportée où les valeurs de facteur F sont enregistrées respectivement de (F=1681,00 ; P=0,000) et (F=1156,00 ; P=0,000) (Annexe 2).

L'extrait aqueux de *P. tomentosa* montre un effet inhibiteur modéré. Il se traduit par des taux d'inhibition de la germination plus faible comparativement aux valeurs rapportées pour l'extrait de *C. arabica* et par des anomalies de croissances diverses.

Les anomalies de croissance se traduisent par un ralentissement de la croissance radiculaire, un gonflement de la graine. Pour certaines graines, la germination s'arrête au début de l'apparition de la radicule ou l'absence totale de la germination. Celle-ci est probablement liée aux molécules allélochimiques présentes au niveau des extraits. L'effet de ces substances se manifeste par des variations morphologiques qui sont observées aux premiers stades de développement, des effets sur l'allongement de la tigelle et de la radicule (KRUSE *et al.*, 2000).

BENMEDDOUR et FENNI (2018), notent que le pourcentage d'inhibition atteint 100% pour les graines de *K. scoparia* et *T. durum* traitées par la concentration de 5% du *Peganeum harmala* (*Zygophyllaceae*). Pour la concentration 5% de Laurier rose *Nerium oleander* (*Apocynaceae*), l'inhibition est enregistrée respectivement à 97% et 83% pour *K. scoparia* et *T. durum*. Pour la même dose (5%) d'*Ailanthus altissima* (*Simaroubaceae*), le taux d'inhibition est de 93% et 78% pour *K. scoparia* et *T. durum*.

Des études réalisées en Palestine par SALMAN *et al.* (2017) sur l'effet inhibiteur de l'extrait aqueux d'*Artemisia annua* L. (*Asteraceae*) sur la germination des graines d'orge, ils ont signalé pour la concentration 0,5% que l'extrait foliaire d'*A. annua* inhibe la germination des graines d'orge avec un taux enregistré d'ordre de 6% et 18,5% pour les graines aspergées par l'extrait racinaire d'*A. annua*.

L'effet de l'extrait de *C. arabica* et *P. tomentosa* aux faibles doses montre un effet simulateur de la croissance des graines d'orge et de blé dur. BEN MEDDOUR et FENNI (2018) ont trouvé que les extraits de *Peganeum harmala* (Zygophyllaceae) et *Ailanthus altissima* (Simaroubaceae) à faibles doses présentent un effet simulateur pour la croissance des plantules de blé. Aussi HEGAB *et al.* (2008) ont mentionné que l'extrait de *Beta vulgaris* (*Amaranthaceae*) à concentration 1% stimule la germination des graines et le développement des plantules de blé. Pour les fortes doses soient 8% et 12%, elles inhibent significativement la germination des graines de blé dur. La stimulation de germination des graines d'orge et de blé dur est due à la faible quantité des substances allélochimiques pour les concentrations appliquées. Ceci explique les pourcentages de stimulation élevés observés uniquement pour les pots ensemencés par l'orge et le blé dur et traités par les extraits de *C. arabica* et *P. tomentosa* à de faibles concentrations.

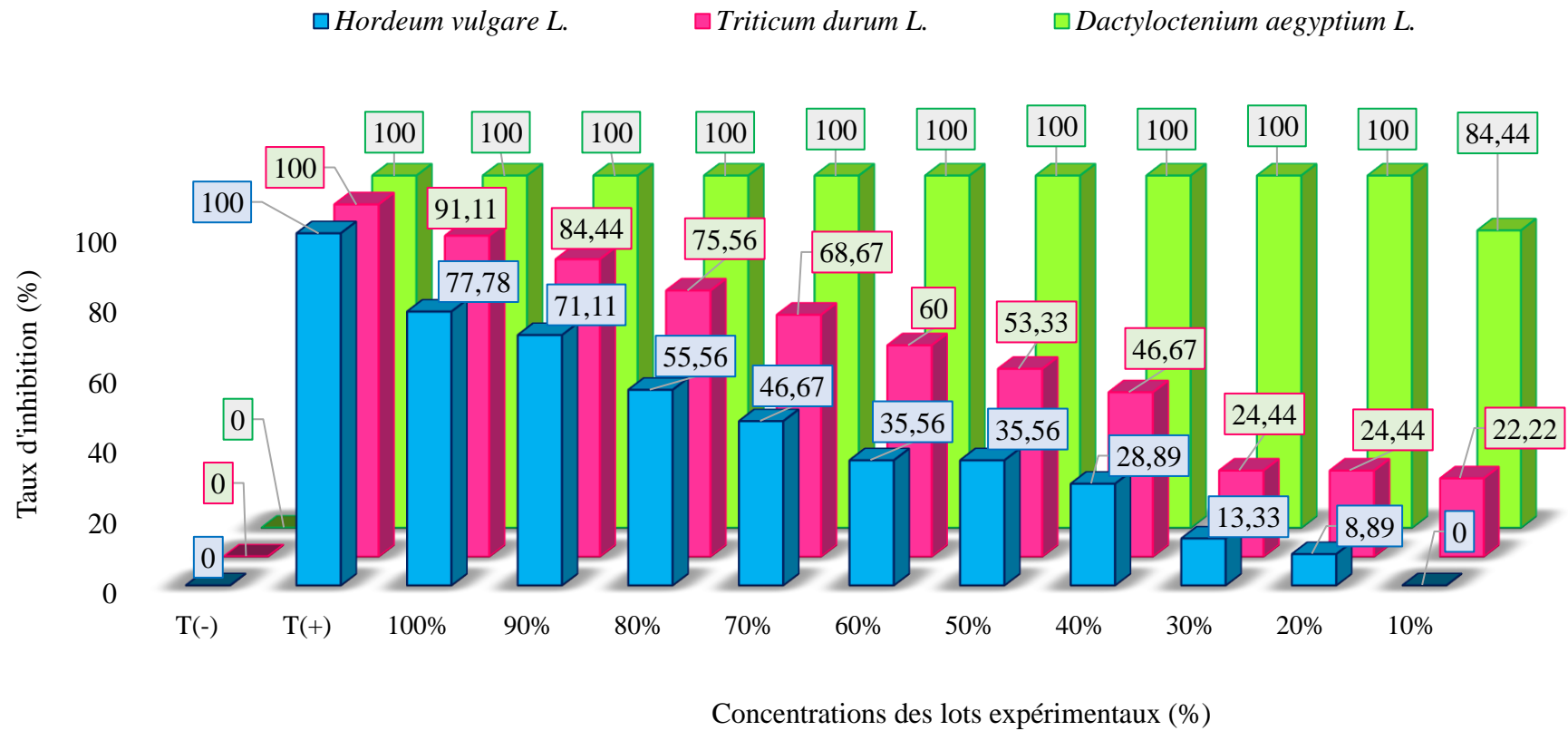

**Figure 8.-** Taux d'inhibition enregistrés au pour l'extrait aqueux de *C. arabica* sur la germination des graines d'orge, le blé et le Dactylocténion

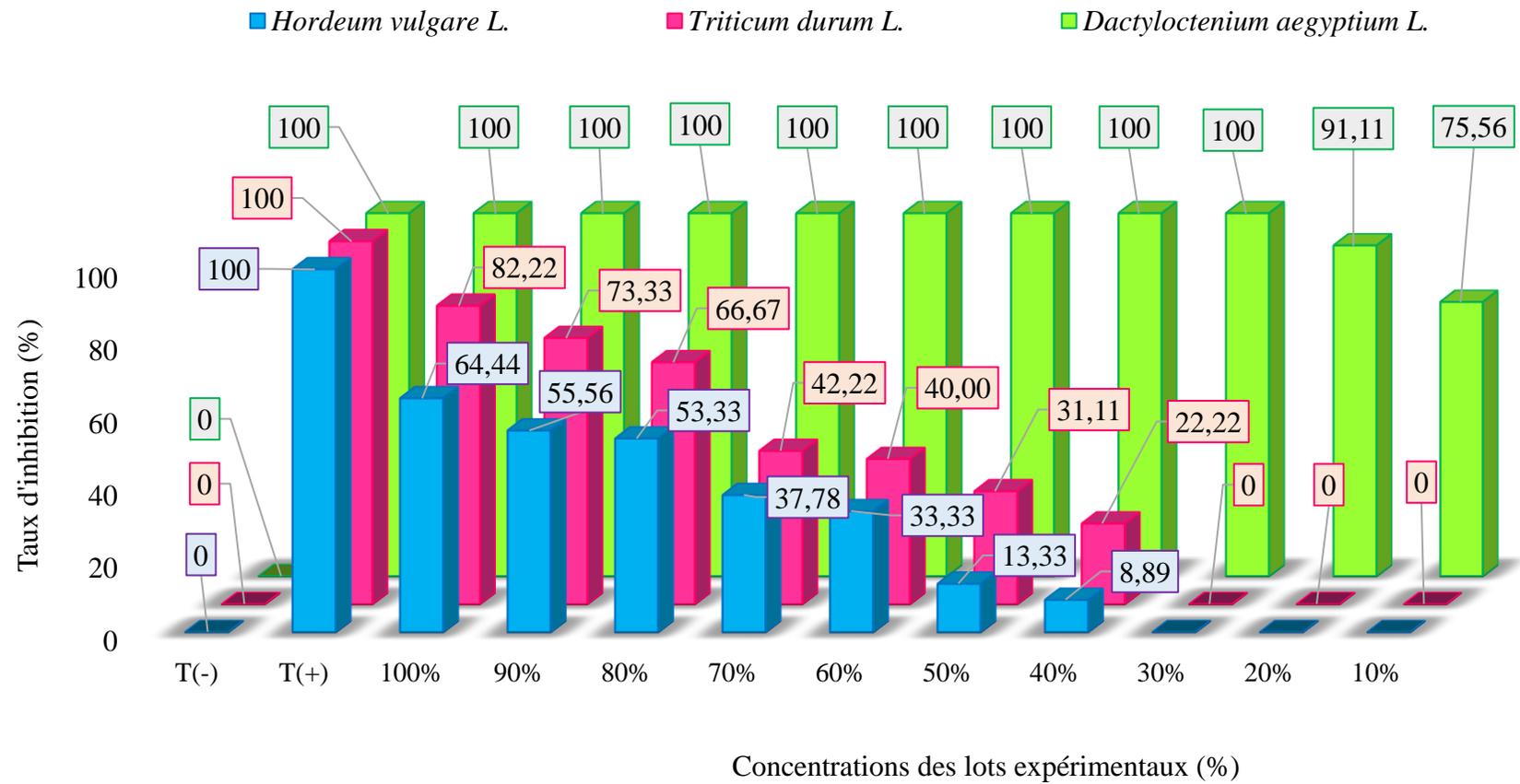

**Figure 7.-** Taux d'inhibition enregistré pour l'extrait aqueux de *P. tomentosa* sur la germination des graines d'Orge, de Blé et de Dactylocténion

## II.3.2.- Cinétique de germination

Les figures 10 et 11 représentent l'évolution dans le temps de la germination des graines de l'orge, du blé dur et du dactylocténion. Les courbes relatives aux taux de germination des graines traitées, sont situées au-dessous de celles des courbes témoins.

La figure 8 présente la cinétique de germination des graines (orge, blé dur, dactylocténion) vis-à-vis les doses de l'extrait *C. arabica*. En effet, les graines d'orge du témoin négatif germent à partir de troisième jour après le semis avec un taux 6,67%. Dès le cinquième jour, la cinétique présente un palier traduisant une valeur maximale de 100% du taux de germination (fig. $10_A$).

Une première phase de latence de très courte durée, nécessaire à l'apparition des premières germinations, au cours de laquelle le taux de germination reste faible. Néanmoins, les graines exposées aux concentrations 40%, 30%, 20% et 10%, ne réagissent qu'au bout du quatrième jour de semis avec un taux de germination respectivement de 2,02%, 2,22%, 2,22% et 2,22%. A partir du cinquième jour, les taux de germination soient 8,89%, 8,89%, 24,44%, 26,67%, 28,89% et 40% sont notés successivement pour les lots 100%, 90%, 80%, 70%, 60% et 50% (fig. $10_A$).

Une deuxième phase exponentielle où l'on assiste à une accélération de la germination, le taux de germination des graines d'orge traitées à 40%, 30%, 20% et 10%, se stabilise dès le septième jour avec un taux de germination respectivement de 68,89%, 75,56%, 86,67%, 95,56% et 100%. Il s'ensuit une évolution exponentielle lente pour les graines aspergées par des concentrations 100%, 90%, 80%, 70%, 60% et 50% pour arriver à un taux de germination maximal l'ordre de 22,22%, 28,89%, 42,22%, 53,33%, 64,44% et 64,44% respectivement. Puis se stabilise à partir du huitième jour indiquant un arrêt de la germination qui représente le pourcentage optimal de la germination et traduisant la capacité germinative de chaque concentration. Le pourcentage maximal de germination des graines d'orge, est enregistré successivement pour 22,22%, 28,89%, 44,44%, 53,33%, 64,44%, 71,11%, 86,67%, 91,11% et 100% pour les lots de l'extrait *C. arabica* à 100%, 90%, 80%, 70%, 60%, 50%, 40%, 30%, 20% et 10% (fig. $10_A$).

Le processus de germination des graines de blé dur aspergées par des concentrations de 40%, 30%, 20% et 10%, est déclenché au bout du quatrième jour de semis où le taux de germination enregistré est de 2,22%, 2,22%, 4,44% et 13,33% respectivement. Dès le cinquième jour, le taux de germination est de 2,22% pour les graines de blé dur arrosées par la dose 50%. Les graines traitées aux fortes doses 100%, 90%, 80%, 70%et 60%, ne réagissent qu'au bout du septième jour, avec un taux de germination enregistré l'ordre de 6,67%, 8,89%, 8,89%, 15,56% et 22,22% (fig. $10_B$).

La deuxième phase se caractérise par une augmentation rapide de germination qui évolue proportionnellement en fonction du temps, où les taux de germination enregistrées sont de 6,67%,

13,33%, 24,44%, 31,11%, 33,33%, 46,67%, 53,33%, 75,56%, 75,56% et 77,78%, pour les doses 100%, 90%, 80%, 70%, 60%, 50%, 40%, 30%, 20% et 10%, puis se stabilise à un pourcentage maximal de germination (fig. 10$_B$).

Pour les graines dactylocténion imprégnées aux concentrations 100%, 90%, 80%, 70%, 60%, 50%, 40%, 30% et 20%, aucune graine n'a germé durant la période de suivi. A l'exception des graines irriguées par la concentration 10%, la germination débute dès le quatrième jour avec un pourcentage de 2,02% et suit une évolution linéaire pour arriver un taux maximal de 15,56% (fig. 10$_C$).

La figure 11 illustre la cinétique de germination des graines (orge, blé dur, dactylocténion) vis-à-vis de l'extrait de *P. tomentosa* à différentes concentrations. La lecture des résultats dressés, permet de ressortir que la capacité germinative diffère selon la concentration appliquée et l'espèce utilisée. Durant dix jours de suivi, aucune germination des graines dactylocténion ensemencées pour les pots à concentration 100%, 90%, 80%, 70%, 60%, 50%, 40% et 30%. Pour les graines traitées par la concentration 20% et 10%, la germination apparait dès le quatrième jour avec un pourcentage seulement de 2,22% et 4,44%. Puis s'évolue lentement jusqu'à atteindre 8,89% et 24,44% dès le septième jour, puis se stabilise à un taux maximal de germination (fig. 11$_C$).

En effet, pour toutes les différentes concentrations, la majorité des graines d'orge germent dès le quatrième jour mais avec des taux différents. Plus la dose est élevée, plus la capacité germinative des semences diminue, ce qui explique l'augmentation exponentielle des taux de germination en fonction du temps. Ensuite, le taux de germination se stabilise et attient 100% pour les doses 10%, 20% et 30%. Les doses 40%, 50%, 60%, 70%, 80%, 90% et 100% induisent un taux de germination de 77,78%, 73,33%, 64,44%, 57,78%, 46,67%, 44,44% et 35,56% respectivement (fig. 11$_A$).

D'après les résultats retenus de l'évolution des pourcentages de germination des graines de blé dur, il est observé que la germination des graines commence dès le quatrième jour pour toutes les concentrations appliquées. A partir du huitième jour, le taux de germination s'accélère puis se stabilise à un taux maximal de germination soit 17,78%, 26,67%, 33,33%, 57,78%, 57,78%, 62,22% et 77,78% pour les doses successivement de 100%, 90%, 80%, 70%, 60%, 50% et 40%. Pour les doses soient 10%, 20% et 30%, le pourcentage de germination est atteint 100% (fig. 11$_B$).

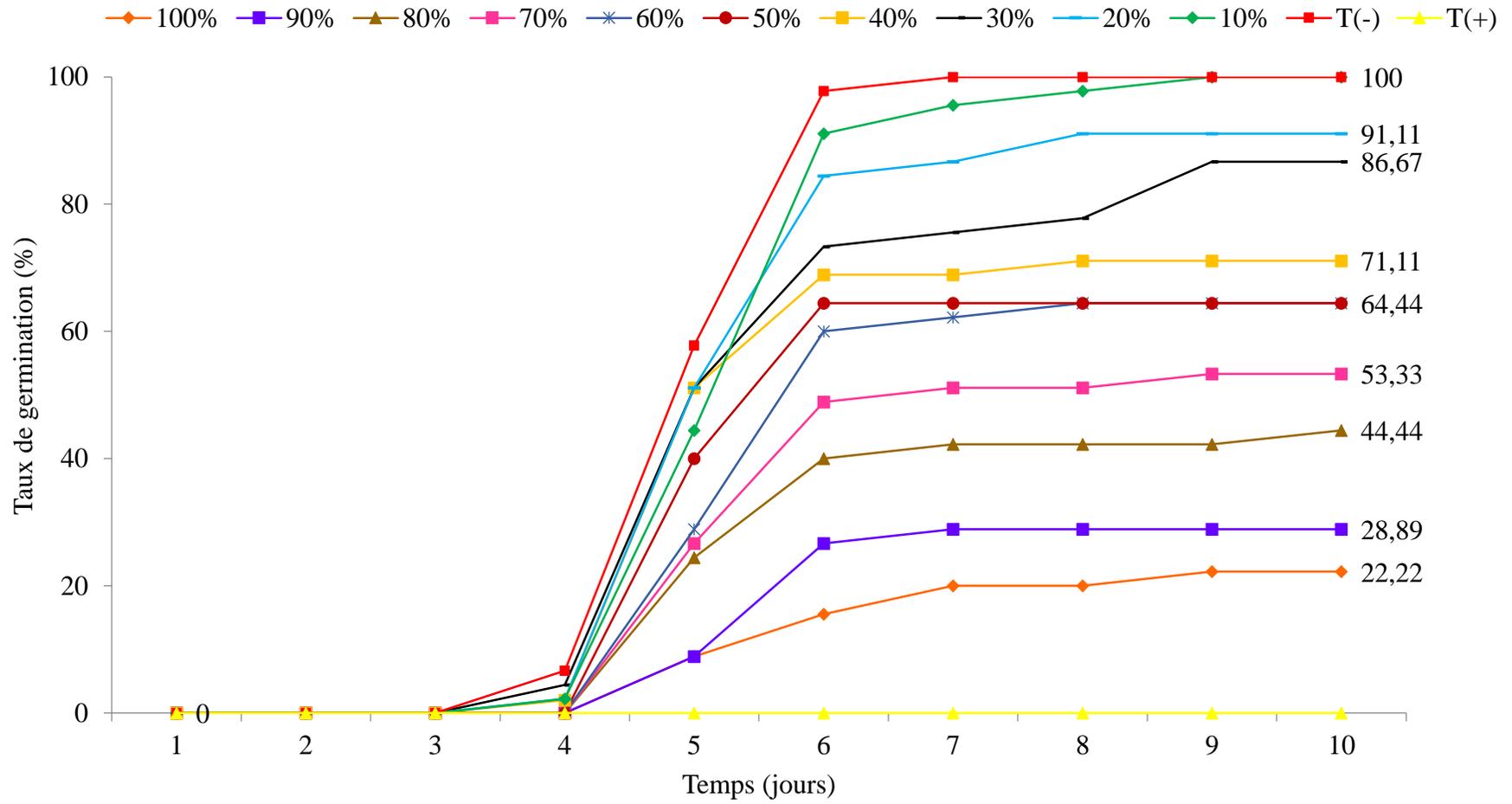

**A :** Cinétique de germination enregistré pour l'extrait aqueux de *C. arabica* sur les graines d'orge

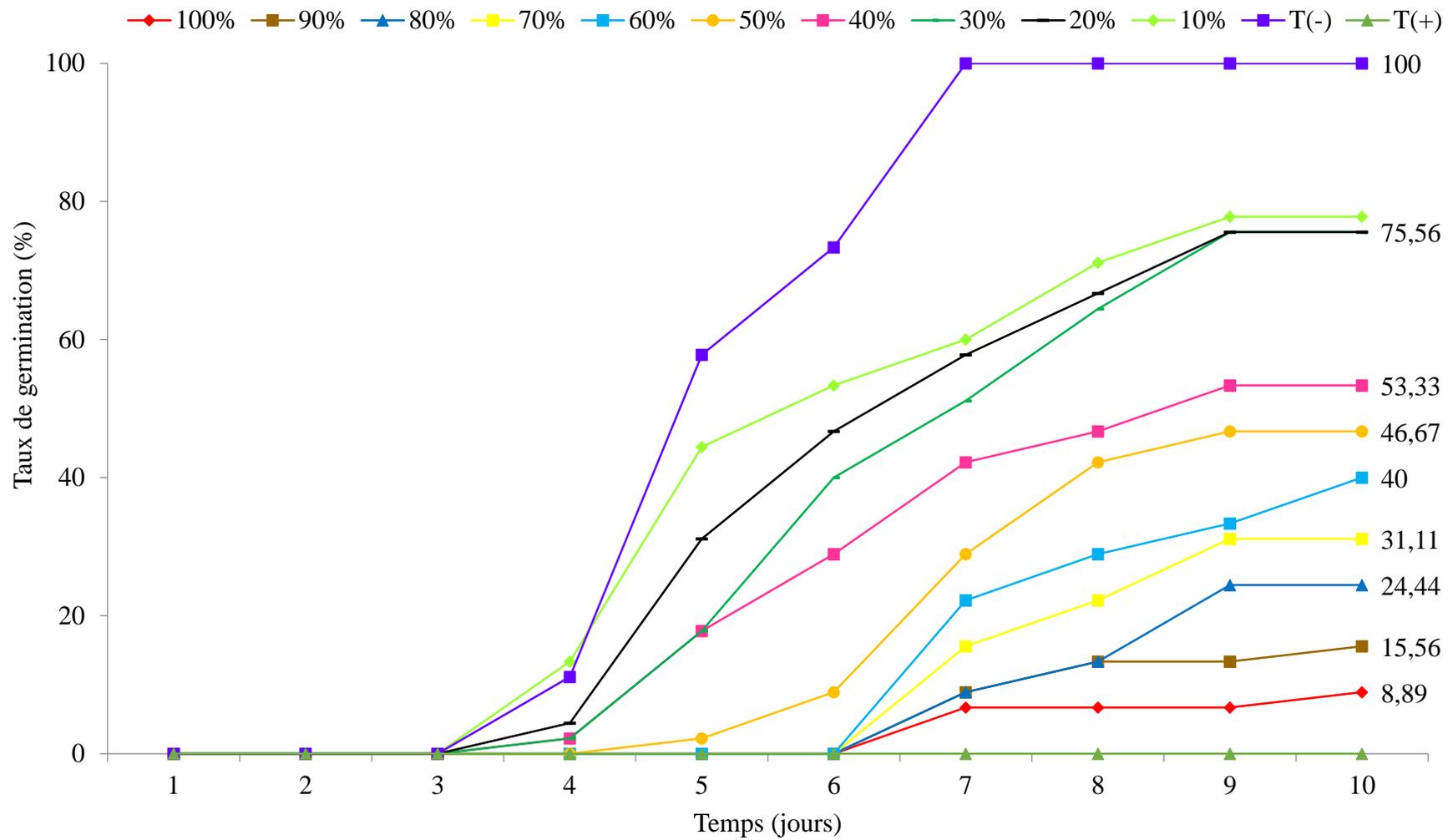

**B :** Cinétique de germination enregistré pour l'extrait aqueux de *C. arabica* sur les graines de Blé dur

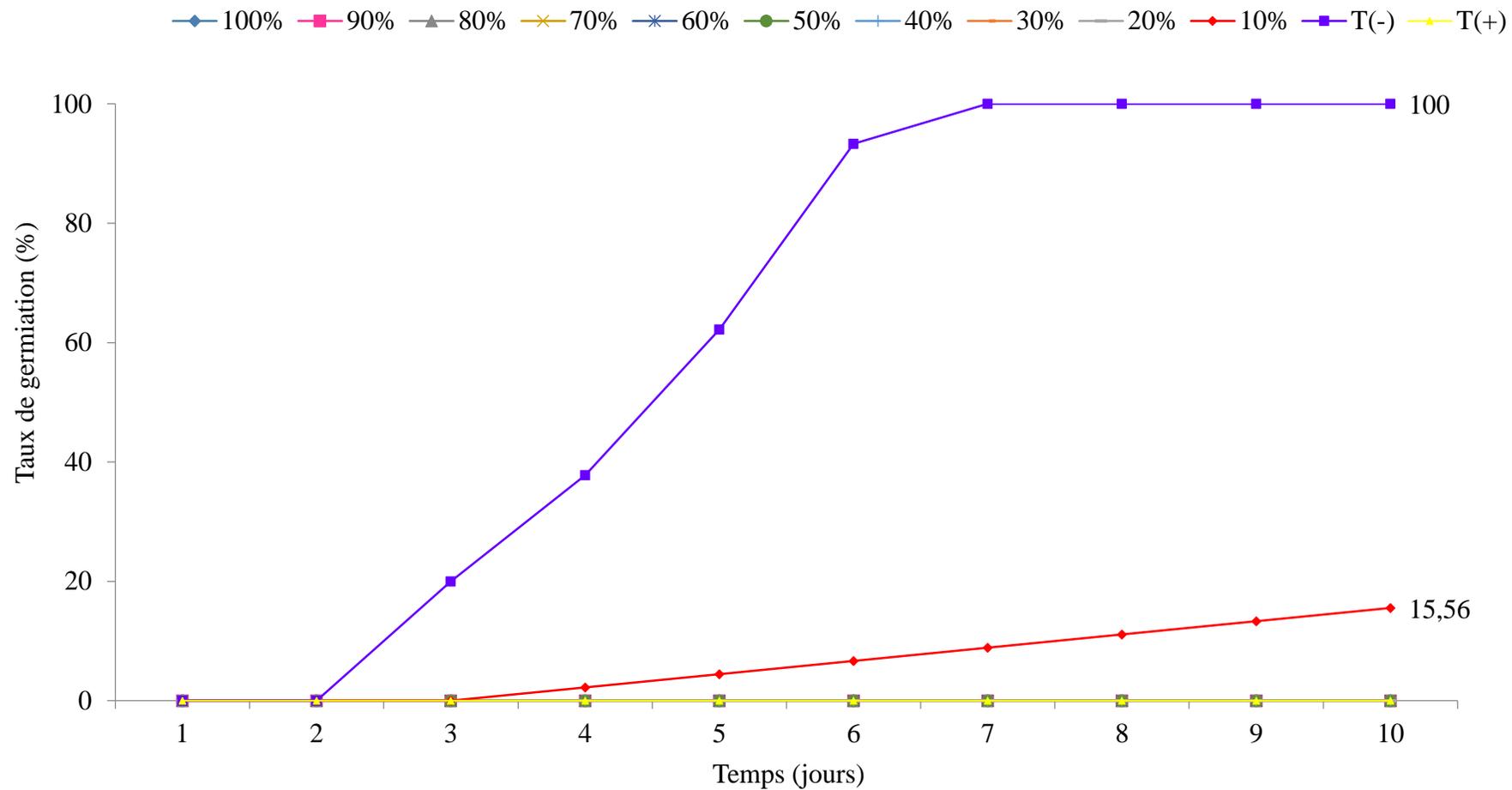

**C :** Cinétique de germination pour l'extrait aqueux de *C. arabica* sur les graines dactylocténion

**Figure 10 A, B, C:** Cinétiques de germination pour l'extrait aqueux de *C. arabica* sur les graines de trois espèces végétales testées

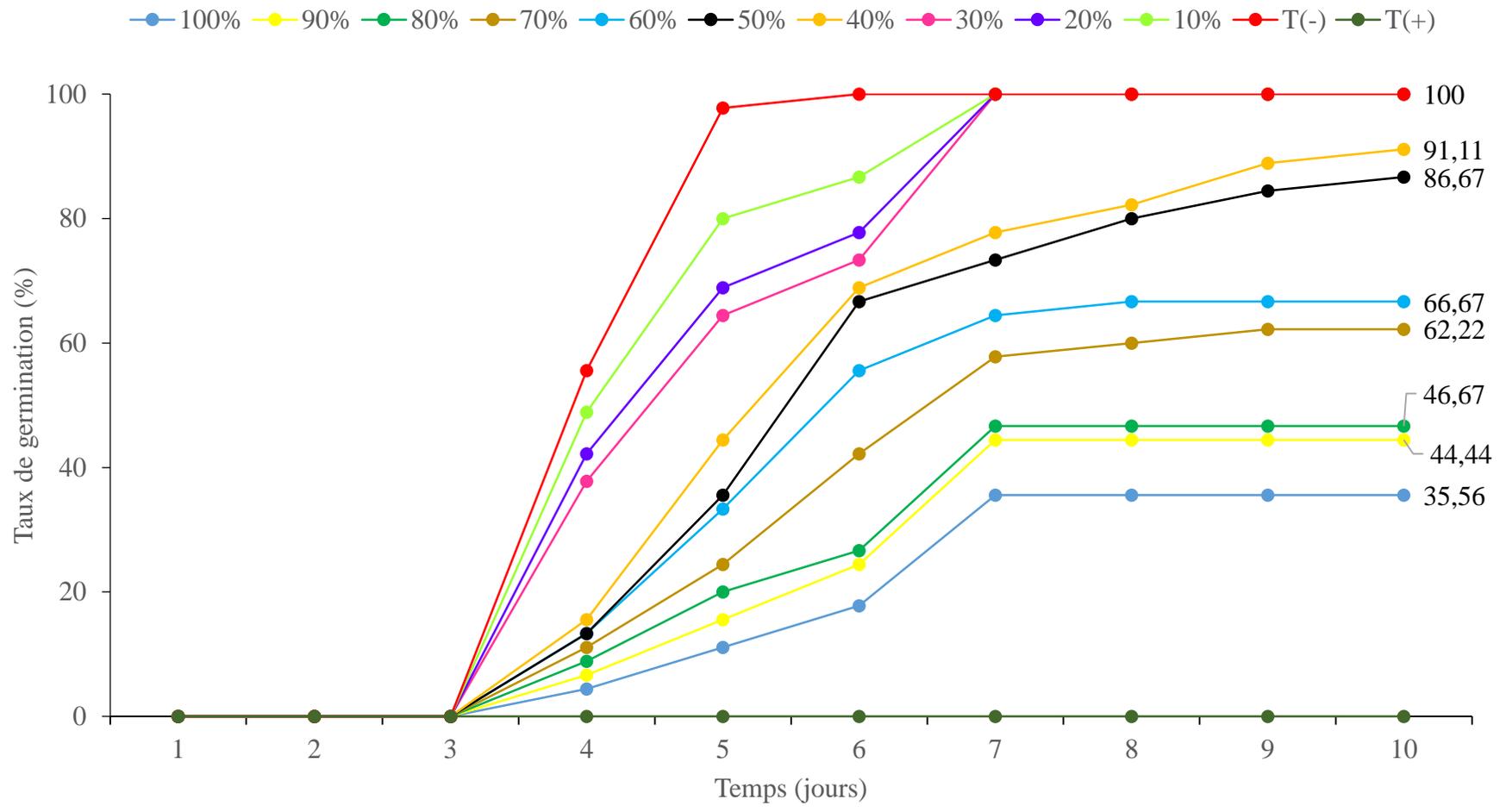

**A :** Cinétique de germination pour l'extrait aqueux de *P. tomentosa* sur les graines d'orge

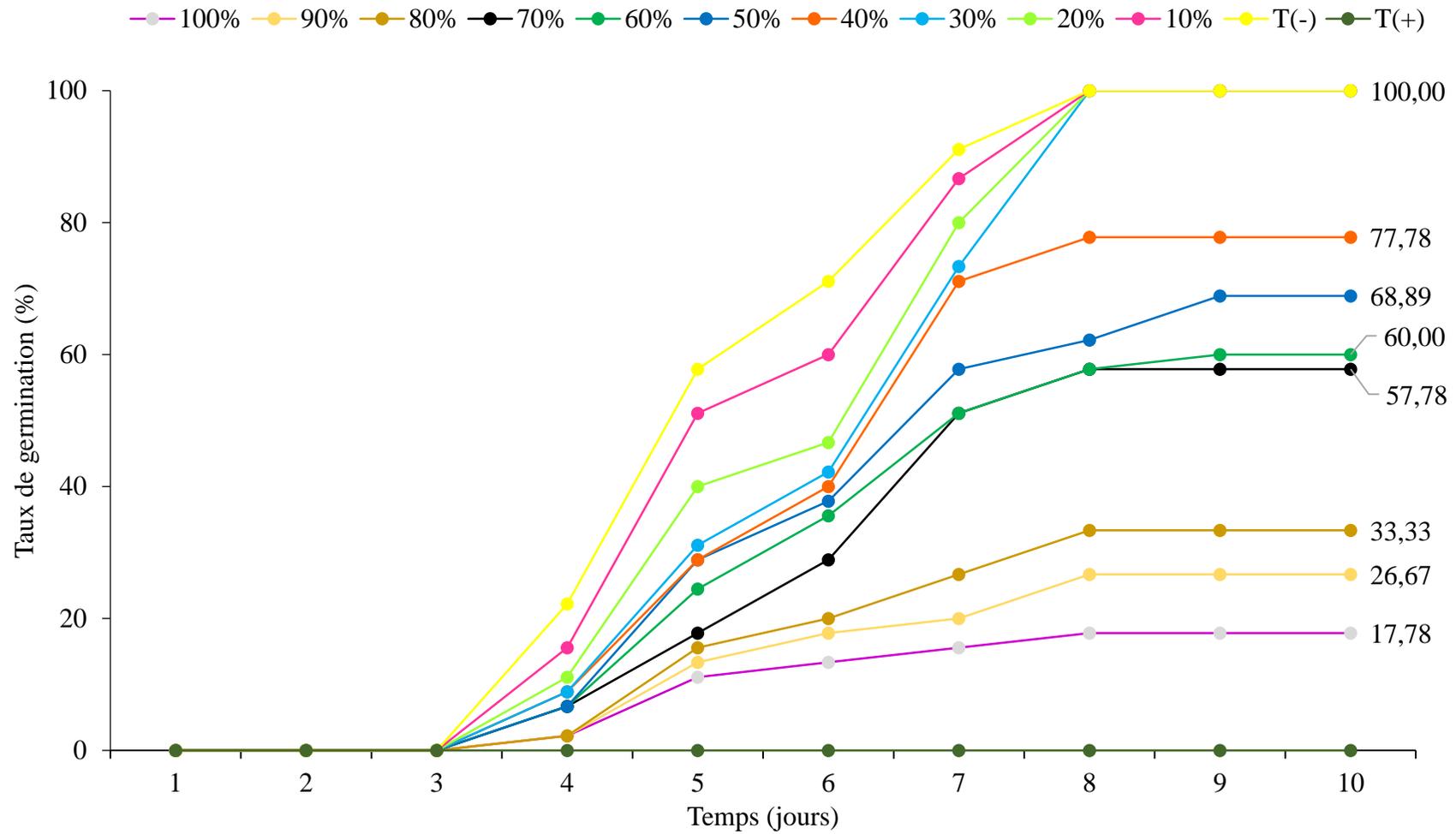

**B :** Cinétique de germination pour l'extrait aqueux de *P. tomentosa* sur les graines de Blé dur

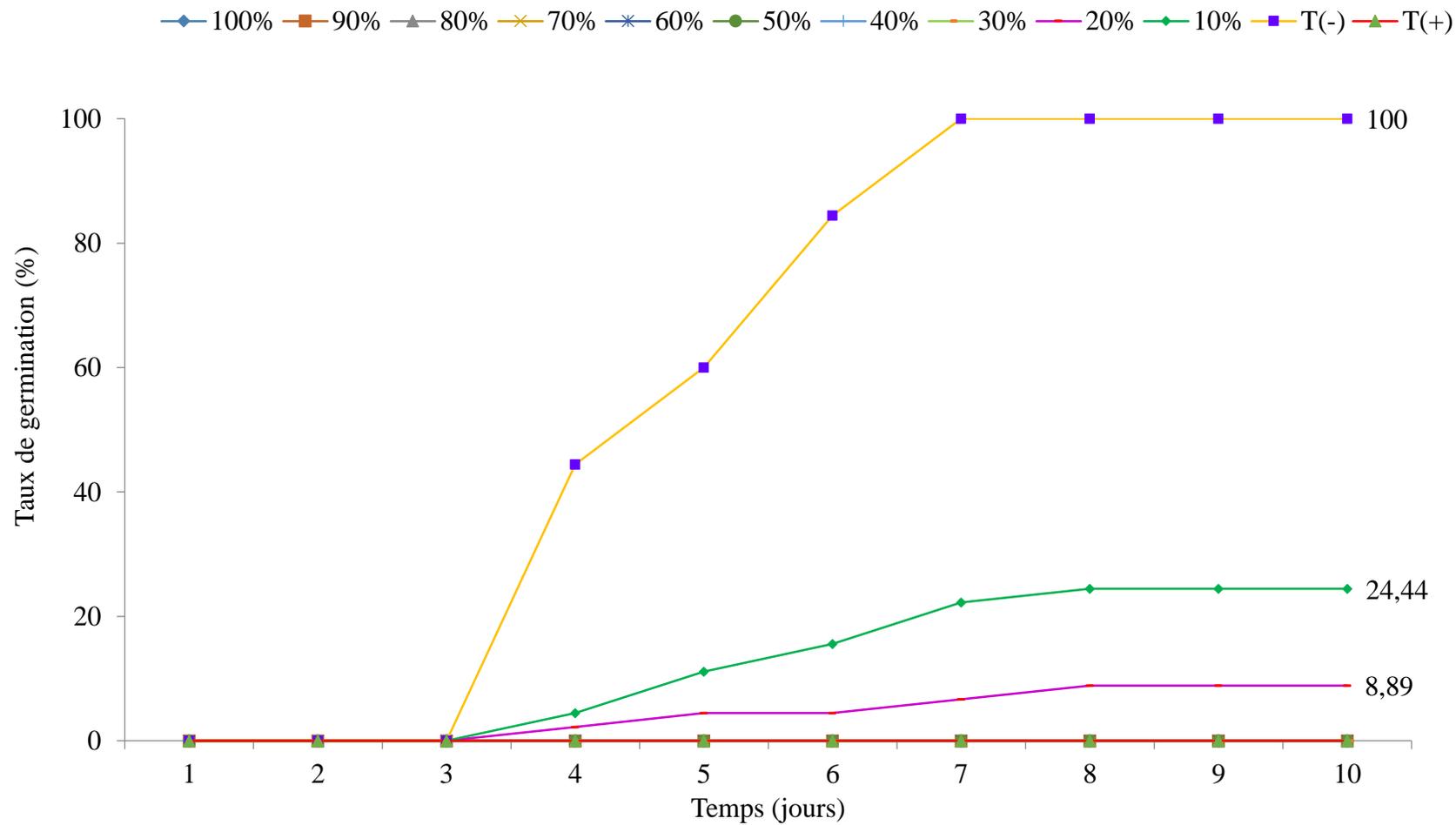

**C :** Cinétique de germination pour l'extrait aqueux de *P. tomentosa* sur les graines dactylocténion

**Figure 11**A, B, C **:** Cinétique de germination pour l'extrait aqueux de *P. tomentosa* sur les graines de trois espèces végétales traitées

Les résultats rapportés montrent que les trois espèces végétales testées (orge, blé dur, dactylocténion) sont sensibles à l'action des extraits aqueux de *C. arabica* et *P. tomentosa*, au stade de germination. Les graines *D. aegyptium* sont les plus sensibles, suivies par les graines de blé dur, puis les graines d'orge. En effet, la capacité germinative des variétés étudiées diminue avec l'augmentation de la concentration appliquée. En outre, l'effet inhibiteur de ces deux extraits (*C. arabica* et *P. tomentosa*) sur la germination et le développement des plantules peut être expliqué par la différence de la dose utilisée (concentration) et les caractéristiques physicochimiques de l'espèce allélopathique qui probablement mettent en jeu des substances allélochimiques spécifiques.

Selon FRIEDMAN (1995), les métabolites secondaires sont généralement inhibiteurs de la croissance des racines, des tiges, des feuilles et de la croissance globale de la plante cible. Durant la période de suivi expérimental, l'application des fortes doses sur les graines dactylocténion, provoquent un effet inhibiteur significatif sur l'émergence de la radicule, ce qui explique les effets négatifs des extraits allélopathiques qui conduisent à l'inhibition ou à un blocage total de la croissance.

Il est aussi testé les deux espèces végétales (blé dur et orge) en culture, afin de vérifier si les espèces allélopathiques (*C. arabica* et *P. tomentosa*) affectent également les plants de culture ou non. Les résultats obtenus montrent que les croissances des plantes de blé dur et l'orge ne sont pas affectées par les différentes concentrations d'extraits.

ENNACERIE *et al*. (2018), ont signalé que le taux de germination enregistrée respectivement 57,5% et 72% après 48 heures pour les graines de *Lepidium sativum* (*Brassicaceae*) irriguées par l'extrait aqueux des fleurs et des fruits de *Capparis spinosa* (*Capparidaceae*). Après 48 heures, le taux germination est noté respectivement de 72% et 50% pour les graines aspergées par l'extrait alcoolique des fleurs et des fruits *Capparis spinosa*.

## II.3.3.-Concentrations d'efficacités $CE_{50\%}$ et $CE_{90\%}$

Les essais préliminaires de l'effet herbicide du l'extrait *C. arabica* et *P. tomentosa* ont été effectués sur l'orge, le blé dur et le dactylocténion, afin d'estimer les concentrations entrainant l'inhibition de 50% et 90% selon la méthode des Probit. Plusieurs auteurs ont utilisés cette même méthode pour l'évaluation quantitative de la toxicité de leurs extraits végétaux dont (DJEUGAP *et al*., 2011 ; KEMASSI *et al*., 2014; 2015).

Les concentrations d'efficacités ($CE_{50\%}$ et $CE_{90\%}$) de l'extrait *C. arabica* et *P. tomentosa* vis-à-vis les trois espèces végétales traitées sont regroupés dans le tableau 10. Les concentrations qui engendrent une inhibition de 50% et 90% des graines d'orge sont respectivement de l'ordre de 0,020mg/ml et 0,037mg/ml. Les $CE_{50\%}$ et $CE_{90\%}$ sont de 0,012mg/ml et 0,028mg/ml pour les graines de blé dur et de 0,0001mg/ml et 0,0014mg/ml pour les graines de dactylocténion. Les graines *D. aegyptium* sont plus sensibles à l'action de l'extrait foliaire de *C. arabica* comparativement aux graines d'orge et le blé dur.

Les concentrations d'efficacités ($CE_{50\%}$ et $CE_{90\%}$) de l'extrait *P. tomentosa* à l'égard des trois espèces sont dressées dans le tableau 13. D'après les résultats retenus, les concentrations d'efficacités 50% et 90% obtenues sont de 0,026mg/ml et 0,041mg/ml. Pour les graines de blé dur, les $CE_{50\%}$ et $CE_{90\%}$ mentionnées sont de 0,023mg/ml et 0,035mg/ml respectivement. Les $CE_{50\%}$ et $CE_{90\%}$ sont de 0,0009mg/ml et 0,004mg/ml pour les graines dactylocténion.

L'extrait aqueux de *C. arabica* a révélé un effet inhibiteur significatif sur la germination des graines semis par rapport l'extrait aqueux de *P. tomentosa*. Cela est probablement dû à la présence et la distribution des molécules allélochimiques dans les feuilles de la plante utilisé. La répartition de ces substances selon BUBEL (1988) dans la plante diffère d'un organe à un autre et d'une plante à une autre.

Les résultats de criblage phytochimique des extraits foliaires bruts de *C. arabica* et *P. tomentosa*, mentionnent que la partie aérienne (feuilles) des plantes choisies est riche en métabolites secondaires. RICE (1984) et PUTNAM (1985) rapportent que les feuilles peuvent être une source importante des composés chimiques. ALDRICH (1984) rapporte que les substances actives sont concentrées dans les feuilles, les tiges ou les racines plutôt que dans les fruits ou les fleurs. ALGHANEM et EL-AMIER (2017), ont entamé le potentiel allélopathique de l'extrait de *P. tomentosa* sur la germination de *Chenopodium murale* L. (*Chenopodioideae*) pendant quatre jours. Ils ont observé que la phyto-toxicité des extraits de *P. tomentosa* augmente de manière significative avec l'augmentation des concentrations d'extrait. La dose 20g/l engendre l'inhibition de 100% de graine de *C. murale* et était de l'ordre de 10,42%, pour la concentration la plus faible 2,5 g/l.

Les résultats d'ALGHANEM et EL-AMIER (2017), sur le criblage phytochimique de l'extrait brut de *C. arabica* montrent la présence de certains groupes phytochimiques dont les alcaloïdes, les flavonoïdes, les saponosides, etc. ALGHANEM et EL-AMIER (2017), signalent la présence de tanins, de saponines, de flavonoïdes et d'alcaloïdes dans l'extrait de *P. tomentosa*. GOHAR *et al.*,

(2000) et GREEN *et al*., (2011) ont trouvé des cardénolides, glycosides, des β-sitosterolglucoside, des alcaloïdes, des tanins, des flavonoïdes et des composés phénoliques dans les extraits de *C. arabica*.

LADHARI *et al*., (2013) révélent un effet phytotoxique pour différentes parties dont les racines, pousses, siliques, graines de *C. arabica* sur la germination et la croissance des graines de laitue *Lactuca sativa* L. (*Asteraceae*) et du *Peganum harmala* L. (*Zygophyllaceae*). L'extrait aqueux de la racine était le plus toxique pour la croissance des plantules de laitue, qui présentait une inhibition significative de 94%, suivis des extraits de siliques et des graines induisant une inhibition de 74%, puis de l'extrait de pousses avec une inhibition de 53% à la concentration la plus faible.

LADHARI *et al*. (2013) rapportent que la majorité des extraits aqueux de *C. arabica* ont nettement réduit la germination et la croissance des espèces cibles avec un mécanisme dépendant de la concentration. Un certain nombre d'études antérieures CHUNG et MILLER (1995) et LAOSINWATTANA et *al*., (2009) ont indiqué que le degré d'inhibition augmentait avec l'augmentation des concentrations d'extrait. Les extraits aqueux de *C. arabica* contiennent des substances hydrosolubles qui inhibent la germination et la croissance des espèces cibles.

BOUAFIANE *et al*., (2014) notent que la concentration d'efficacité ($CE_{50\%}$) estimée pour *Eucalyptus occidentalis* (Myrtaceae) est de l'ordre de 0,0145mg/ml vis-à-vis des graines d'une plante adventice *Polygonum monspeliensis* L. (*Poaceae*) associée à la culture de pomme de terre au Sahara Algérien. BELEAIDI (2014) a révélé que la concentration d'efficacité ($CE_{50\%}$) est de l'ordre de 0,007mg/ml pour les graines de *D. aegyptium aegyptium* (*Poaceae*) et de 0,008mg/ml pour *Hordeum vulgare* (*Poaceae*). BELAIDI (2014), mentionne que les graines de *D. aegyptium* sont plus sensibles à l'action de l'extrait foliaire de *Nerium oleander* L.

**Tableau 08-** Taux d'inhibition et Probit correspondants en fonction de la concentration de l'extrait aqueux de la plante *C. arabica*.

| | Concentrations | | Taux d'inhibition maximal d'Orge | | Taux maximal d'inhibition du Blé | | Taux maximal d'inhibition du Dactylocténion | |
|---|---|---|---|---|---|---|---|---|
| (%) | *C. arabica* (mg/ml) | Log [mg/ml] | (%) | Probit | (%) | Probit | (%) | Probit |
| **100** | 0,0315 | -1,50168945 | 77,78 | 5,77 | 91,11 | 6,34 | 100 | 7,614 |
| **90** | 0,0284 | -1,54668166 | 71,11 | 5,55 | 84,44 | 5,99 | 100 | 7,614 |
| **80** | 0,0252 | -1,59859946 | 55,56 | 5,15 | 75,56 | 5,71 | 100 | 7,614 |
| **70** | 0,0220 | -1,65757732 | 46,67 | 4,92 | 68,67 | 5,5 | 100 | 7,614 |
| **60** | 0,0189 | -1,7235382 | 35,56 | 4,64 | 60 | 5,25 | 100 | 7,614 |
| **50** | 0,0157 | -1,80410035 | 35,56 | 4,64 | 53,33 | 5,08 | 100 | 7,614 |
| **40** | 0,0126 | -1,89962945 | 28,89 | 4,45 | 46,67 | 4,92 | 100 | 7,614 |
| **30** | 0,00945 | -2,02456819 | 13,33 | 3,87 | 24,44 | 4,29 | 100 | 7,614 |
| **20** | 0,0063 | -2,20065945 | 8,89 | 3,66 | 24,44 | 4,29 | 100 | 7,614 |
| **10** | 0,00315 | -2,50168945 | 0 | 0 | 22,22 | 4,23 | 84,44 | 5,99 |

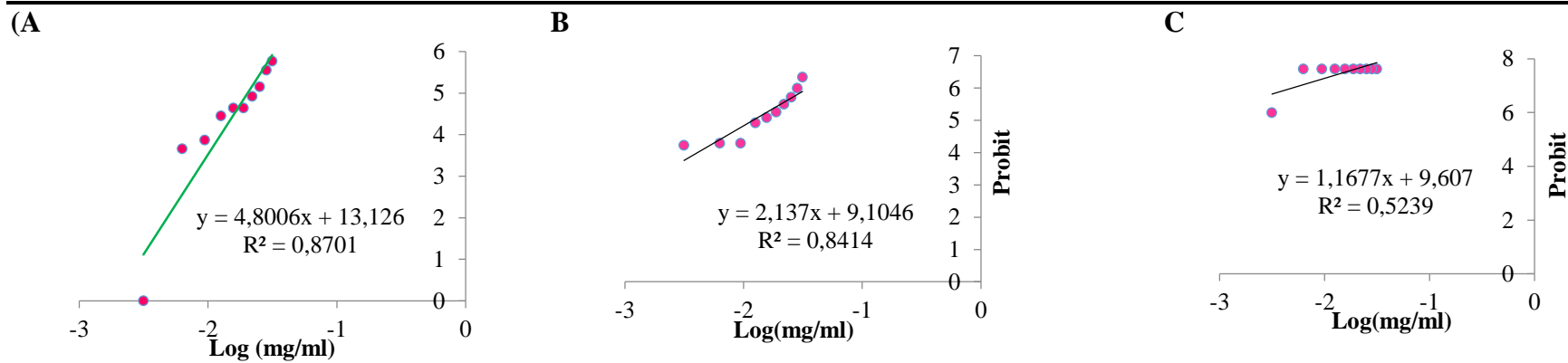

**Figure 12 A, B, C-** Droites de régression des doses de trois espèces testes traitées par l'extrait de *C. arabica*.
**A :** Orge, **B :** Blé dur, **C:** Dactylocténion

**Tableau 09-** Taux d'inhibition et Probit correspondants en fonction de la concentration de l'extrait aqueux de la plante *P. tomentosa*.

| | Concentrations | | Taux d'inhibition maximal d'orge | | Taux d'inhibition maximal de Blé | | Taux maximal d'inhibition du Dactylocténion | |
|---|---|---|---|---|---|---|---|---|
| (%) | P. tomentosa (mg/ml) | log [mg/ml] | (%) | Probits | (%) | Probit | (%) | Probit |
| **100** | 0,0315 | -1,50168945 | 64,44 | 5,36 | 82,22 | 5,92 | 100 | 7,614 |
| **90** | 0,0284 | -1,54668166 | 55,56 | 5,13 | 73,33 | 5,61 | 100 | 7,614 |
| **80** | 0,0252 | -1,59859946 | 53,33 | 5,08 | 66,67 | 5,41 | 100 | 7,614 |
| **70** | 0,022 | -1,65757732 | 37,78 | 4,67 | 42,22 | 4,8 | 100 | 7,614 |
| **60** | 0,0189 | -1,72353820 | 33,33 | 4,56 | 40 | 4,75 | 100 | 7,614 |
| **50** | 0,0157 | -1,80410035 | 13,33 | 3,87 | 31,11 | 4,5 | 100 | 7,614 |
| **40** | 0,0126 | -1,89962945 | 8,89 | 3,59 | 22,22 | 4,23 | 100 | 7,614 |
| **30** | 0,00945 | -2,02456819 | 0,00 | / | 0 | / | 100 | 7,614 |
| **20** | 0,0063 | -2,20065945 | 0,00 | / | 0 | / | 91,11 | 6,34 |
| **10** | 0,00315 | -2,50168945 | 0,00 | / | 0 | / | 75,56 | 5,67 |

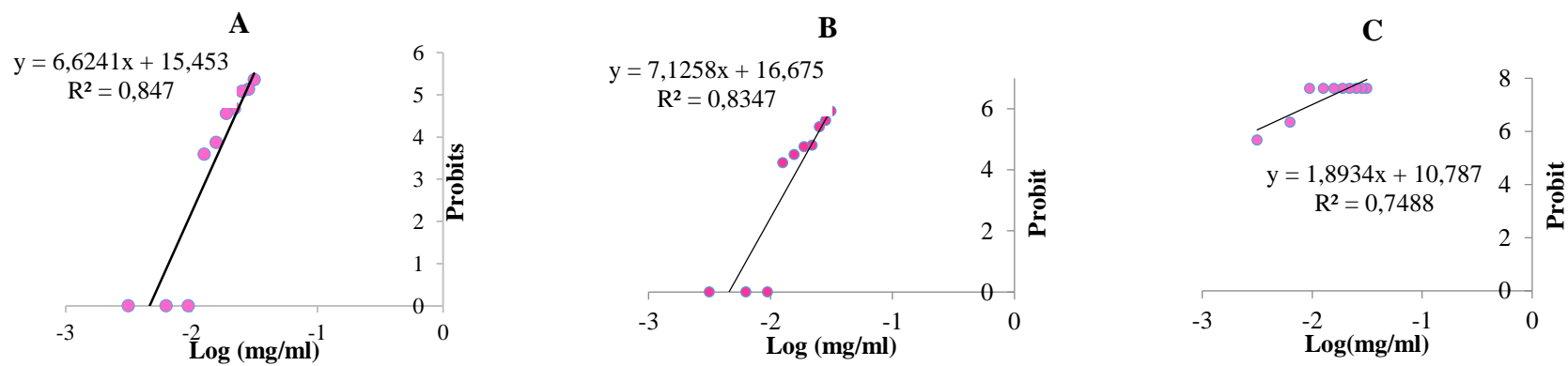

**Figure 13<sub>A, B,C</sub>-** Droites de régression des doses de trois espèces testes traitées par l'extrait de *P. tomentosa*
**A :** Orge, **B :** Blé dur, **C:** Dactylocténion

**Tableau 10-** Valeurs des $CE_{50\%}$ et des $CE_{90\%}$ des extraits testés *C. arabica* et *P. tomentosa* sur les graines de trois espèces végétales traitées

| Extraits appliqués | Espèces tests | $CE_{50\%}$ (mg/ml) | $CE_{90\%}$ (mg/ml) |
|---|---|---|---|
| *C. arabica* | Orge | 0,020 | 0,037 |
| | Blé dur | 0,012 | 0,028 |
| | Dactylocténion | 0,0001 | 0,0014 |
| *P. tomentosa* | Orge | 0,026 | 0,041 |
| | Blé dur | 0,023 | 0,035 |
| | Dactylocténion | 0,0009 | 0,0040 |

Les résultats obtenus ont montré le pouvoir allélopathique de l'extrait de *C. arabica* et *C. arabica* sur la germination des graines des espèces cibles aux fortes doses, qui pourrait être attribuée à la teneur élevée en composés phénoliques, tanins et alcaloïdes. La réduction de la germination des graines testées au cours du traitement pourrait être attribuée à l'action diversifiée des extraits végétaux appliqués et les espèces tests choisies. Ils affectent un grand nombre de réactions biochimiques au niveau de la graine semée et aussi, il provoque des modifications de différentes fonctions physiologiques.

Selon AN (1998) et ELISANTE (2013), les composés allélochimiques ont des effets importants sur les activités enzymatiques, la division et l'ultra-structure des cellules, la perméabilité des membranes, la capture d'ions et, en conséquence, la germination, la croissance et le développement des plantes sont modifiés.

## II.3.4.- Actions des extraits végétaux sur certains paramètres de croissance des plantules d'orge, blé dur et le dactylocténion

Les tableaux 11 et 12, regroupent les variations de la longueur de la partie aérienne et souterraine ainsi que la variation du poids frais en fonction des différents traitements de l'extrait *C. arabica* et *P. tomentosa*.

Au vus des résultats du tableau 11, la longueur de la partie aérienne et souterraine des plantules d'orge du lot témoin négatif est notée de 11,00±0,58cm/10,00±1,00cm et le poids frais est de l'ordre de 0,59±0,01mg /0,57±0,02mg.

En parallèle, les graines exposées à 100%, 90% et 80%, la longueur de la partie aérienne et souterraine est mesurée respectivement soit 8,00±2,52cm/7,00±1,00cm, 8,00±2,08cm/6,25±0,00cm et 8,75±1,53cm/8,25±2,31cm. Pour les poids frais des parties aériennes et souterraines, ils sont notés successivement de 0,55±0,02mg/0,54±0,01mg, 0,54±0,02mg/0,54±0,00mg et 0,55±0,01mg/0,53±0,01mg (Tab. 11).

Pour les graines d'orge aspergées à des concentrations 70%, 60%, 50%, 40%, 30%, 20% et 10%, la variation de la longueur de la partie aérienne par rapport à la longueur de la partie souterraine est enregistrée successivement de 12,25±1,00cm/8,50±1.00cm, 10,25±1,53cm/14,75±1,53cm, 11,00±2,52cm/13,75±0,00, 12,50±1,53cm/6,00±0,58, 10,50±1,15cm/12,25±2,65cm, 12,25±2,65cm/11,75±2,08cm et 16,75±1,00cm/12,25±2,65cm. Pour les poids frais des plantules d'orge sont consignées successivement pour les mêmes concentrations de 10% à 70% soient 0,60±0,02mg/0,55±1,00mg, 0,55±0,05mg/0,61±0,01mg, 0,59±0,03mg/0,61±0,00mg, 0,59±0,02/0,54±0,01mg, 0,58±0,01mg/0,59±0,03mg, 0,60±0,03mg/0,59±0,03mg et 0,63±0,03mg/0,60±0,03mg.

L'analyse de la variance indique que la croissance de la partie aérienne des plantules d'orge n'est guère affectée par l'effet de l'extrait de *C. arabica* aux concentrations 100%, 60%, 50% et 40%. Les valeurs de facteur F estimées sont dans leur ensemble pour une probabilité P>0,05 (Annexe 02).

Pour le lot traité à 90% de concentration, il affecte significativement la longueur de la partie aérienne des plantules d'orge, la valeur de facteur F observée est pour une probabilité 0,05<P<0,01. Une différence très hautement significativement est signalée dans la partie souterraine à une probabilité inférieure à P≤0,001 (Annexe 2).

L'effet de l'extrait de *P. tomentosa* à dose 80%, inhibe significativement la croissance de la partie aérienne (F=75,00 ; P=0,001) par apport la croissance des racines (P>0,05) (tab. 17).

L'application de l'extrait de *C. arabica* aux concentrations 70%, 30% et 20%, n'affecte peu la croissance de la partie aérienne et souterraine ; l'analyse de la variance effectuée montre un effet non significatif des traitements réalisées sur la croissance ; les valeurs de facteur F sont estimées pour une probabilité (P>0,05) (Annexe 2).

L'effet de l'extrait de *C. arabica* à dose 60%, n'est pas inhibé significativement (P>0,05) la croissance de la partie aérienne, est hautement significative (F=36,10 ; P=0,004) sur la croissance de la partie souterraine (Annexe 2).

L'effet inhibiteur de l'extrait de *C. arabica* à concentration 10% induit une différence très hautement significative sur la longueur de la partie aérienne (F= 132,25 ; P=0,000) et un effet non significatif sur la longueur de la partie souterraine (F=3,38 ; P=0,140).

Pour le lot traité à 100% de l'extrait *C. arabica*, une différence hautement significative est notée sur le poids de la partie aérienne (F=28,12 ; P=0,006), elle affecte significativement (F=16,90 ; F=0,015) le poids de la partie aérienne des plantules d'orge, par apport aux lots témoins négatifs.

Une différence hautement significative est notée sur les poids de la partie aérienne et souterraine des plantules d'orge traitées par les concentrations 90% et 80%. Les valeurs de facteur F observées pour une probabilité inferieure (P≤0,01) (Annexe 2).

Au niveau les lots traités à 70%, 30%, 20% et 10% de concentration, n'affectent pas significativement le poids de la partie aérienne et souterraine des plantules d'orge, pour une probabilité supérieure de (P>0,05) (Annexe 2).

L'effet de l'extrait *C. arabica* aux concentrations 60%, 50% et 40%, montre une différence hautement significative sur le poids de la partie souterraine (P≤0,01). Il n'affecte pas significativement le poids de la partie aérienne (P>0,05) (Annexe 02).

Pour les pots ensemencés par des graines de Blé dur, les concentrations 100%, 90%, 80% et 70% provoquent une élongation de la partie aérienne (LF) par rapport à la longueur de la partie souterraine (LR) où les mesures sont successivement enregistrées soit 20,17±2,57cm/17,83±2,47cm, 24,00±1,00cm/12,33±2,08cm, 20,93±2,93cm/16,33±1,15cm et 21,00±5,29cm/17,17±2,25cm. Le poids frais des parties aériennes et souterraines est mesuré respectivement pour les mêmes concentrations soit 0,12±0,02mg/0,12±0,02mg, 0,14±0,03g/0,11±0,02mg, 0,13±0,02mg/0,13±0,06mg et 0,15±0,04mg/0,13±0,03mg.

Pour les graines de blé dur traitées par 60%, 50%, 40%, 30%, 20% et 10%, la croissance de la partie souterraine dépasse nettement la longueur de la partie foliaire où LF/LR est déterminée successivement de 15,00±2,65/21,33±3,21cm, 17,67±3,21/22,00±1,73cm,

17,00±3,46/26,67±6,66cm, 16,00±1,00/21,67±2,52cm et 14,17±1,89/17,50±3,62cm. Pour les mêmes doses de 60%, 50%, 40%, 30%, 20%, 10%, le poids frais des parties aériennes et souterraines des plantules de blé dur sont notés successivement soit 0,12±0,01mg/0,14±0,01mg, 0,16±0,02mg/0,21±0,01mg, 0,16±0,05mg/0,21± 0,01mg, 0,15±0,01mg/0,19±0,04mg, 0,14±0,02mg/0,22±0,02mg et 0,11±0,02mg/0,18±0,04mg.

L'effet inhibiteur de l'extrait *C. arabica* à dose 100%, n'a pas un effet significatif sur la croissance de la partie aérienne et souterraine (P>0,05). Alors que, cette différence est hautement significative dans le lot traité par l'extrait à 90% (F=33,80 ; P=0,004) et (F=52,07 ; P=0,002) (Annexe 2).

L'effet de l'extrait de *C. arabica* est hautement significatif sur la longueur des racines et significatif sur la longueur de partie aérienne de blé dur à concentration 80% (F=45,00 ; P=0,003) et (F=8,25 ; P=0,045) (Annexe 2).

Pour le lot traité par l'extrait à 70%, une différence hautement significative est notée pour la partie souterraine (F=9,62 ; P=0,036) et non significative sur la longueur de la partie aérienne (F= 3,94 ; P=0.118) (Annexe 2).

L'extrait de *C. arabica* aux concentrations 60%, 50%, 40% et 30% ne présentent aucun effet significatif sur la longueur de la partie aérienne et aussi souterraine les valeurs de facteur F sont estimées sont une probabilité (P>0,05) (Annexe 2).

L'effet de l'application de l'extrait à 20% de concentration, ne présent pas un effet significatif sur la croissance de la partie aérienne, bien qu'il affecte significativement la partie souterraine. Les valeurs de F observées sont de (F=1,89 ; P=0,241) et (F=10,12 ; P=0,033) respectivement. Par contre, la faible concentration (10%), engendre une différence très hautement significative de la longueur de la partie souterraine par rapport au témoin négatif, et un effet non significatif sur la croissance de la partie aérienne ; les valeurs de facteur F observées sont (F=75,57 ; P=0,001) pour la partie souterraine et de (F=0,00 ; P=1,000) pour la partie aérienne (Annexe 2).

Les résultats de la variance montre que tous les concentrations n'inhibent pas significativement le poids de la partie aérienne de blé dur.

Les résultats de l'analyse de la variance de l'extrait de *C. arabica* à 100% et 90%, montrent une différence très hautement significative sur le poids de la partie souterraine des plantules d'orge, les valeurs de facteur F observées pour une probabilité inferieure ($P \leq 0{,}01$) (Annexe 2).

Pour les lots traités par les concentrations de 70% et 60%, ils affectent significativement le poids de la partie souterraine que la partie aérienne. Elles sont notées pour une probabilité de ($P \leq 0{,}05$) (Annexe 2).

Pour les concentrations dont 80%, 50%, 40%, 30%, 20% et 10%, n'ont pas un effet significatif sur le poids de partie aérienne et souterraine de blé dur. Les valeurs de facteurs F sont enregistrées pour une probabilité supérieure de ($P > 0{,}05$) (Annexe 2).

Pour les graines dactylocténion traitées par les concentrations 100%, 90%, 80%, 70%, 60%, 50%, 40%, 30% et 20%, aucune graine n'a germé, à l'exception, des graines exposée à une concentration 10%, où la longueur de la partie aérienne et souterraine est très réduite, elle est de 1,72±0,05cm/0,12±0,25cm et le poids frais est de 0,05±0,25mg/0,003±0,01mg (Annexe 2).

Pour le lot traité par la concentration 10% de l'extrait *C. arabica*, une différence très hautement significative ($P \leq 0.001$) est rapportée sur la croissance de la partie aérienne et souterraine de l'espèce dactylocténion (Annexe 2).

En comparaison avec le témoin négatif, l'effet inhibiteur de l'extrait *C. arabica* n'est pas affectée significativement le poids de la partie aérienne et souterraine des graines dactylocténion traitées par la dose 10%. Les valeurs de facteur F observées pour une probabilité de (F=0,40 ; P=0,561) et (F=0,00 P=1,000) (Annexe 2).

Le tableau 12 regroupe les variations de la longueur de partie aérienne et souterraine des plantules traitées aux différentes concentrations de *P. tomentosa*.

L'effet de l'extrait de *P. tomentosa* à concentrations de 100%, 90%, 70%, 50%, 40%, 30%, 20% et 10%, provoque une élongation de la partie aérienne que la partie souterraine des plantules d'orge, elle est de l'ordre 26,17±2,52/14,00±1,00cm, 23,83±4,04/10,50±1,50cm, 25,33±4,65/15,17±0,76cm, 27,67±2,47/16,17±4,19cm, 28,83±1,89/11,50±2,18cm, 24,50±2,18/12,33±4,04cm et 25,83±1,76/14,33±3,55cm respectivement (Annexe 2).

Pour les mêmes concentrations, le poids frais des plantules d'orge est noté respectivement de 0,21±0,04mg/0,18±0,03mg, 0,25±0,04mg/0,21±0,02mg, 0,21±0,06mg/0,15±0,06mg, 0,26±0,04mg/0,20±0,03mg, 0,29±0,05mg/0,21±0,03mg, 0,25±0,05mg/11,5±2,18mg, 0,22±0,02mg/0,19±0,01mg et 0,26±0,06mg/0,22±0,07mg (Annexe 2).

Pour les doses 80% et 60%, la longueur de la partie souterraine est plus importante que la partie arienne des plantules d'orge, il est enregistré respectivement de 10,83±0,15/20,67±3,53cm et 12,00±1,00/19,50±2,50cm et le poids frais est d'ordre de 0,13±0,04/0,21±0,04mg et 0,11±0,02/0,20±0,01mg respectivement (tab.12).

L'analyse de la variance de l'extrait aqueux de *P. tomentosa* à 100% et à 70% de concentration, montre un effet significatif sur la croissance de la partie souterraine par apport la partie aérienne, les valeurs de facteurs F observées pour une probabilité inferieure ($P \leq 0,05$) (Annexe 2).

Pour les lots traités par les concentrations 90%, 80%, 50%, 40%, 20% et 10%, ils ne présentent aucun effet significatif sur la croissance de la partie aérienne et souterraine des plantules d'orge. Elles sont enregistrées pour une probabilité supérieure ($P>0.05$) (Annexe 2).

Une différence très hautement significative est rapportée sur la longueur de la partie aérienne et souterraine chez les lots traités à dose 60%. Elle est notée pour une probabilité de ($F=144,60$ ; $P=0,000$) et ($F=21,13$ ; $F=0,010$). La concentration 30%, ne présente aucun effet significatif sur la croissance de la partie souterraine ($P>0,05$), mais significatif sur la croissance de la partie aérienne ($F=8,80$ ; $P=0,041$), par apport au témoin négatif.

L'effet inhibiteur de l'extrait de *P. tomentosa* aux concentrations 100%, 90%, 80%, 70%, 50%, 40%, 30%, 20%, et 10%, ne présent aucun effet significatif sur le poids de la partie aérienne et souterraine de l'espèce d'orge. Les valeurs de facteurs F sont enregistrées pour une probabilité supérieure ($P>0,05$) (Annexe 2).

Une différence très hautement significative ($F=112,00$ ; $F=0,00$) pour le poids de la racine et significative ($F=15,52$ ; $P=0,017$) pour la partie aérienne à dose 60%.

En comparaison avec les graines de blé dur aspergées par l'eau distillée (témoin), la longueur de la partie aérienne est de 14,17±2,47cm et la longueur de la partie souterraine est de 12,83±1,04cm, Le poids frais des plantules du blé dur est noté de 0,16±0,01mg/0,13±0,02mg.

Selon les valeurs dressées dans les tableaux 12 et 11, il apparait que les graines exposées à 100% et 90%, la longueur de la partie souterraine dépasse la longueur de la partie aérienne. Elle est respectivement d'ordre 7,33±0,29/8,17±3,44cm et 5,67±1,53/11,83±3,18cm. Pour les mêmes doses, le poids frais est enregistré de 0,04±0,01mg/0,05±0,01mg et 0,06±0,03mg/0,07±0,07mg.

Les graines de blé dur traitées par 80% et 70%, il apparait que la croissance de la partie aérienne dépasse nettement la longueur de la partie souterraine où les mesures sont notées respectivement de 17,5±0,87cm/15,00±0,77cm et 16,5±0,50cm/15,17±1,53cm. Pour les mêmes doses, le poids frais est respectivement de 0,16±0,05mg/0,14±0,06mg et 0,13±0,02mg/0,16±0,04mg.

Pour les graines du blé dur exposées à 60%, la longueur de la partie aérienne égale à la longueur de la partie souterraine où elle est notée de 7,50±1,32/7,5±1,32cm, et le poids frais est enregistrée de 0,08±0,04mg/0,09±0,02mg.

Les concentrations 50%, 40%, 30%, 20% et 10%, provoquent une élongation de la partie souterraine et réduisent la longueur de la partie aérienne des plantules blé dur, où les valeurs sont mentionnées successivement de 8,33±0,58cm/12,17±5,25cm, 3,67±0,29cm/10,83±1,89cm, 6,33±2,36cm/8,67±4,51cm, 10,67±0,58cm/13,17±1,61cm et 7,50±6,50cm/6,67±5,55cm. Pour les mêmes concentrations, le poids frais des plantules blé dur est noté respectivement de 0,06±0,02mg/0,08±0,06mg, 0,06±0,05g/0,09±0,05mg, 0,05±0,01mg/0,09±0,06mg, 0,10±0,02mg/0,10±0,01mg et 0,09±0,04mg/0,10±0.06mg.

La croissance de la partie aérienne et souterraine des plantules d'orge, est affectée hautement significatif voir très hautement significatif par l'effet de l'extrait aqueux de *P. tomentosa* à concentration de 100%, pour une probabilité inferieure ($P \leq 0,01$) (Annexe 2).

Pour les lots traités à 90% et 40%, ils provoquent une différence hautement significative sur la croissance de la partie aérienne et non significative sur la longueur de la partie souterraine, le facteur F observé pour une probabilité supérieure de ($P > 0,05$) (Annexe 2).

Une différence hautement significative de la longueur de la partie aérienne et souterraine au niveau des lots traités par les concentrations de 90% et 40%, le facteur F observé respectivement pour une probabilité de F=25,75 ; P=0,007, F=0,27 ; P=0,632 et F=53,64 ; P=0,002, F=2,57 ; P=0,184.

La longueur de la partie aérienne et souterraine des plantules d'orge, ne sont guère affectés par l'effet de l'extrait de *P. tomentosa* aux concentrations de 80%, 70%, 20% et 10%, les valeurs sont citées pour une probabilité supérieure P>0,05 (Annexe 2).

Une différence hautement significative est rapportée à F=30,12 et P=0,005 pour la croissance de la partie souterraine et une différence significative F=17,02 ; P=0,015 pour la longueur de la partie aérienne au niveau de lot traité par la dose 60%.

Pour le lot traité à 50% de concentration, il ne présente aucun effet significatif sur la longueur de la racine F=0,05 ; P=0,840, et significativement sur la longueur de la partie aérienne de blé dur F=15,91 ; P=0,016.

Une différence très hautement significative pour la croissance de la partie aérienne et souterraine de blé dur, le facteur F observé pour une probabilité inferieure de ($P \leq 0.001$) (Annexe 2).

Pour les lots traités à 90%, 50% et 20% de concentration, ils présentent une différence hautement significativement sur le poids de partie aérienne F=28,13 ; P=0,006, F=46,12 ; P=0,002 et F=29,45 ; P=0,006, et significative pour le poids de la racine F=8,10 ; P=0,047, F=10,47 ; P=0,032 et F=9,85 ; P=0,035.

L'effet de l'extrait *P. tomentosa* à concentration 80% et 70%, ne présentent aucun effet significatif sur le poids de partie aérienne et souterraine (P>0,05) (Annexe 2).

Pour les concentrations 60% et 10%, présentent un effet significatif sur le poids de partie aérienne F=8,64 ; P=0 et F=10,76 ; P=0,031 et hautement significatif sur le poids de la partie souterraine F=38,44 ; P=0,003 et F=49,47 ; P=0,002.

Pour la concentration 40%, une différence significativement sur le poids de la partie aérienne et souterraine de blé dur. Les valeurs sont citées respectivement de F=9,56 ; F=0,037 et F= 16,49 ; P=0,015.

Pour les graines dactylocténion, aucune graine germée pour les pots arrosés par les concentrations 100%, 90%, 80%, 70%, 60%,50%, 40% et 30%. Par contre, pour les pots de 20% et 10%, la longueur de la partie aérienne et souterraine est notée respectivement de 1,51±0,022/1,02±0,50cm, 0,71±1,07/0,12±0,20cm et les valeurs de poids frais est respectivement d'ordre 0,03±0,01/0,03±0,01mg et 0,01±0,01/0,01±0,002mg.

L'effet inhibiteur de l'extrait de *P. tomentosa* à concentration 100%, 90%, 80%, 70%, 60%, 50%, 40% et 30%, se manifeste par l'absence totale de la germination des graines de dactylocténion.

La croissance de la partie aérienne et souterraine des lots traités par 20% et 10%, est hautement significative voir à très hautement significative, les valeurs de facteur F observée pour une probabilité inferieure (P≤0,01) (Annexe 2).

L'effet inhibiteur de l'extrait de *P. tomentosa* à concentration 20%, ne présente aucun effet significatif sur le poids de la partie aérienne et souterraine de dactylocténion. Elle est de F=2,58 ; P=0,184 et F=4,50 ; P=0,101 respectivement et un effet significatif sur le poids de la partie souterraine à concentration 10%. La valeur F est de F=4,92 ; P=0,091 et F=9,80 ; P=0,035.

Les extraits végétaux de *C. arabica* et *P. tomentosa* sont influencés d'une manière variable sur la croissance des plantules d'orge, de blé due et dactylocténion soit par stimulation et ou inhibition de la germination des graines traitées. Les anomalies de croissance se traduisent par l'élongation de la partie aérienne que la partie souterraine ou l'inverse, ainsi que le flétrissement et le jaunissement de partie foliaire.

**Tableau 11 : Lots expérimentaux de *C. arabica L.***

| Espèces tests | | | T(-) | 100% | 90% | 80% | 70% | 60% | 50% | 40% | 30% | 20% | 10% |
|---|---|---|---|---|---|---|---|---|---|---|---|---|---|
| **Orge** | PA | LT | 11.00±0.58 | 8.00±2.52 | 8.00±2.08 | 8.75±1.53 | 12.25±1.00 | 10.25±1.53 | 11.00±2.52 | 12.50±1.53 | 10.50±1.15 | 12.25±2.65 | 16.75±1.00 |
| | | PT | 0.59±0.01 | 0.55±0.02 | 0.54±0.02 | 0.55±0.01 | 0.60±0.02 | 0.55±0.05 | 0.59±0.03 | 0.59±0.02 | 0.58±0.01 | 0.60±0.03 | 0.63±0.03 |
| | PS | LR | 10.00±1.00 | 7.00±1.00 | 6.25±0.00 | 8.25±2.31 | 8.50±1.00 | 14.75±1.53 | 13.75±0.00 | 6.00±0.58 | 12.25±2.65 | 11.75±2.08 | 12.25±2.65 |
| | | PR | 0.57±0.02 | 0.54±0.01 | 0.54±0.00 | 0.53±0.01 | 0.55±1.00 | 0.61±0.01 | 0.61±0.00 | 0.54±0.01 | 0.59±0.03 | 0.59±0.03 | 0.60±0.03 |
| **Blé dur** | PA | LT | 14.17±2.75 | 20.17±2.57 | 24.00±1.00 | 20.83±2.93 | 21.00±5.29 | 15.00±2.65 | 17.67±3.21 | 17.00±3.46 | 16.00±1.00 | 16.67±1.53 | 14.17±1.89 |
| | | PT | 0.12±0.04 | 0.12±0.02 | 0.14±0.03 | 0.13±0.02 | 0.15±0.04 | 0.12±0.01 | 0.16±0.02 | 0.16±0.05 | 0.15±0.01 | 0.14±0.02 | 0.11±0.02 |
| | PS | LR | 21.33±0.58 | 17.83±2.47 | 12.33±2.08 | 16.33±1.15 | 17.17±2.25 | 21.33±3.21 | 22.00±1.73 | 26.67±6.66 | 21.67±2.52 | 24.33±1.53 | 17.50±3.62 |
| | | PR | 0.21±0.01 | 0.12±0.02 | 0.11±0.02 | 0.13±0.06 | 0.13±0.03 | 0.14±0.01 | 0.21±0.01 | 0.21±0.01 | 0.19±0.04 | 0.22±0.02 | 0.18±0.04 |
| **Dactylocténion** | PA | LT | 12,73±0,15 | / | / | / | / | / | / | / | / | / | 1,72±0,05 |
| | | PT | 0,12±0,04 | / | / | / | / | / | / | / | / | / | 0,05±0,01 |
| | PS | LR | 12,13±0,22 | / | / | / | / | / | / | / | / | / | 0,12±0,25 |
| | | PR | 0,12±0,03 | / | / | / | / | / | / | / | / | / | 0,002±0,001 |

PA : partie aérienne, PS : partie souterraine, LF : longueur de feuille, PT : poids de tige, LR : longueur de racine, PR : poids de racine.

**Tableau 12 : Lots expérimentaux de Plante *C. arabica L.***

| Espèces tests | | | T(-) | 100% | 90% | 80% | 70% | 60% | 50% | 40% | 30% | 20% | 10% |
|---|---|---|---|---|---|---|---|---|---|---|---|---|---|
| **Orge** | PS | LT | 24.67±1.53 | 23.00±3.91 | 26.17±2.52 | 10.83±0.15 | 23.83±4.04 | 12.00±1.00 | 25.33±4.65 | 27.67±2.47 | 28.83±1.89 | 25.50±2.18 | 25.83±1.76 |
| | | PT | 0.24±0.06 | 0.21±0.04 | 0.25±0.04 | 0.13±0.04 | 0.21±0.06 | 0.11±0.02 | 0.26±0.04 | 0.29±0.05 | 0.25±0.05 | 0.22±0.02 | 0.26±0.06 |
| | PS | LR | 16.50±3.04 | 11.00±1.32 | 14.00±1.00 | 20.67±3.58 | 10.50±1.50 | 19.50±2.50 | 15.17±0.76 | 16.17±4.19 | 11.50±2.18 | 12.33±4.04 | 14.33±3.55 |
| | | PR | 0.19±0.01 | 0.18±0.03 | 0.21±0.02 | 0.21±0.04 | 0.15±0.06 | 0.20±0.01 | 0.20±0.03 | 0.21±0.03 | 0.17±0.01 | 0.19±0.01 | 0.22±0.07 |
| **Blé dur** | PA | LT | 14.17±2.47 | 7.33±0.29 | 5.67±1.53 | 17.5±0.87 | 16.5±0.50 | 7.50±1.32 | 8.33±0.58 | 3.67±0.29 | 6.33±2.36 | 10.67±0.58 | 7.50±6.50 |
| | | PT | 0.16±0.01 | 0.04±0.01 | 0.06±0.03 | 0.16±0.05 | 0.13±0.02 | 0.08±0.04 | 0.06±0.02 | 0.06±0.05 | 0.05±0.01 | 0.10±0.02 | 0.09±0.04 |
| | PS | LR | 12.83±1.04 | 8.17±3.44 | 11.83±3.18 | 15.00±0.77 | 15.17±1.53 | 7.50±1.32 | 12.17±5.25 | 10.83±1.89 | 8.67±4.51 | 13.17±1.61 | 6.67±5.51 |
| | | PR | 0.13±0.02 | 0.05±0.01 | 0.07±0.07 | 0.14±0.06 | 0.16±0.04 | 0.09±0.02 | 0.08±0.06 | 0.09±0.05 | 0.09±0.06 | 0.10±0.01 | 0.10±0.06 |
| **Dactylocténion** | PA | LT | 12,75±2,07 | / | / | / | / | / | / | / | / | 1,51±0,02 | 0,71±1,07 |
| | | PT | 0,14±0,01 | / | / | / | / | / | / | / | / | 0,03±0,01 | 0,01±0,01 |
| | PS | LR | 11,10±0,20 | / | / | / | / | / | / | / | / | 1.02±0,50 | 0,12±0,20 |
| | | PR | 0,12±0,01 | / | / | / | / | / | / | / | / | 0,03±0,01 | 0,01±0,002 |

PA : partie aérienne, PS : partie souterraine, LF : longueur de feuille, PT : poids de tige, LR : longueur de racine, PR : poids de racine.

D'après les résultats recensés dans les tableaux 11 et 12, il est signalé que l'action des extraits appliqués à différentes concentrations engendre des variations morphologiques de la partie aérienne et souterraine des plantules d'orge, blé dur et le dactylocténion. L'effet simulateur de l'extrait *C. arabica* sur la germination les graines d'orge se manifeste par l'élongation de la partie aérienne et souterraine par rapport aux plantules du lot témoin négatif, suivant les plantules de blé dur, où la longueur de la partie aérienne est plus importante que la partie souterraine.

Il apparaît que les effets inhibiteurs des extraits de *P. tomentosa* montrent que la longueur des radicules de plantules d'orge était relativement sensible mais la partie aérienne augmente, alors que pour les plantules de blé dur la longueur de la radicelle est plus importante et la partie aérienne réduite.

CHUNG et Miller (1995), TURK et TAWAHA (2002) rapportent que des extraits aqueux des plantes allélopathiques avaient des effets plus prononcés sur la croissance des radicules que sur la croissance des coléoptiles. Cela est probable car les racines constituent la première partie des plantes à absorber les substances allélochimiques présentes dans le milieu arrosé.

SAZADA *et al*. (2009), ont signalé que l'effet des extraits *Prosopsis juliflora* (*Mimosaceae*) diminuerait la longueur des racines de *Triticum aestivum* (*Poaceae*) dans la concentration (50g/500ml) soit 1,30±0,08cm. La réduction la plus importante de la longueur des racines de *Triticum* a révélé que les substances allélochimiques de l'extrait aqueux entraînaient une réduction maximale de la longueur des racines.

BELAIDI (2014) montre l'effet inhibiteur des extraits *Datura stramonium* L. (*Solanaceae*) sur la croissance des plantes *Hordeum vulgare* L. (*poaceae*) et *D. aegyptium*. Les résultats illustrent que l'effet allélopathique induise des anomalies de croissance pour plantules *D. aegyptium* où la longueur des radicelles est nettement plus long que les coléoptiles, bien que chez les plantules *H. vulgare* la longueur des coléoptiles sont nettement plus long que les radicelles.

CHON *et al*. (2004) ont réalisé des études sur l'utilisation d'extraits aqueux de feuilles de luzerne que le retard de la germination des semences et en particulier la réduction de l'élongation des racines étaient principalement dus à des molécules toxiques de l'extrait de feuilles de luzerne.

Ces molécules toxiques sont principalement des métabolites secondaires soient les terpènes, alcaloïdes, molécules aromatiques, qui sont impliqués dans des interactions allélopathiques. Selon WALLER (1991), SIQUEIRA *et al.* (1989), ces composés secondaires ont d'abord été caractérisés par leur rôle protecteur contre les bio-agresseurs (insectes, bactéries, champignons) mais ils peuvent également affecter la croissance d'autre plante. Ces composés phyto-toxiques agissent par de multiples voies comme l'inhibition de la division et de l'élongation cellulaire et de la synthèse des protéines, la modification de la perméabilité membranaire et de l'absorption minérale, la modification de la photosynthèse et la respiration, ou bien encore par des interférences négatives avec les hormones de croissance (RICE, 1984; EINHELLIG, 1986 ; SIQUEIRA *et al.*, 1991). De même, HARPER et BALKE, (1981) notent que leurs cibles principales sont les phytohormones dont l'auxine, gibbérellines et l'acide abscissque.

L'élongation de la partie aérienne chez certaines plantules observées est probablement dûe aux phytohormones dont l'auxine qui est fabriquée au niveau des zones de croissance soit les méristèmes apicaux des tiges, jeunes feuilles ou bourgeon. Toutefois, son rôle primordial s'exerce sur la division cellulaire, en contrôlant à la fois le déclenchement des mitoses, l'élongation cellulaire et la formation des parois (WEIL, 1994). Cette variation dans la croissance soit l'élongation de la tigelle et la radicelle est en fonction des régulateurs de croissance (phytohormones) soient l'auxine (Acide indolacétique AIA.), cytokinines, l'éthylène, les gibbérellines (GA) et l'acide abscissique (ABA) (GOLDSMITH, 1993).

Les cytokinines sont responsables à la division cellulaire dans les graines, les fruits et les feuilles. Elles régulent la division et la différenciation cellulaire. Elles sont susceptibles de lever la dormance de certaines graines. D'une façon générale, elles activent la production chlorophyllienne, l'ouverture des feuilles, favorisent la croissance cellulaire et la formation des jeunes pousses. Lorsque la cytokinines est associée à l'auxine, elles stimulent les divisions cellulaires (WEIL, 1994).

L'élongation de la partie aérienne et souterraine des plantules d'orge est illustrée par la présence de l'hormone gibbérelline dans les parties des plantules. Elle est responsable de la croissance des feuilles et des tiges en stimulant l'allongement et la division cellulaire. Elle favorise la germination et la fructification des graines. Les gibbérellines agissent essentiellement sur les cellules des entrenœuds qu'elles allongent et contribuent aussi à la levée de dormance des graines. Les gibbérellines provoquent une exaltation de la croissance. On note que la gibbérelline et l'auxine sont les principales phytohormones (GHRHARD, 1993).

SINGH et THAPAR (2003), BELZ et HURLE (2004) ont démontré que les composés allélochimiques ont des effets importants sur le métabolisme des phytohormones. Ces composés allélochimiques peuvent être utilisés dans le développement de nouveaux types d'herbicides ainsi que comme outils de gestion biologique pour lutter contre les mauvaises herbes sur les cultures.

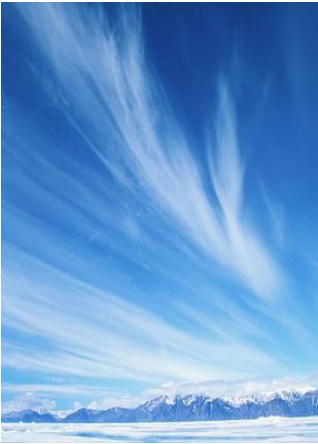

*Conclusion*

## Conclusion

La présente étude porte sur l'évaluation de pouvoir biocide des extraits aqueux de *Cleome arabica* et *Pergularia tomentosa* récoltées au Sahara Est-Algérien, vis-à-vis des espèces tests. L'expérimentation est réalisée au niveau de laboratoire de biologie, Université de Ghardaïa.

Les résultats de criblage phytochimique laisse apparaître que l'extrait brut de *Cleome arabica* et de *Pergularia tomentosa* sont riches en métabolites secondaires notamment en alcaloïdes, flavonoïdes, saponosides, stérols et terpènes…etc., toutefois, l'extrait de *C. arabica* s'est révélé plus riche en ces métabolites.

L'étude du pouvoir insecticide des extraits de deux plantes testées montrent qu'ils présentent un fort effet insecticide vis-à-vis les imagos de *T. confusum*; les pourcentages de mortalités obtenus avoisines les 90% pour les deux plantes. En outre, l'application par ingestion des deux extraits, montre l'influence de ces préparations sur le métabolisme de cet insecte; des troubles de mouvement et du métabolique sont constatés. En revanche, l'application de ces extraits à de faibles concentrations ne présente aucun effet sur les imagos *T. confusum,* aucun signe d'intoxication n'était observé sur les accouplements, les pontes et même des larves néonates sont observés.

L'étude des effets des extraits de ces deux espèces sur la germination des graines de l'orge, blé dur et le dactylocténion laisse apparaitre le fort pouvoir inhibiteur de la germination et sur la croissance. Les pourcentages d'inhibition de la germination sont élevés pour tous les lots traités, particulièrement les lots de l'adventice dactylocténion qui semble très sensible aux effets des extraits végétaux appliqués même à de faibles doses.

En outre, en comparant les résultats des différents traitements, il ressort que l'orge et le blé dur sont plus résistants aux effets inhibiteurs de la germination des extraits végétaux testés que l'espèce adventice (dactylocténion), du fait que l'application des extraits à de doses faibles n'engendrent aucun effet sur la germination des graines d'orge et de blé dur mais il provoque des taux acceptables d'inhibition au niveau des lots de dactylocténion ou bien ils retardent la germination des graines de celle-ci.

S'ajoute aux effets inhibiteurs de la germination des extraits végétaux, des effets indésirables sur la croissance des plantules ; des anomalies de croissances sont observées, soit des incohérences dans la croissance entre les deux parties aérienne et souterraine, ceux-ci se manifeste par des dimensions anormaux des organes végétaux. Ces variations morphologiques peuvent êtres justifiées par l'action des métabolites secondaires contenus dans les extraits végétaux sur les processus de croissance à différents niveaux notamment au niveau tissulaire.

L'étude comparative des activités biologiques des extraits aqueux de *C. arabica* et *P. tomentosa* récoltées au Sahara septentrional, Est Algérien, permet de ressortir les possibilités insecticides et herbicides des extraits de deux plantes.

Le travail réalisé permet de mettre en exergue le fort potentiel pesticide des molécules d'origine végétale, ce qui peut inciter la collectivité scientifique mondiale à déployer les efforts pour la recherche et le développement des méthodes alternatives de lutte peu ou pas nocives sur l'environnement et sur l'homme.

En perspective, pour poursuivre cette étude, il est souhaitable de:

o Caractériser quantitativement des molécules bioactives des extraits végétaux étudiés;
o Tester les extraits en application directe sur les espèces adventices en plein champs;
o Etudier l'effet de la pulvérisation de ces extraits sur les denrées alimentaires entreposées dont les semoules, les farines, les légumes secs, etc, afin de rechercher leur phytotoxicité sur l'être humain;
o Rechercher les activités biologiques des extraits aqueux, notamment « activité antibactérienne, antifongique, etc »;
o Etudier l'effet biocide des extraits aqueux de la partie souterraine de ces deux plantes choisies.

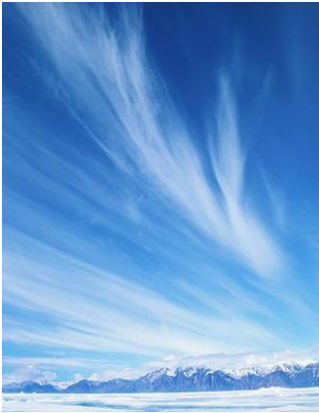
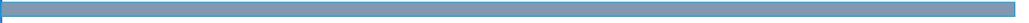

*Références bibliographiques*

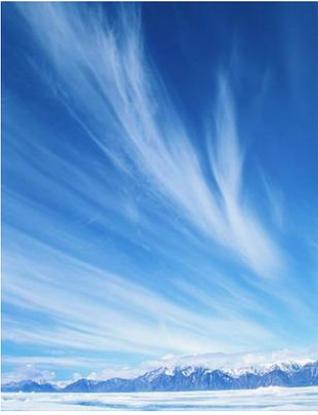

*Annexe I*

# Annexe I.

## 1. Préparation des concentrations

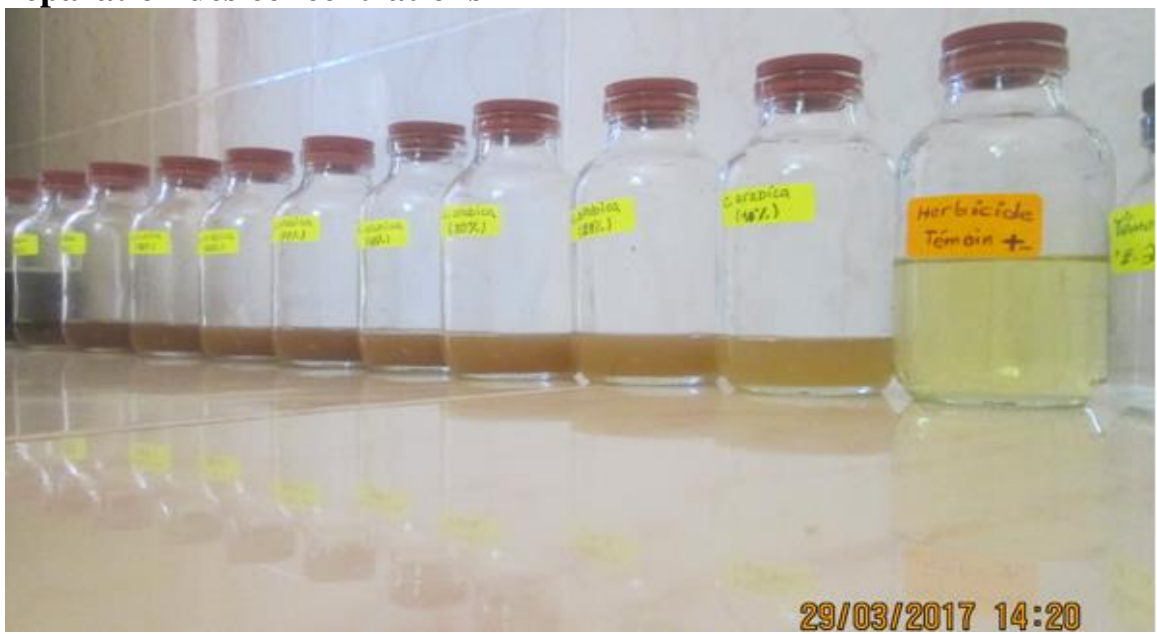

Différentes concentrations de l'extrait foliaire de *Cleome arabica* L.

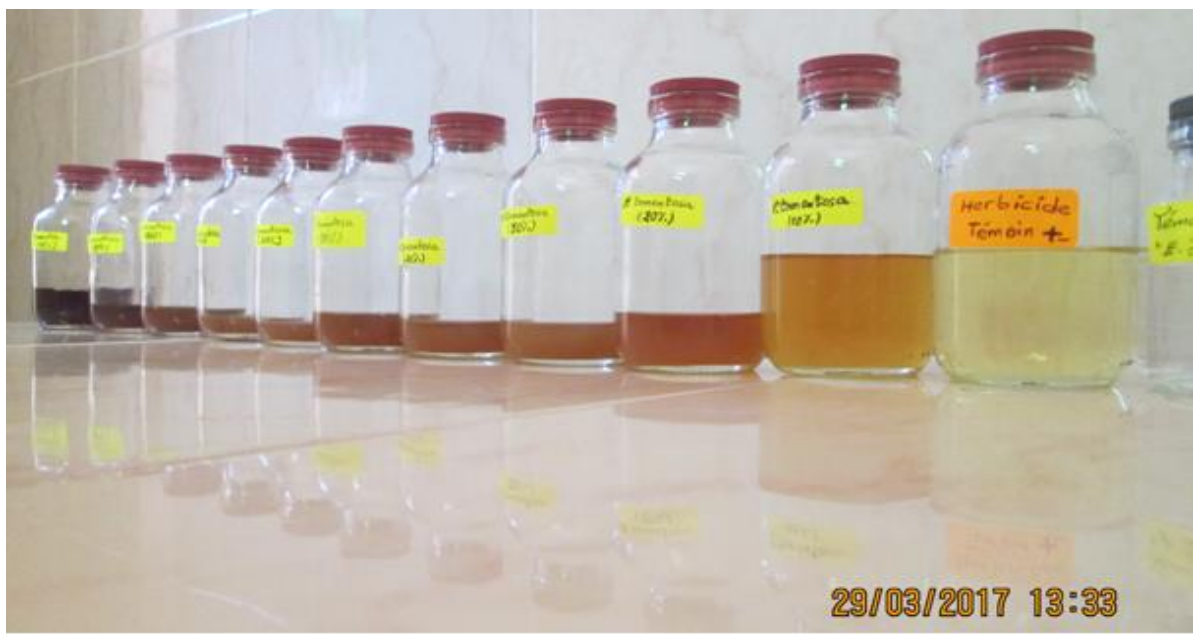

Différentes concentrations de l'extrait foliaire de *Pergularia tomentosa* L.

## 2. Effet insecticides des extraits aqueux sur la mortalité des insectes

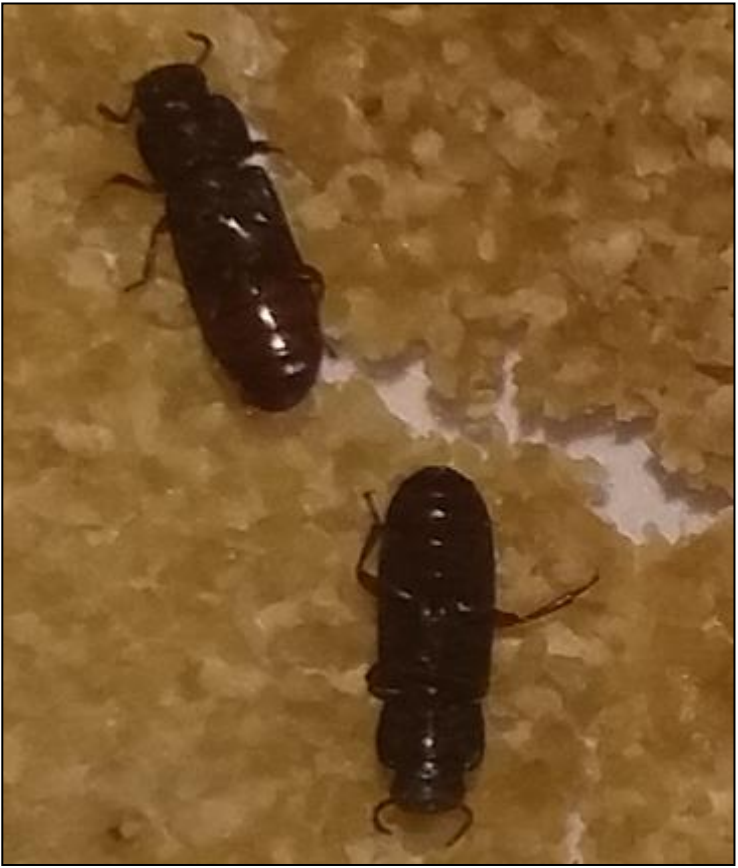
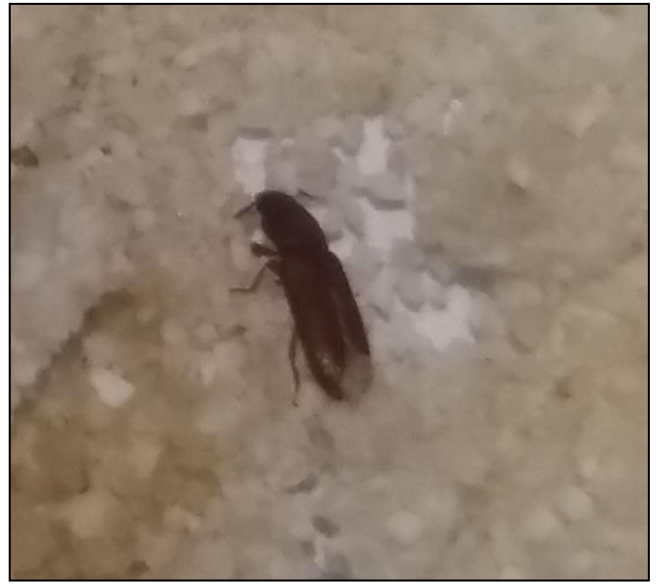
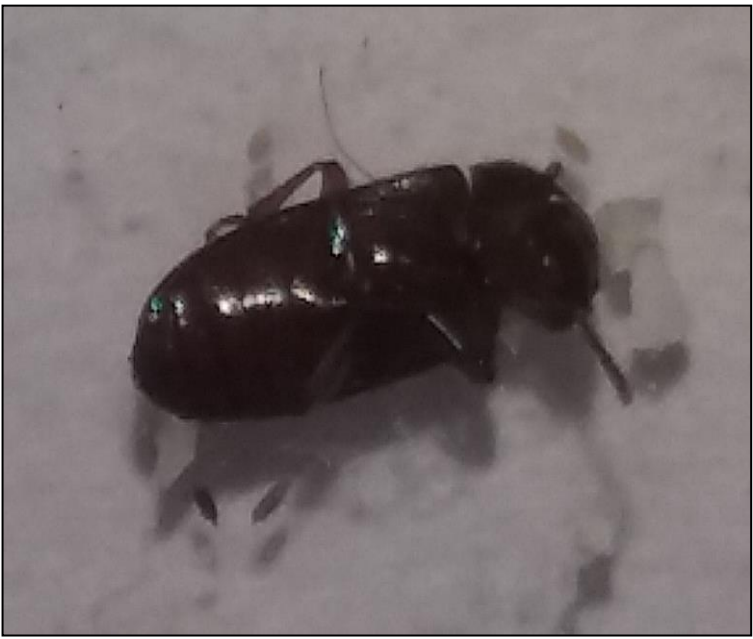
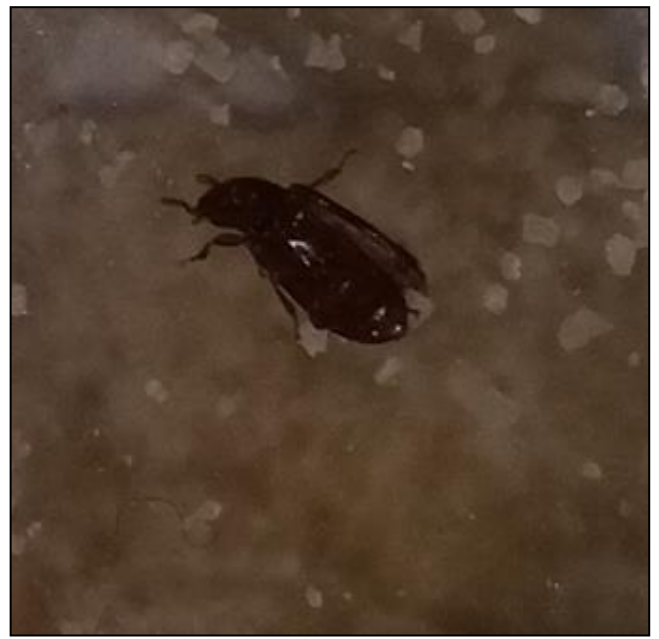

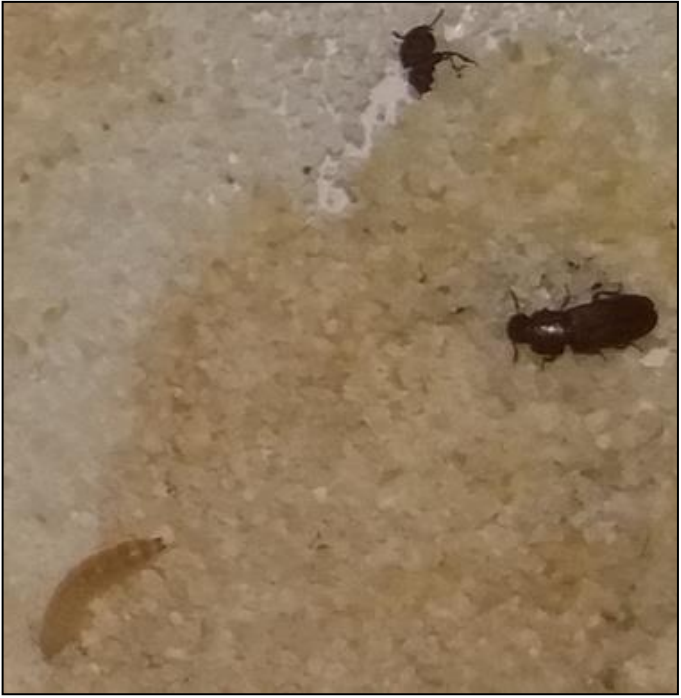
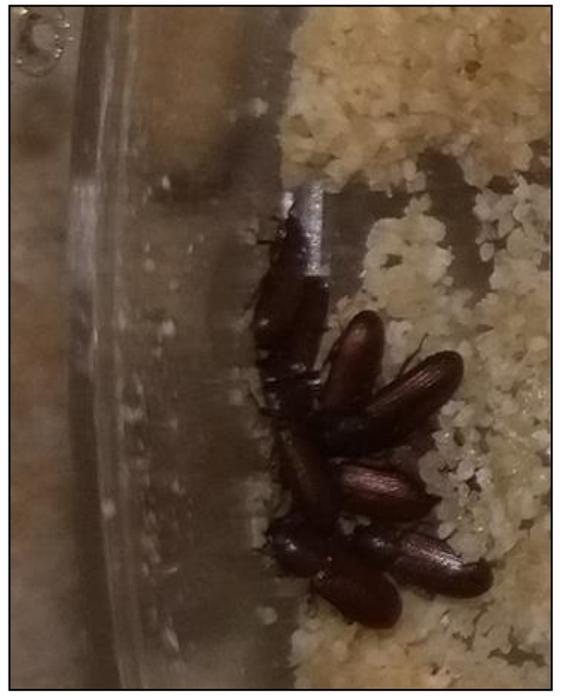
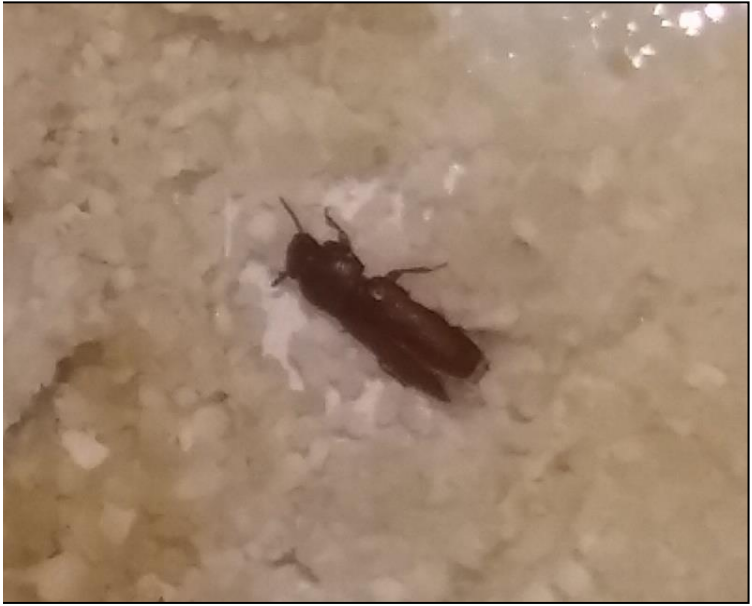
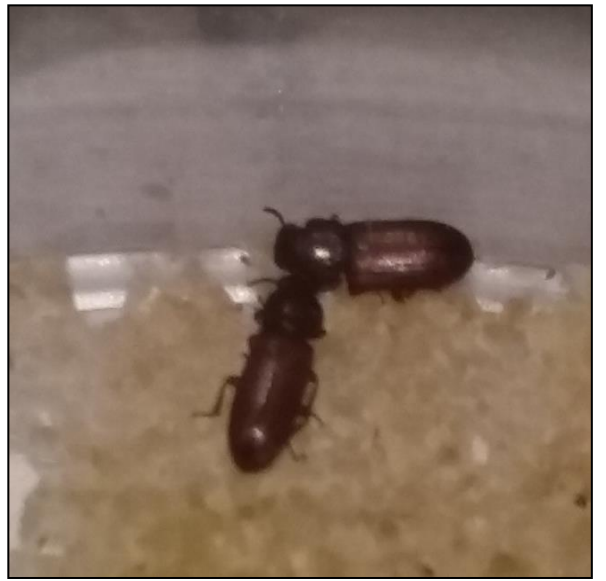

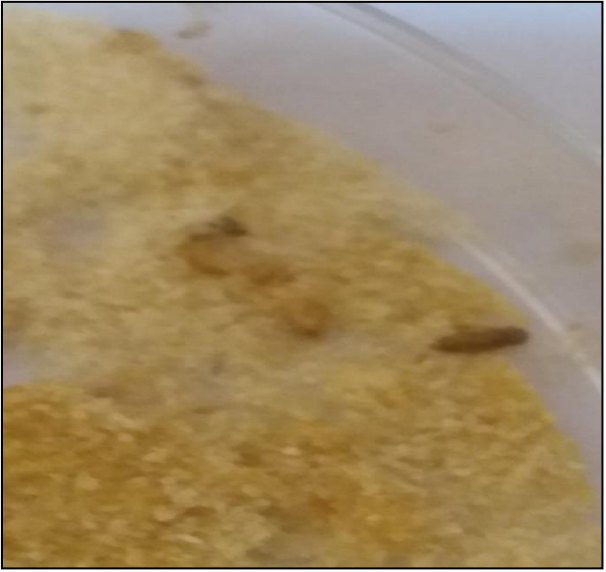
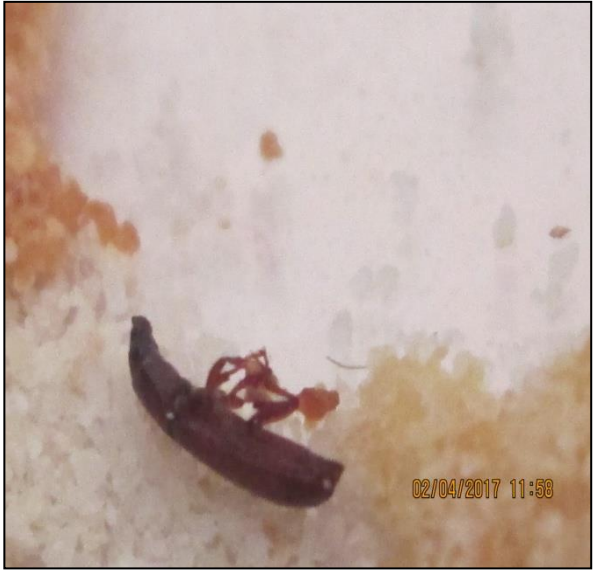
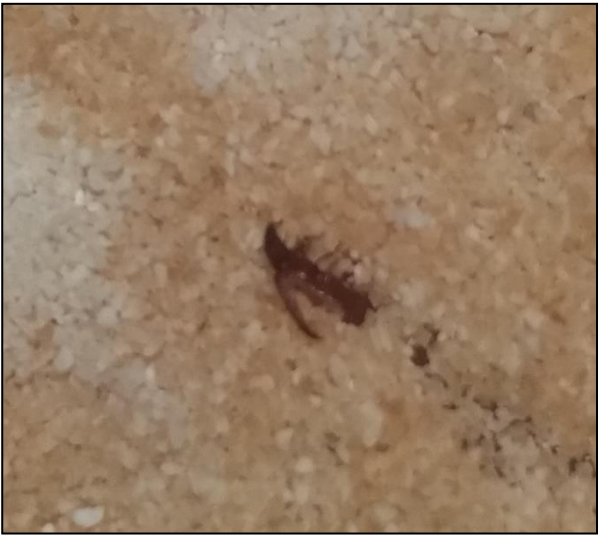
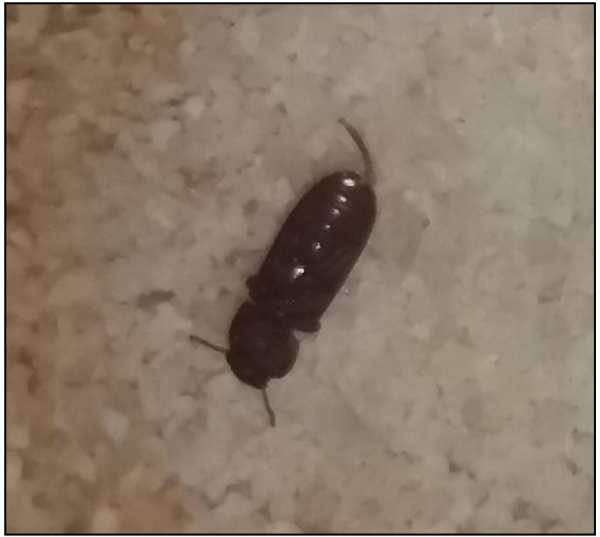
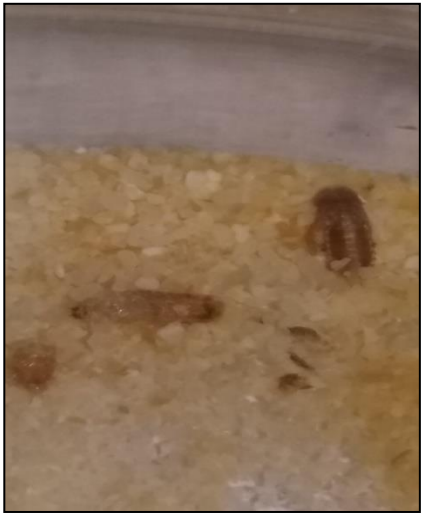
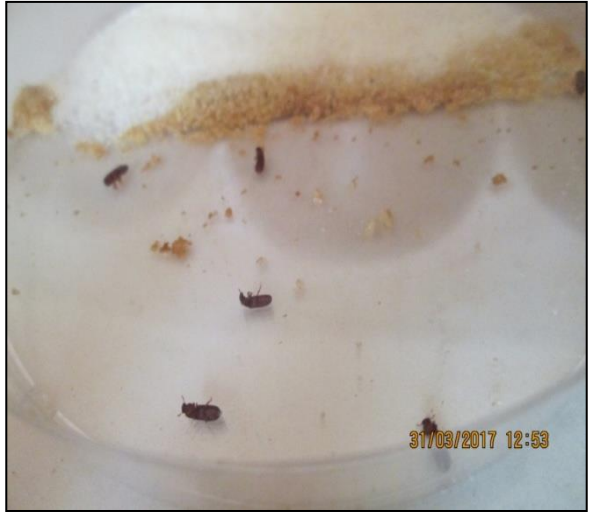

### 3. Morphologie de l'insecte *Tribolium confusum*.

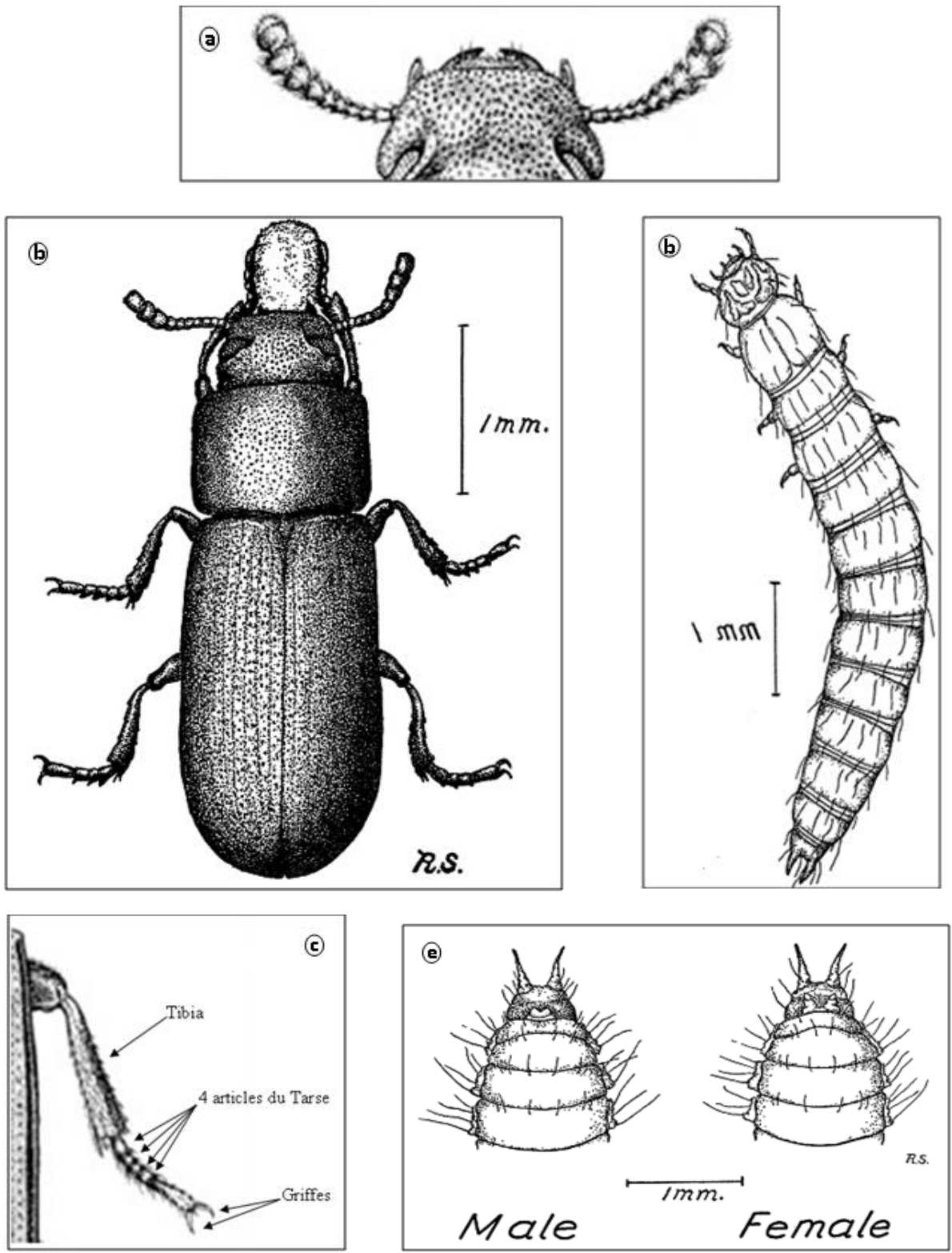

Dessin représentatif du l'adulte *Tribolium confusum* Duval (BRINDELY, 1930)

## 4. Cycle de développement de *T. confusum*

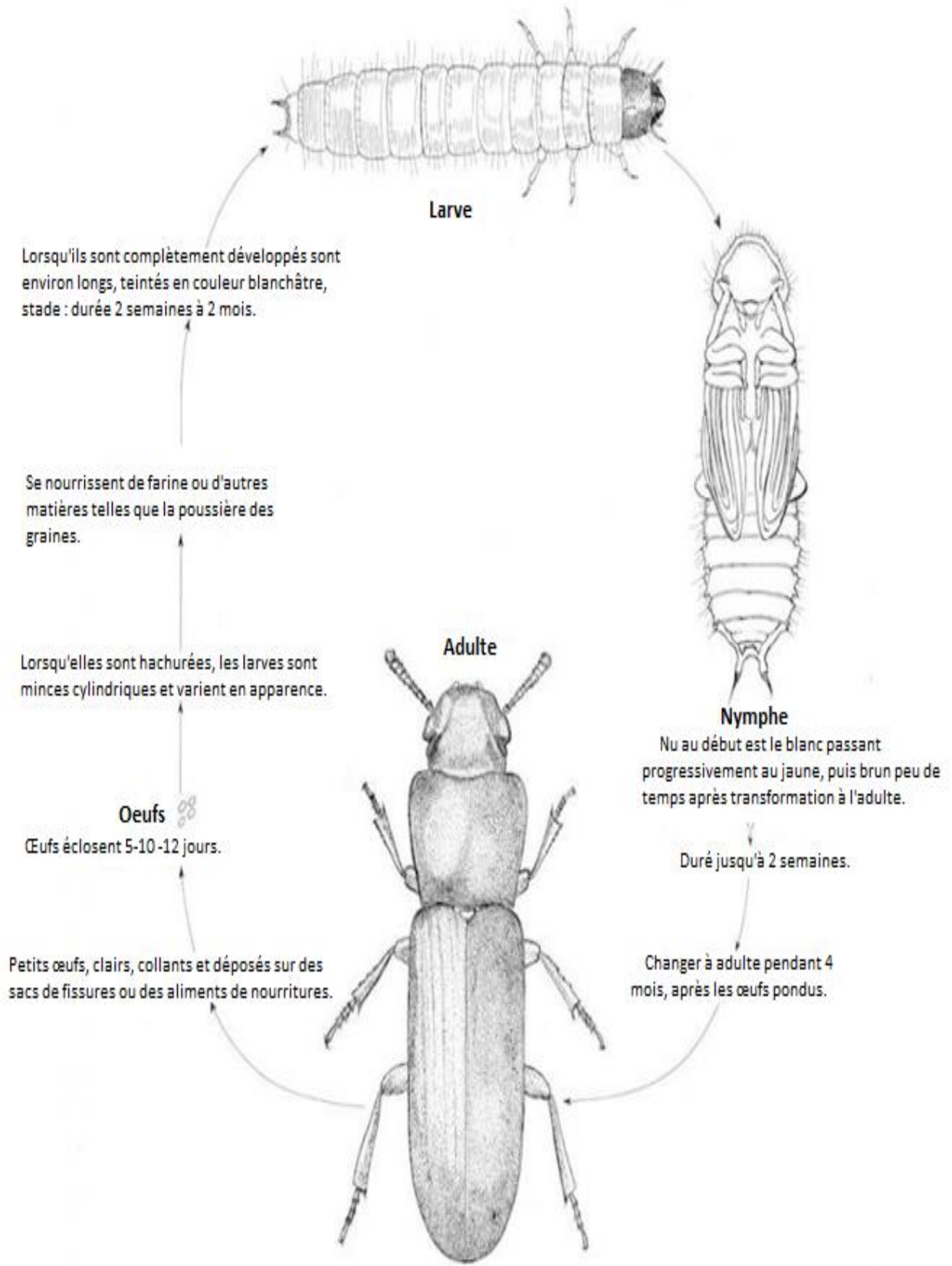

Cycle de vie de coléoptère *Tribolium confusum* Duval (BRINDELY, 1930)

## 5. Insecticide HILAC et l'Herbicide industriel

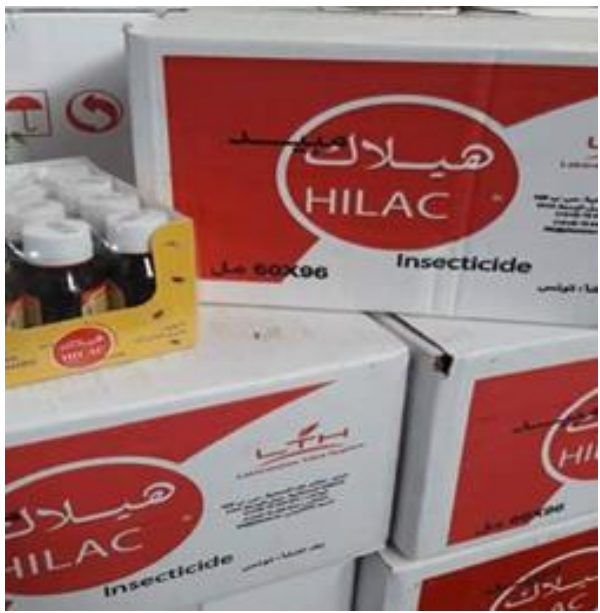 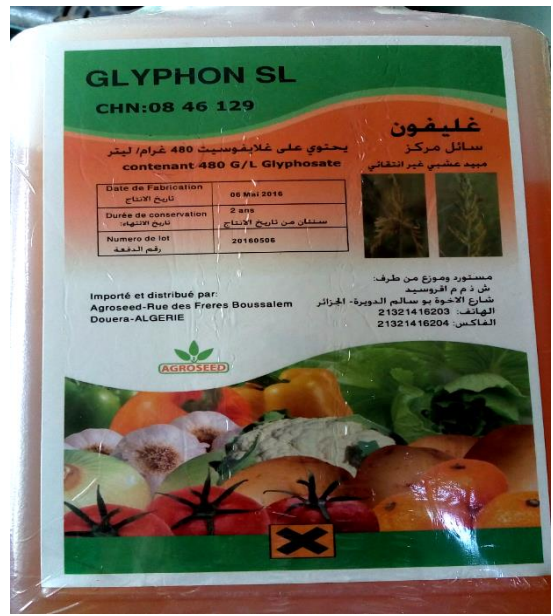

Insecticide HILAC, produit tunisien, l'emballage (60ml, 100ml, 250ml et 25 litres)

Herbicide de synthèse Glyphon. (80 g/L de glyphosate)

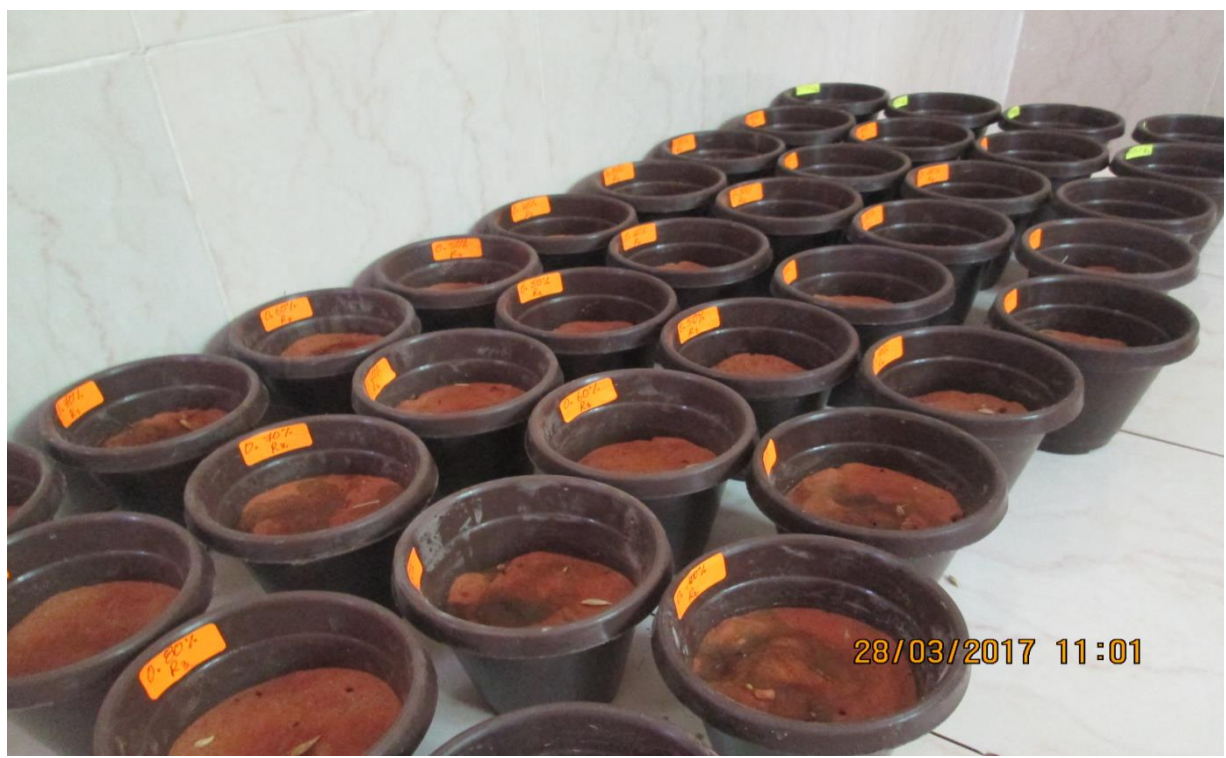

Préparation des pots ensemencés par les graines d'orge et traités par l'extrait aqueux de *C. arabica*.

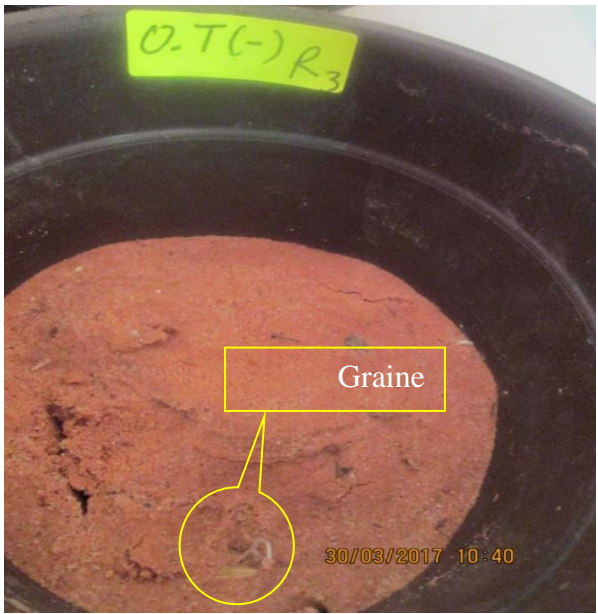 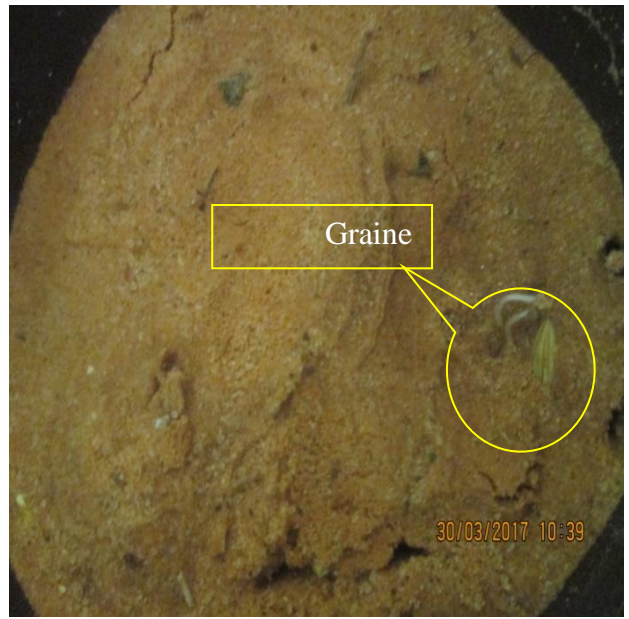

Graines d'orge traitées par l'eau distillé après 2 jours d'expérimentaion.

## 6. Conditions expérimentales pour réalisations de test

**Tableau 01 :** Conditions d'essai pour le test d'inhibition de la germination et de la croissance.

| | |
|---|---|
| Espèces tests | Les semences d'Orge, Blé dur et le chiendent de patte-poule. |
| Température de suivie | 24± 2°C |
| Sol | Sol des dunes de sables : 70% sable de silice, 22% limon, 5% kaolin, 3% terre noire. pH : 6 ± 0,2. |
| Taux d'humidité | 100ml d'eau de robinet pour assurer la rétention en eau pour un sol hydraté au début de test seulement. |
| Type de contenant | Pots spéciale pour l'ensemencement des grains, en plastique de hauteur 10cm et de diamètre 4cm, contenant un trou en bas. |
| Masse du sol | 400g de sable humide. |
| Nombre des grains/pot | 15 grains par pot. |
| Nombre de répétition | Trois répétitions pour chaque concentration. |
| Extraits utilisés | Extrait de la plante *Cleome arabica L.* et *Pergularia tomentosa L.* |
| Nature d'extrait | Extrait aqueux. |
| Références test | Témoin négatif : eau de robinet, Témoin positif : herbicide. |
| Dose de traitement | 4ml pour les groupes tests et 3ml pour les groupes positifs. |
| Durée du test | 10 jours. |
| Mesures biologiques | Nombre des grains germés, longueur des tiges, longueur des racines, poids humide des tiges et des racines après 10 jours d'exposition, anomalies de croissance observés au cours de test. |
| Paramètre de mesures et expressions des résultats | Taux maximal d'inhibition (%), Cinétique de germination (% germination/temps) et Concentration d'efficacité $CE_{50\%}$ et $CE_{90\%}$ en (mg/ml). |
| Statistiques | Logiciel d'ANOVA, tableau de Probit, courbes de tendances (coefficient de corrélation et équation de régression). |

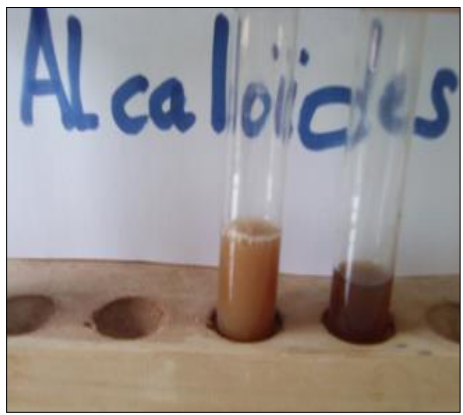 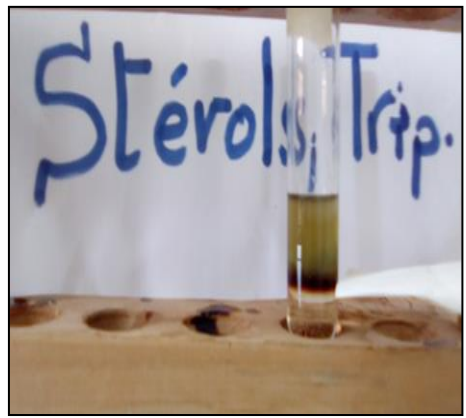 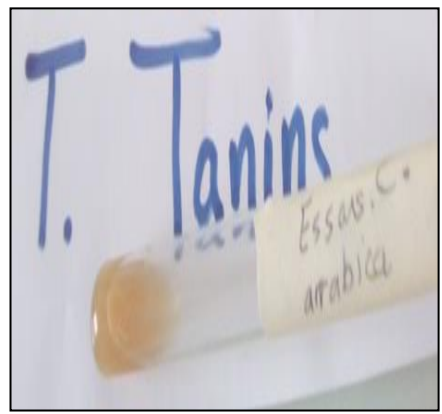 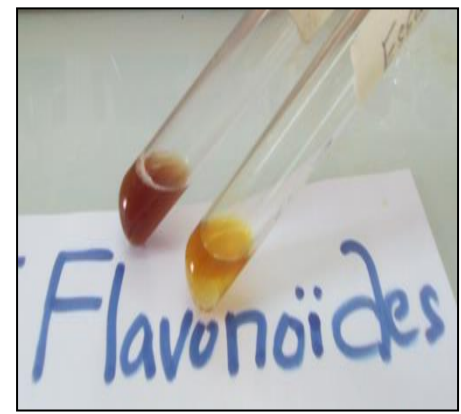

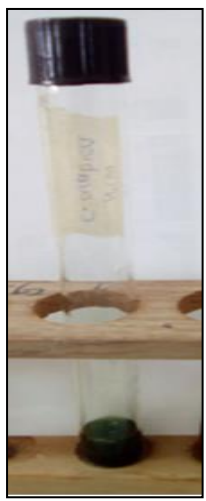 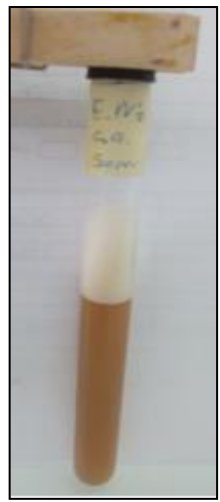 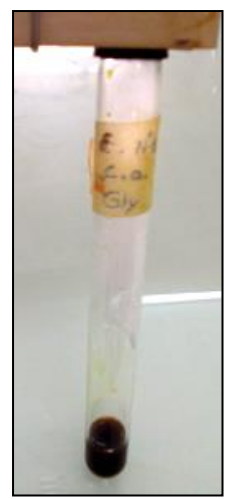 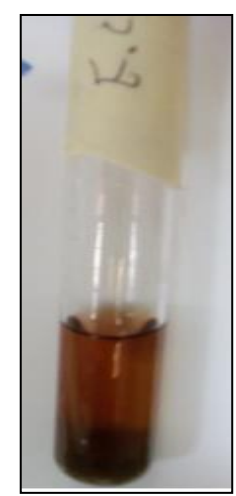

| **T. Désoxyose** | **T. Saponosides** | **T. Glycoside** | **T. Glycoside** |

**Criblage phytochimique des extraits aqueux de *Cleome arabica* L.**

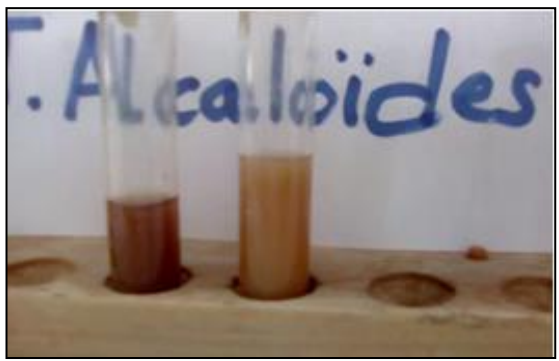
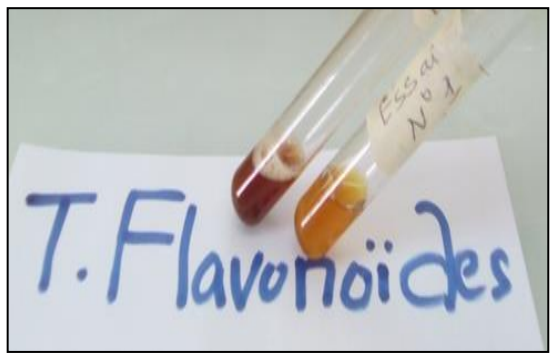
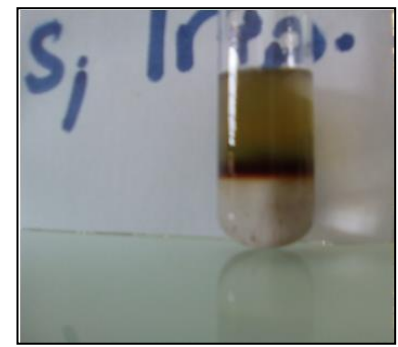

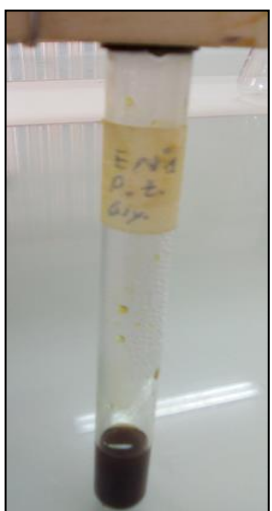
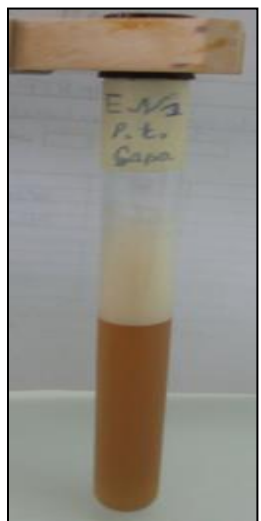
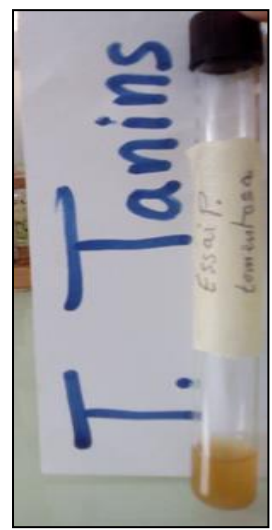
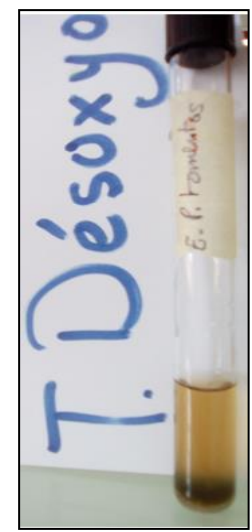

**T. Glycoside**   **T. saponosides**

**Criblage phytochimique des extraits aqueux de *Pergularia tomentosa* L.**

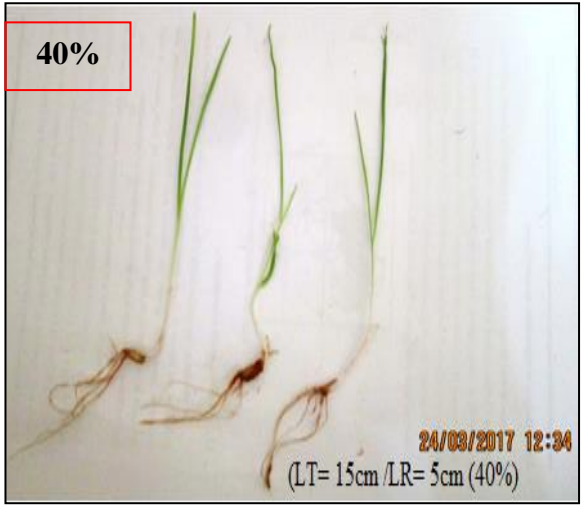
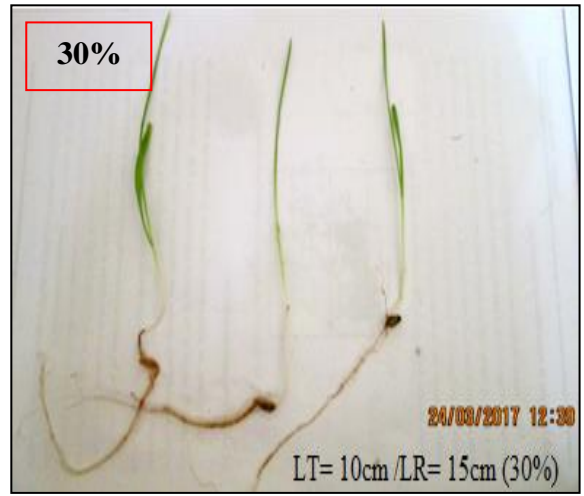
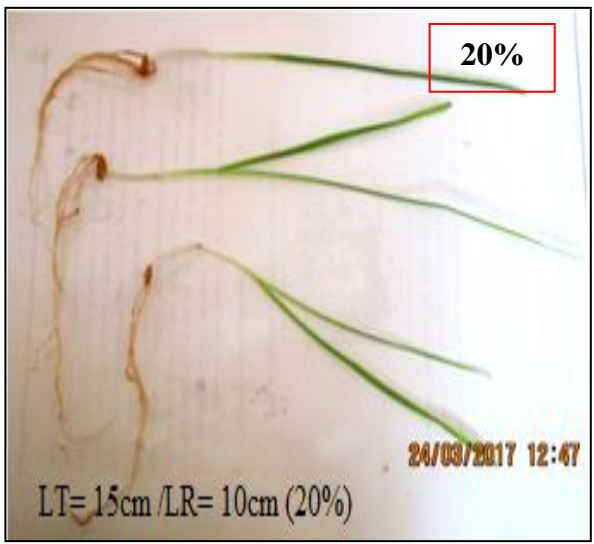
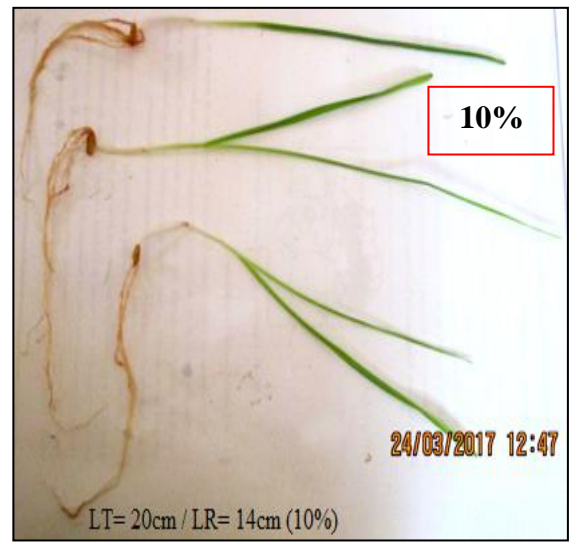
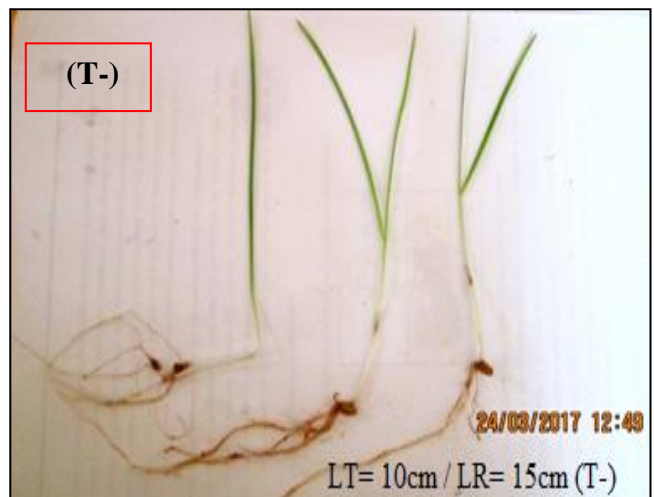

Longueur de la partie aérienne et souterraine des plantules de blé dur traitées par l'extrait aqueux de *Cleome arabica* à différentes concentrations et témoin négatif.

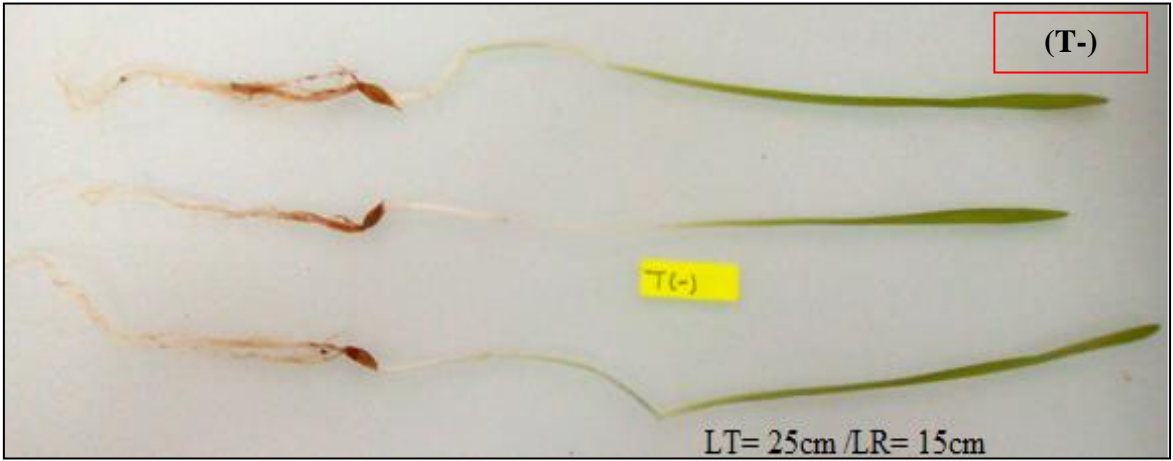

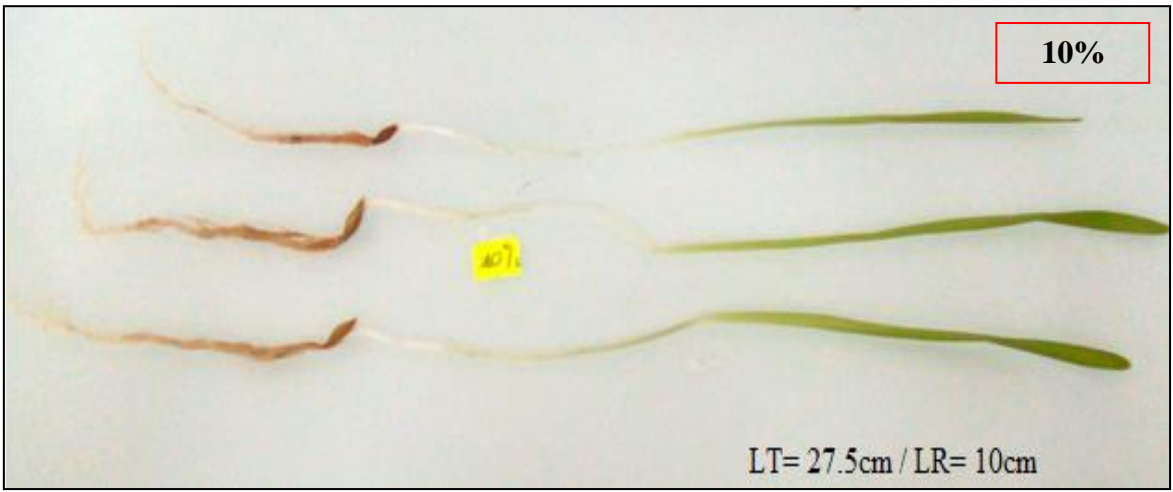

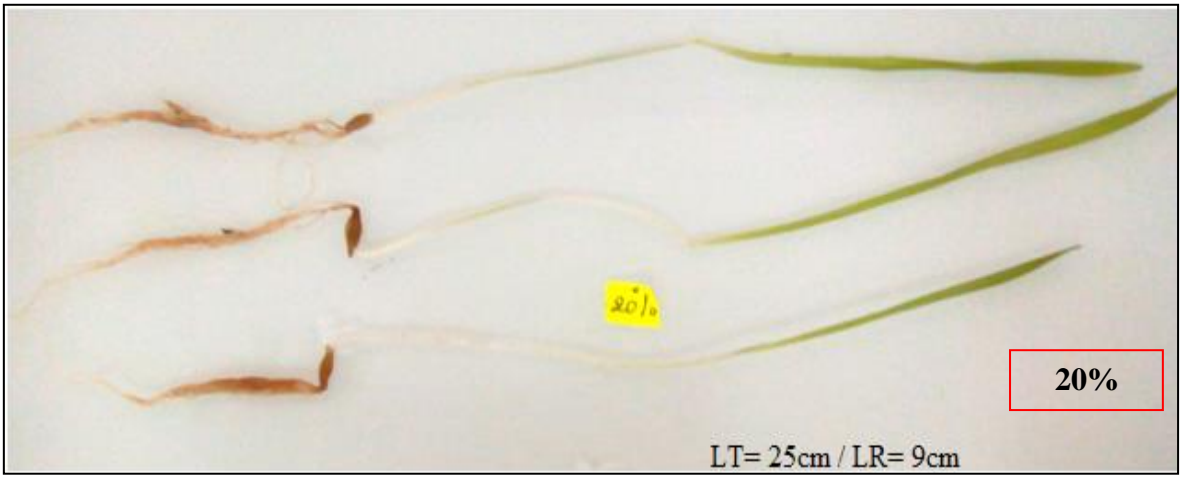

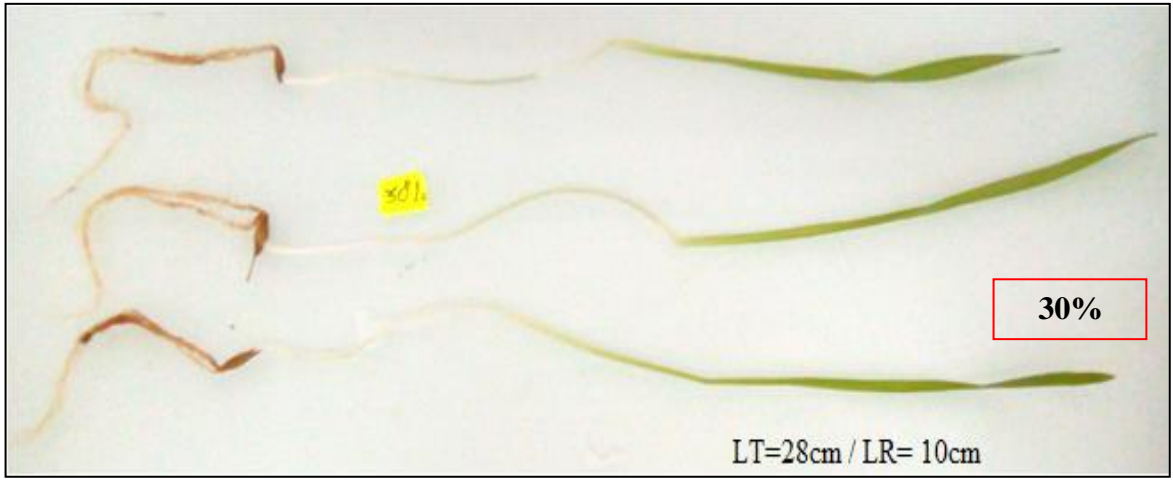

LT=28cm / LR= 10cm — 30%

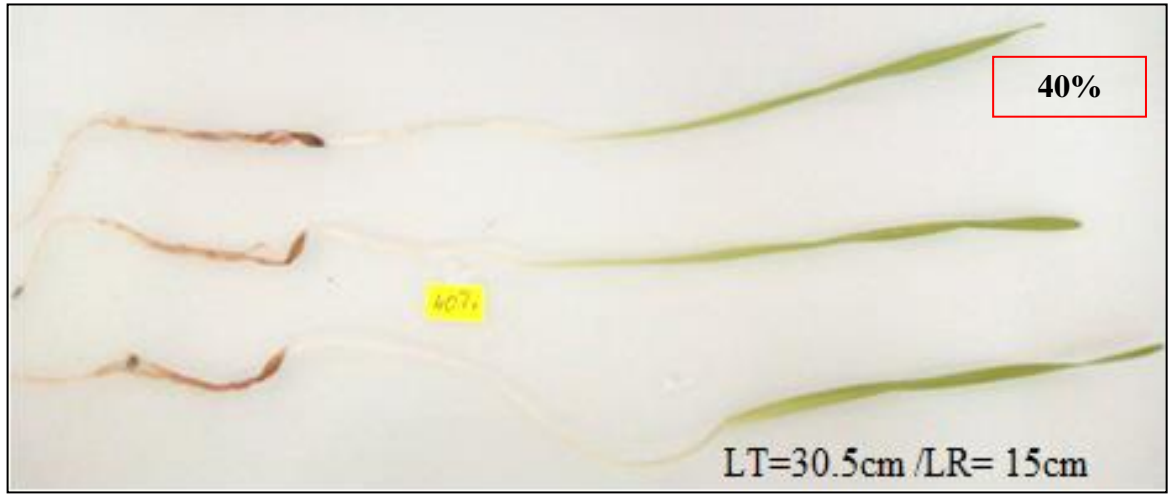

LT=30.5cm /LR= 15cm — 40%

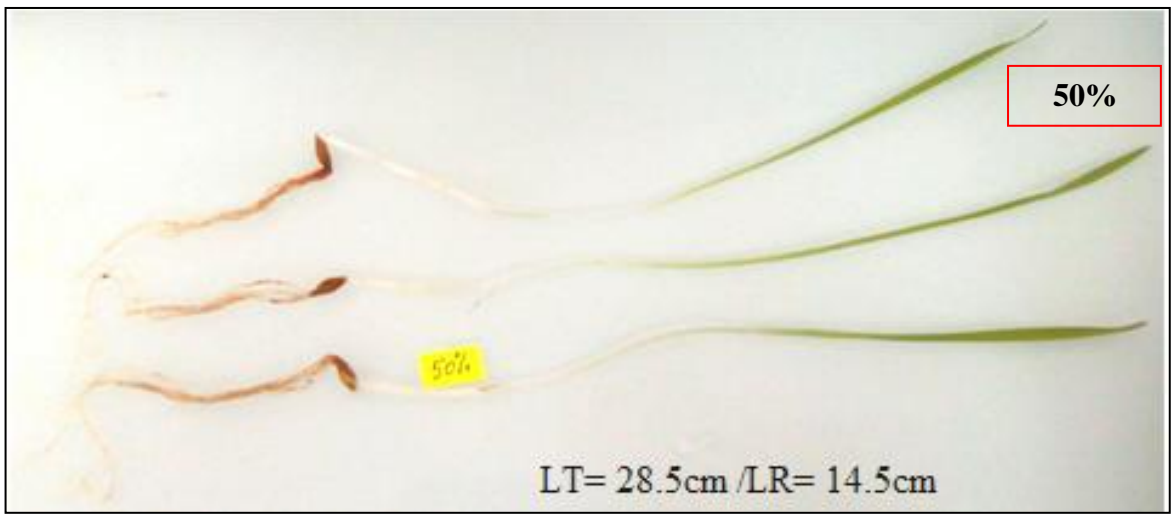

LT= 28.5cm /LR= 14.5cm — 50%

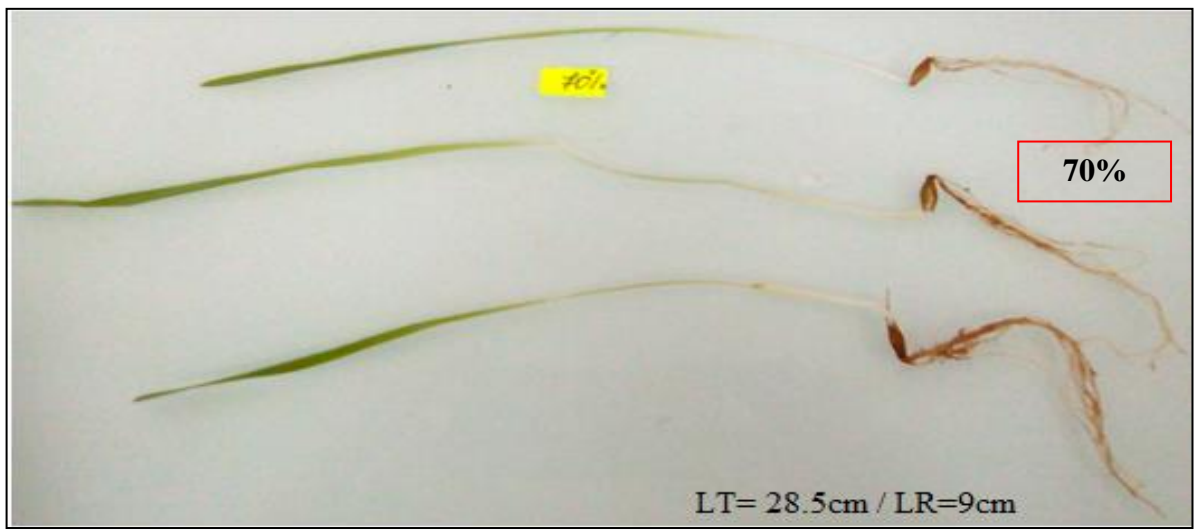

70%

LT= 28.5cm / LR=9cm

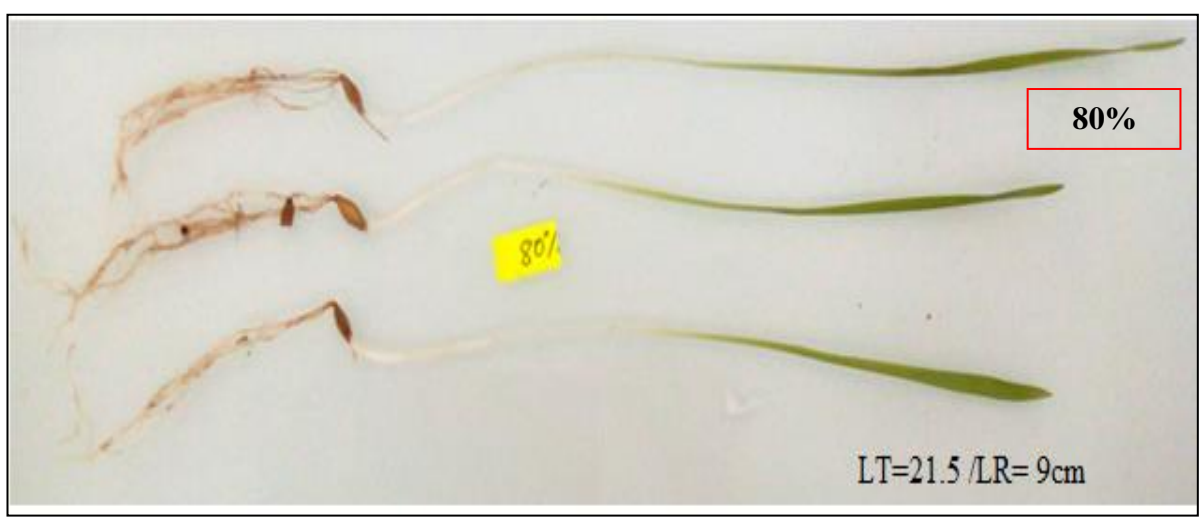

80%

LT=21.5 /LR= 9cm

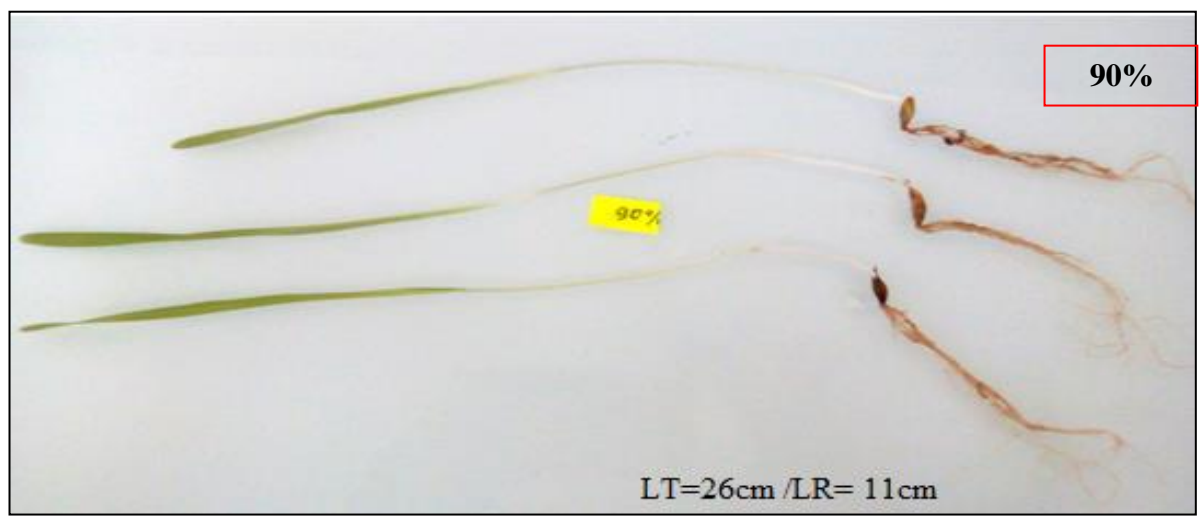

90%

LT=26cm /LR= 11cm

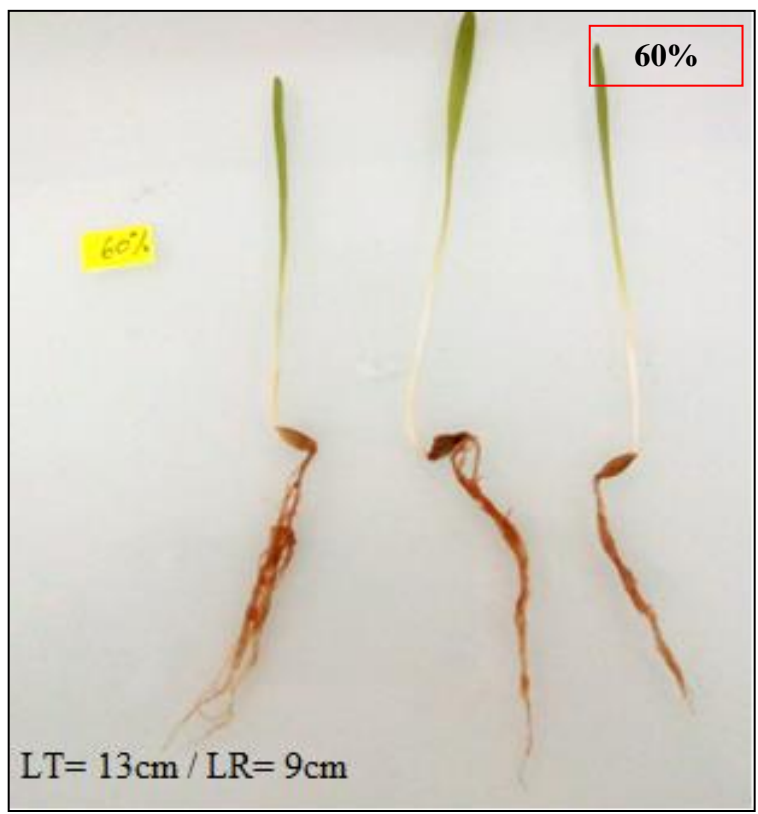
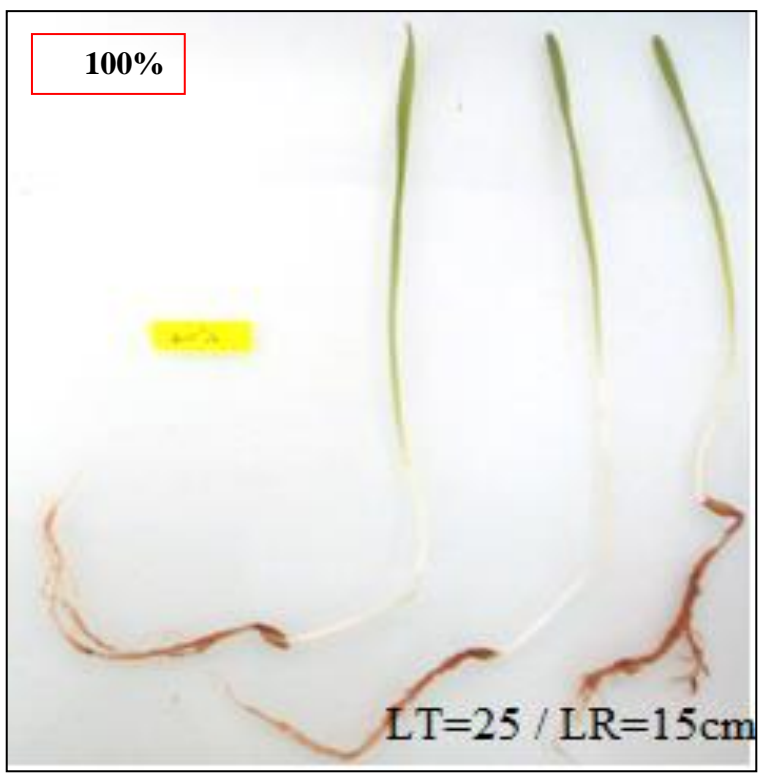

Longueur de la partie aérienne et souterraine des plantules d'orge traitées par l'extrait aqueux de *Pergularia tomentosa* à différentes concentrations et témoin.

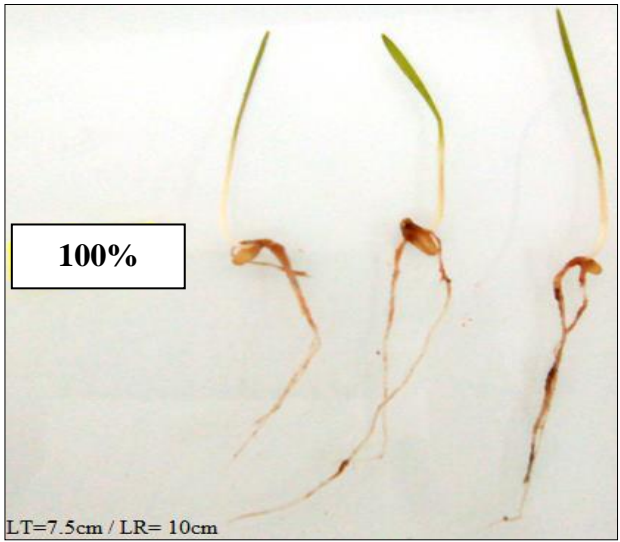
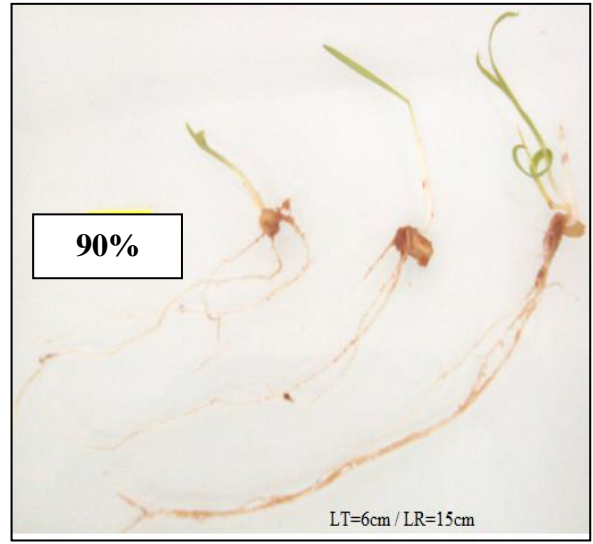
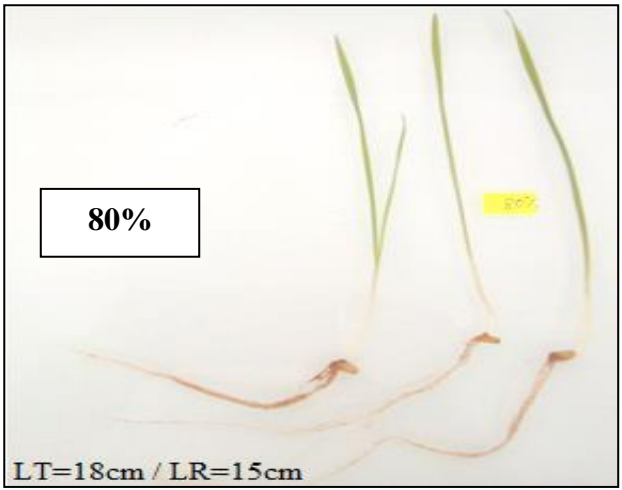
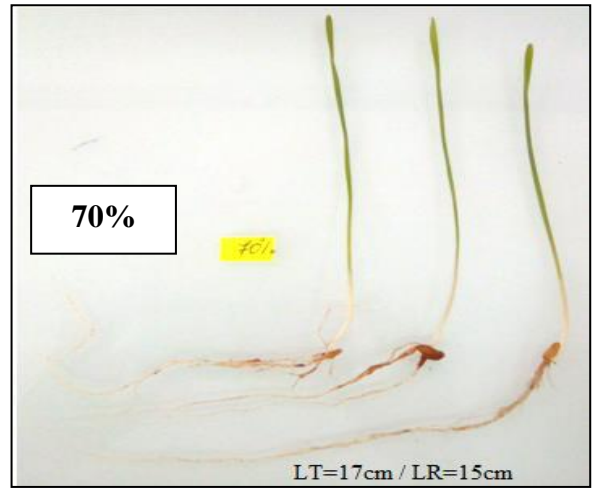
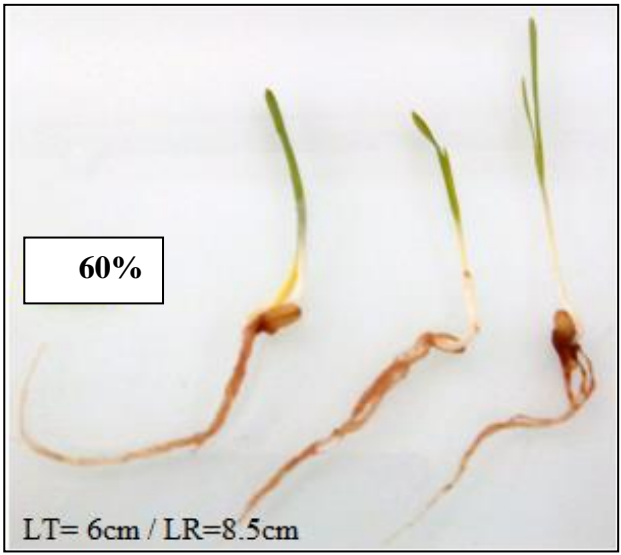
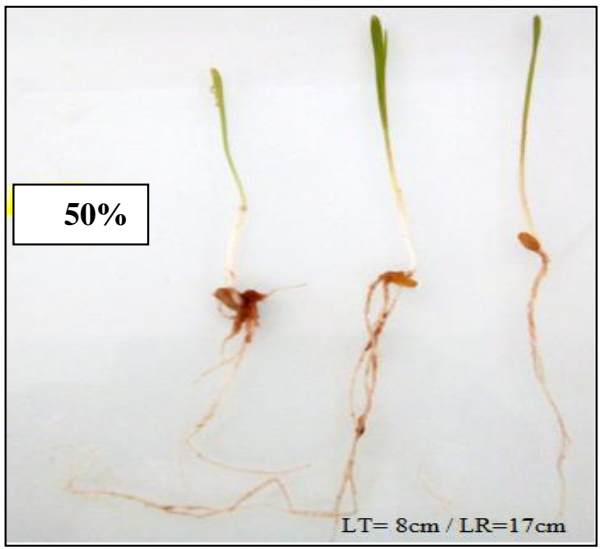

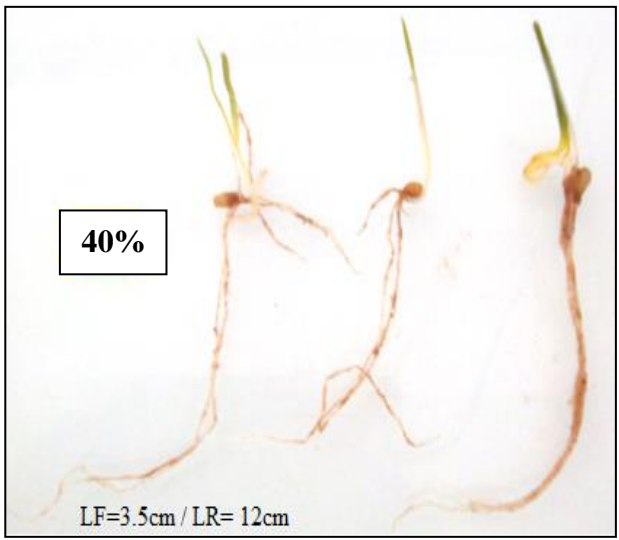
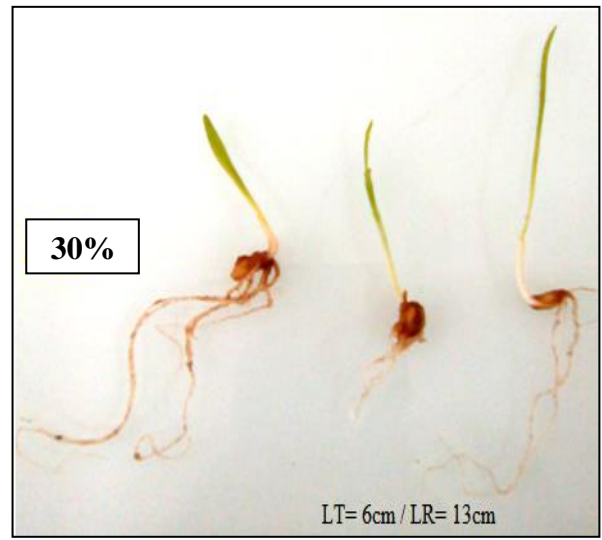
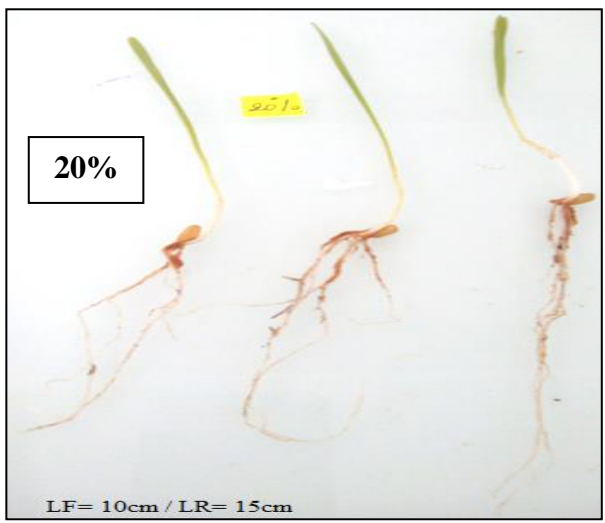
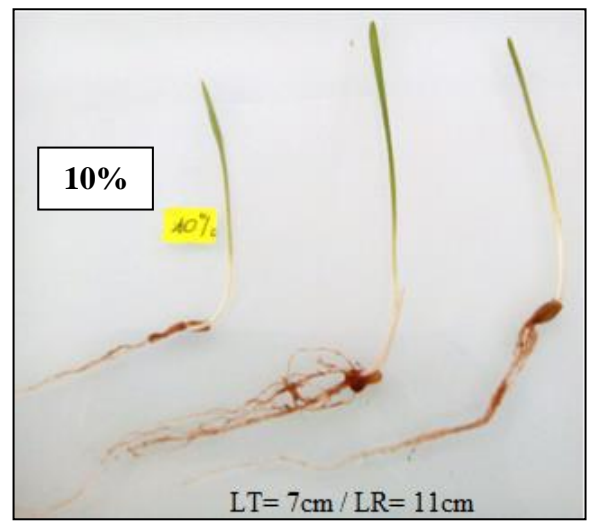
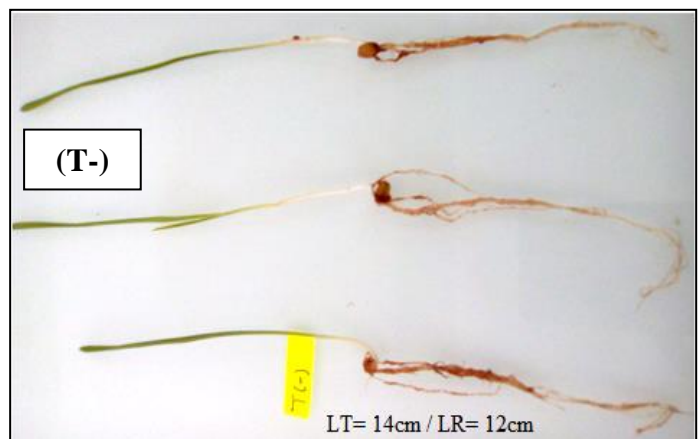

Longueur de la partie aérienne et souterraine des plantules de blé dur traitées par l'extrait aqueux de *P. tomentosa* à différentes concentrations et témoin.

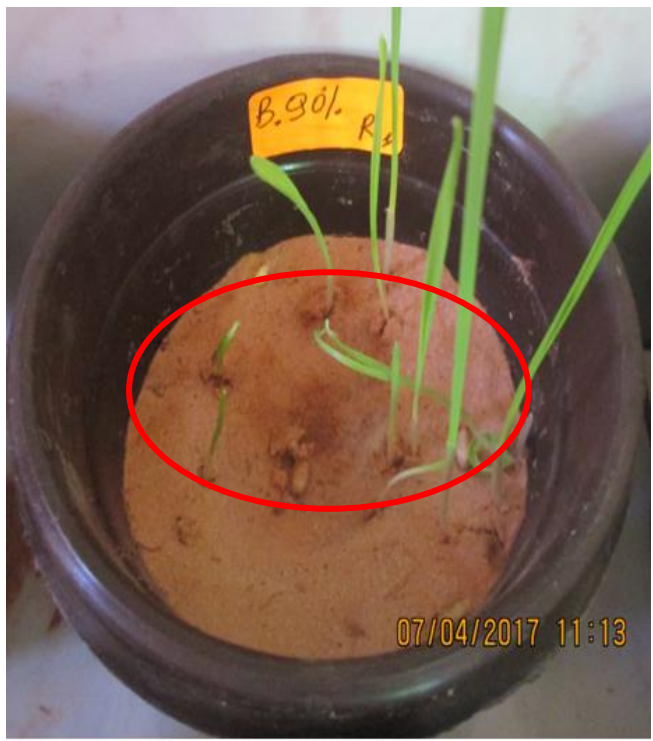 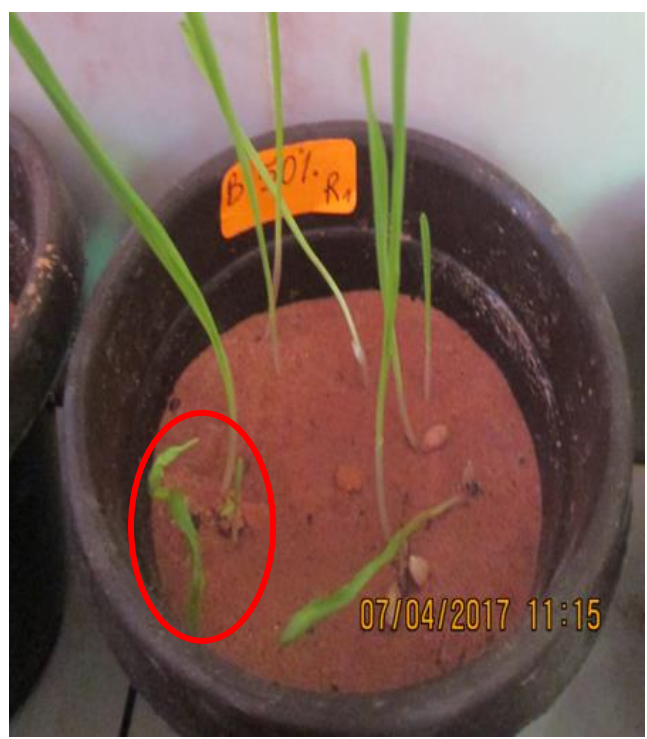
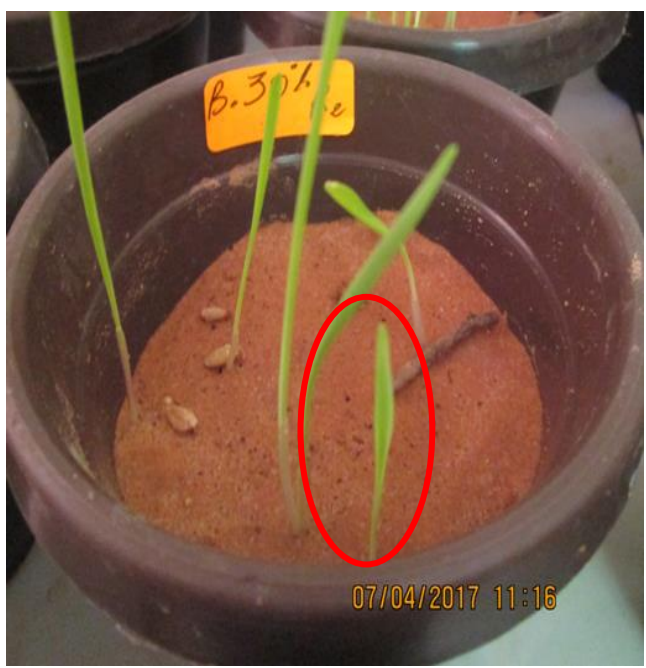 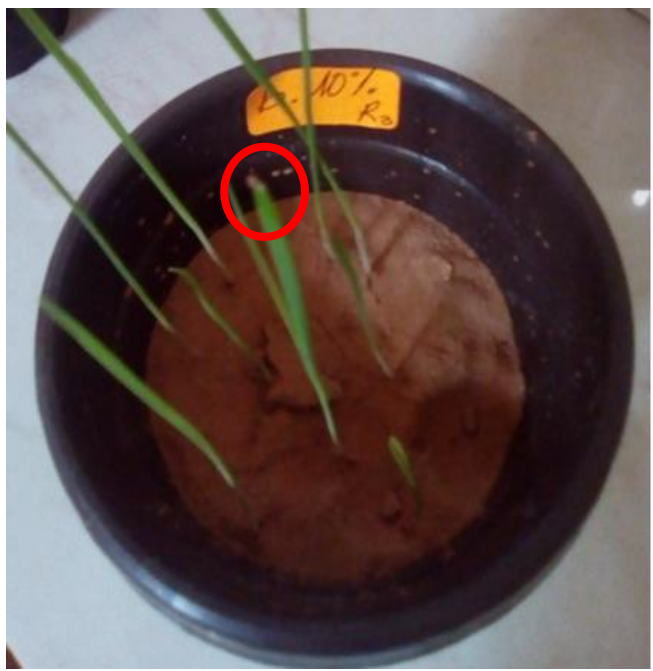

Quelques anomalies de croissance (flétrissement, jaunissement ou noircissement des extrémités du feuillage et le dessèchement) apparue pendant la duré de suivi

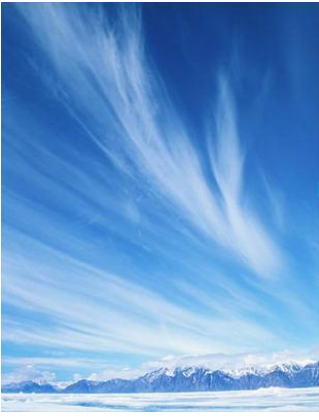

# Annexe II

# Annexe II.
## 1.-Taux de mortalité (*Cleome arabica* et *Pergularia tomentosa*)

Tableau 1 : Valeurs de facteur F et la probabilité Taux de mortalité

| Concentration | *Cleome arabica* | | *Pergularia tomentosa* | |
|---|---|---|---|---|
| | Facteur F | Probabilité | Facteur F | Probabilité |
| 100% | F=841.00 | P= 0.000 | F= 676.00 | P= 0.000 |
| 90% | F=243.00 | P= 0.000 | F=361.00 | P= 0.000 |
| 80% | F= 484.00 | P= 0.000 | F=121.00 | P= 0.000 |
| 70% | F=196.00 | P= 0.000 | F=100.00 | P= 0.001 |
| 60% | F=169.00 | P= 0.000 | F= 25.00 | P= 0.007 |
| 50% | F=100.00 | P= 0.001 | F=16.00 | P=0.016 |
| 40% | F=64.00 | P= 0.001 | F=16.00 | P=0.016 |
| 30% | F=* | P=* | F=* | P=* |
| 20% | F=25.00 | P= 0.007 | F=4.00 | P=0.116 |
| 10% | F=* | P=* | F=1.00 | P=0/374 |

## 1.-Taux d'inhibition (*Cleome arabica*)

Tableau 2 : Valeurs de facteur F et la probabilité P (Taux d'inhibition *Cleome arabica*)

| (%) | Dactylocténion | | Blé dur | | Orge | |
|---|---|---|---|---|---|---|
| | Facteur F | Probabilité | Facteur F | Probabilité | Facteur F | Probabilité |
| 100% | F=* | P=* | F=1681.00 | P= 0.000 | F=1225.00 | P= 0.000 |
| 90% | F=* | P=* | F=1444.00 | P= 0.000 | F=1024.00 | P= 0.000 |
| 80% | F=* | P=* | F=289.00 | P= 0.000 | F=625.00 | P= 0.000 |
| 70% | F=* | P=* | F=240.00 | P= 0.000 | F=* | P=* |
| 60% | F=* | P=* | F=243.00 | P= 0.000 | F=256.00 | P=0.000 |
| 50% | F=* | P=* | F=64.00 | P= 0.001 | F=169.00 | P= 0.000 |
| 40% | F=* | P=* | F= 49.00 | P= 0.002 | F=42.00 | P=0.003 |
| 30% | F=* | P=* | F=196.00 | P= 0.000 | F=12.00 | P= 0.026 |
| 20% | F=* | P=* | F=121.00 | P= 0.000 | F=4.00 | P=0.116 |
| 10% | F=444.00 | P=0.000 | F=100.00 | P= 0.000 | F=* | P=* |

## 3.-Taux d'inhibition (*Pergularia tomentosa*)

Tableau 3: Valeurs de facteur F et la probabilité P (Taux d'inhibition *P. tomentosa*)

| (%) | Dactylocténion | | Blé dur | | Orge | |
|---|---|---|---|---|---|---|
| | Facteur F | Probabilité | Facteur F | Probabilité | Facteur F | Probabilité |
| 100% | F=* | P=* | F=1369.00 | P=0.000 | F=44.26 | P=0.003 |
| 90% | F=* | P=* | F=363.00 | P=0.000 | F=48.08 | P= 0.002 |
| 80% | F=* | P=* | F=23.08 | P=0.009 | F= 48.00 | P=0.002 |
| 70% | F=* | P=* | F=22.56 | P=0.009 | F=48.08 | P=0.002 |
| 60% | F=* | P=* | F=108.00 | P=0.000 | F=3.95 | P=0.118 |
| 50% | F=* | P=* | F=28.00 | P=0.006 | F=4.00 | P=0.116 |
| 40% | F=* | P=* | F=100.00 | P=0.001 | F=16.00 | P=0.016 |
| 30% | F=* | P=* | F=* | P=* | F=* | P=* |
| 20% | F=1681.00 | P=0.000 | F=* | P=* | F=* | P=* |
| 10% | F=1156.00 | P=0.000 | F=* | P=* | F=* | P=* |

## 4.-Longueur de la partie aérienne et souterraine (*Cleome arabica*)
**Orge :**

Tableau 17 : Valeurs de facteur F et la probabilité P
(LPA/LPS Orge) *Cleome arabica*

| (%) | LPA Facteur F | Probabilité | LPS Facteur F | Probabilité |
|---|---|---|---|---|
| 100% | F=7.20 | P=0.055 | F=24.00 | P=0.008 |
| 90% | F=10.29 | P=0.033 | F=75.00 | P=0.001 |
| 80% | F=10.12 | P=0.033 | F=2.58 | P=0.184 |
| 70% | F=6.25 | P=0.067 | F=6.00 | P=0.070 |
| 60% | F=1.00 | P=0.349 | F=36.10 | P=0.004 |
| 50% | F=0.00 | P=1.000 | F=75.00 | P=0.001 |
| 40% | F=4.50 | P=0.101 | F=64.00 | P=0.001 |
| 30% | F=0.80 | P=0.422 | F=3.38 | P=0.140 |
| 20% | F=1.14 | P=0.346 | F=3.06 | P=0.155 |
| 10% | F=132.25 | P=0.000 | F=3.38 | P=0.140 |

**Blé dur :**

Tableau 18 : Valeurs de facteur F et la probabilité P
(LPA/LPS pour le Blé dur) *Cleome arabica*

| (%) | LPA Facteur F | Probabilité | LPS Facteur F | Probabilité |
|---|---|---|---|---|
| 100% | F=7.62 | P=0.051 | F=5.73 | P=0.075 |
| 90% | F=33.80 | P=0.004 | F=52.07 | P=0.002 |
| 80% | F=45.00 | P=0.003 | F=8.25 | P=0.045 |
| 70% | F=9.62 | P=0.036 | F= 3.94 | P=0.118 |
| 60% | F=0.14 | P=0.725 | F=0.000 | P=1.000 |
| 50% | F=2.05 | P=0.225 | F=0.40 | P=0.561 |
| 40% | F=1.23 | P=0.330 | F=1.91 | P=0.239 |
| 30% | F=1.17 | P=0.339 | F=0.05 | P=0.834 |
| 20% | F= 1.89 | P=0.241 | F=10.12 | P=0.033 |
| 10% | F=75.57 | P=0.001 | F=0.00 | P=1.000 |

**Dactylocténion :**

Tableau 19 : Valeurs de facteur F et la probabilité P
(LPA/LPS Dactylocténion) *Cleome arabica*

| (%) | LPA Facteur F | Probabilité | LPS Facteur F | Probabilité |
|---|---|---|---|---|
| 100% | / | / | / | / |
| 90% | / | / | / | / |
| 80% | / | / | / | / |
| 70% | / | / | / | / |
| 60% | / | / | / | / |
| 50% | / | / | / | / |
| 40% | / | / | / | / |
| 30% | / | / | / | / |
| 20% | / | / | / | / |
| 10% | F=71.04 | P=0.001 | F=121.00 | P=0.000 |

**5.-Poids (Orge / Cleome arabica) :**

Tableau 20 : Valeurs de facteur F et la probabilité P
(Poids Orge) *Cleome arabica*

| | LPA | LPS |
|---|---|---|

| (%) | Facteur F | Probabilité | Facteur F | Probabilité |
|---|---|---|---|---|
| 100% | F=28.12 | P=0.006 | F=16.90 | F=0.015 |
| 90% | F=40.50 | P=0.003 | F=24.14 | P=0.008 |
| 80% | F=42.25 | P=0.003 | F=28.12 | P=0.006 |
| 70% | F=2.00 | P=0.230 | F=4.90 | P=0.091 |
| 60% | F=1.99 | P=0.231 | F=28.90 | P=0.006 |
| 50% | F=0.00 | P=1.00 | F=28.00 | P=0.006 |
| 40% | F=1.13 | P=0.349 | F=24.50 | P=0.008 |
| 30% | F=4.00 | P=0.116 | F=2.29 | P=0.205 |
| 20% | F=0.50 | P=0.519 | F=1.29 | P=0.320 |
| 10% | F=18.05 | P=0.013 | F=2.63 | P=0.180 |

**6.-Poids (Blé dur / Cleome arabica)**

Tableau 21 : Valeurs de facteur F et la probabilité P
(Poids Blé dur) *Cleome arabica*

| | LPA | | LPS | |
|---|---|---|---|---|
| (%) | Facteur F | Probabilité | Facteur F | Probabilité |
| 100% | F=0.16 | P=0.709 | F=66.27 | P=0.001 |
| 90% | F=0.80 | P=0.421 | F=71.27 | P=0.001 |
| 80% | F=1.92 | P=0.238, | F=4.60 | P= 0.099 |
| 70% | F=1.15 | P=0.344 | F=18.00 | P=0.013 |
| 60% | F=2.77 | P=0.171 | F=57.14 | P=0.002 |
| 50% | F=0.02 | P=0.896, | F=0.00 | P=1.000 |
| 40% | F=1.14 | P=0.346, | F=0.14 | P=0.725 |
| 30% | F=0.40 | P=0.650, | F=0.58 | P=0.488 |
| 20% | F=0.70 | P =0.450, | F=0.82 | P=0.417 |
| 10% | F=0.07 | P=0.802 | F=1.45 | P=0.295 |

**7.-Poids (Dactylocténion/ Cleome arabica)**

Tableau 21 : Valeurs de facteur F et la probabilité P
(Poids LPA/LPS Dactylocténion) *Cleome arabica*

| | LPA | | LPS | |
|---|---|---|---|---|
| (%) | Facteur F | Probabilité | Facteur F | Probabilité |
| 100% | / | / | / | / |
| 90% | / | / | / | / |
| 80% | / | / | / | / |
| 70% | / | / | / | / |
| 60% | / | / | / | / |
| 50% | / | / | / | / |
| 40% | / | / | / | / |
| 30% | / | / | / | / |
| 20% | / | / | / | / |
| 10% | F=0.40 | P=0.561 | F=0.00 | P=1.000 |

**8.-Longueur de la partie aérienne et souterraine (Orge / *Pergularia tomentosa*)**

Tableau 22 : Valeurs de facteur F et la probabilité P
(LPA/LPS Orge) *Pergularia tomentosa*

| | LPA | | LPS | |
|---|---|---|---|---|
| (%) | Facteur F | Probabilité | Facteur F | Probabilité |
| 100% | F=0.47 | P=0.529 | F=8.25 | P=0.045 |
| 90% | F=0.78 | P=0.427 | F=1.83 | P=0.248 |
| 80% | F= 1.31 | P=0.316 | F=7.31 | P= 0.054 |
| 70% | F=0.11 | P=0.755 | F=9.39 | P=0.037 |
| 60% | F=144.40 | P=0.000 | F=21.13 | P=0.010 |
| 50% | F=0.06 | P=0.825 | F=0.54 | P=0.502 |
| 40% | F=3.21 | P=0.148 | F=0.01 | P=0.917 |
| 30% | F=8.80 | P=0.041 | F=2.04 | P=0.227 |
| 20% | F=1.99 | P=0.231 | F=2.04 | P=0.227 |
| 10% | F=0.75 | P=0.434 | F=0.65 | P=0.467 |

**9.-Longueur de la partie aérienne et souterraine (Blé dur / *Pergularia tomentosa*)**

Tableau 23 : Valeurs de facteur F et la probabilité P
(LPA/LPS Blé dur) *Pergularia tomentosa*

|  | LPA | | LPS | |
|---|---|---|---|---|
| (%) | Facteur F | Probabilité | Facteur F | Probabilité |
| 100% | F=22.72 | P=0.009 | F= 20.63 | P=0.010 |
| 90% | F=25.75 | P=0.007 | F=0.27 | P=0.632 |
| 80% | F=4.88 | P=0.092 | F=4.79 | P=0.094 |
| 70% | F=2.58 | P =0.184 | F=4.78 | P=0.094 |
| 60% | F=17.02 | P=0.015 | F=30.12 | P=0.005 |
| 50% | F=15.91 | P=0.016 | F=0.05 | P=0.840 |
| 40% | F=53.64 | P=0.002 | F=2.57 | P=0.184 |
| 30% | F=15.78 | P=0.017 | F=2.43 | P=0.194 |
| 20% | F=5.73 | P=0.075 | F= 0.09 | P=0.778 |
| 10% | F=2.76 | P=0.172 | F=3.63 | P=0.129 |

**10.-Longueur de la partie aérienne et souterraine (Dactylocténion / *Pergularia tomentosa*)**

Tableau 24 : Valeurs de facteur F et la probabilité P
(LPA/LPS Dactylocténion) *Pergularia tomentosa*

|  | LPA | | LPS | |
|---|---|---|---|---|
| (%) | Facteur F | Probabilité | Facteur F | Probabilité |
| 100% | / | / | / | / |
| 90% | / | / | / | / |
| 80% | / | / | / | / |
| 70% | / | / | / | / |
| 60% | / | / | / | / |
| 50% | / | / | / | / |
| 40% | / | / | / | / |
| 30% | / | / | / | / |
| 20% | F=52.56 | P=0.002 | F=67.60 | P=0.001 |
| 10% | F=82.57 | P=0.001 | F=120.03 | P=0.000 |

**11.-Poids de la partie aérienne et souterraine (Orge / *Pergularia tomentosa*)**

Tableau 25 : Valeurs de facteur F et la probabilité P
(LPA/LPS Orge) *Pergularia tomentosa*

|  | LPA | | LPS | |
|---|---|---|---|---|
| (%) | Facteur F | Probabilité | Facteur F | Probabilité |
| 100% | / | / | / | / |
| 90% | / | / | / | / |
| 80% | / | / | / | / |
| 70% | / | / | / | / |
| 60% | F=112.00 | P=0.00 | F=15.52 | P=0.017 |
| 50% | / | / | / | / |

| (%) | | | | |
|---|---|---|---|---|
| 40% | / | / | / | / |
| 30% | / | / | / | / |
| 20% | / | / | / | / |
| 10% | / | / | / | / |

**12.-Poids de la partie aérienne et souterraine (Blé dur / *Pergularia tomentosa*)**

Tableau 26 : Valeurs de facteur F et la probabilité P
(LPA/LPS Blé dur) *Pergularia tomentosa*

| | LPA | | LPS | |
|---|---|---|---|---|
| (%) | Facteur F | Probabilité | Facteur F | Probabilité |
| 100% | / | / | / | / |
| 90% | F=28.13 ; P=0.006 | / | F=46.12 ; P=0.002 | / |
| 80% | / | / | / | / |
| 70% | / | / | / | / |
| 60% | F=8.64 | P=0.032 | F=10.76 | P=0.031 |
| 50% | F=29.45 | P=0.006 | F=8.10 | P=0.047 |
| 40% | F=9.56 | F=0.037 | F= 16.49 | P=0.015 |
| 30% | / | / | / | / |
| 20% | F=10.47 | P=0.032 | F=9.85 | P=0.035 |
| 10% | F=38.44 | P=0.003 | F=49.47 | P=0.002 |

**13.-Poids Dactylocténion**

Tableau 27 : Valeurs de facteur F et la probabilité P
(LPA/LPS Dactylocténion) *Pergularia tomentosa*

| | LPA | | LPS | |
|---|---|---|---|---|
| (%) | Facteur F | Probabilité | Facteur F | Probabilité |
| 100% | / | / | / | / |
| 90% | / | / | / | / |
| 80% | / | / | / | / |
| 70% | / | / | / | / |
| 60% | F=112.00 | P=0.00 | F=15.52 | P=0.017 |
| 50% | / | / | / | / |
| 40% | / | / | / | / |
| 30% | / | / | / | / |
| 20% | F=2.58 | P=0.184 | F=4.50 | P=0.101 |
| 10% | F=4.92 | P=0.091 | F=9.80 | P=0.035 |

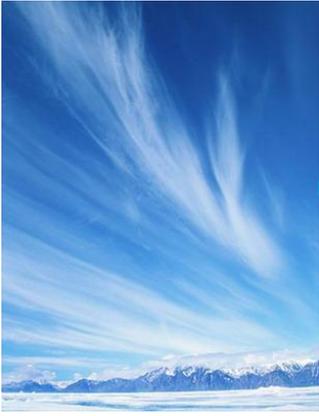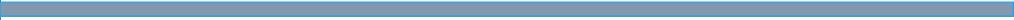

# Productions Scientifiques

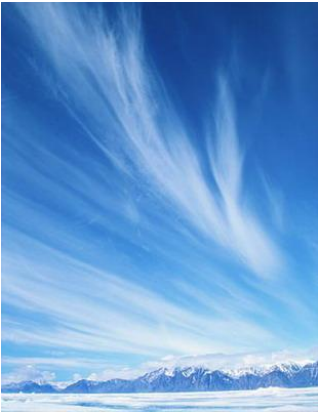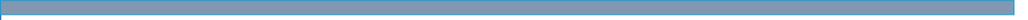

# Publications internationales

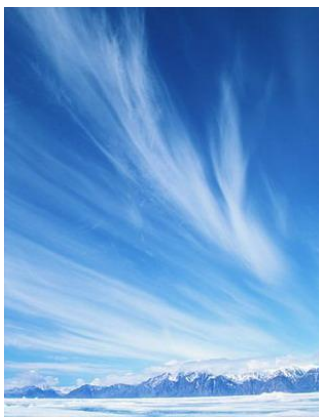

## Publications internationals

- **R. CHERIF**, A. KEMASSI, Z. BOUAL, N. BOUZIANE, F. BENBRAHIM, A. HADJSEYD, T. GHARIB, A. OULD EL HADJ-KHELIL, M.L. SAKEUR ET M.D. OULD EL HADJ, **2016**. Activités biologiques des extraits aqueux de *Pergularia tomentosa* L. (Asclepiadaceae). ***Lebanese Science Journal***, Vol. 17, No. 1, 12pages.
  **Site :**
  https://www.researchgate.net/publication/305223383_Activites_biologiques_des_extraits_aqueux_de_Pergularia_tomentosa_L_Asclepiadaceae

- KEMASSI, A. HEROUINI, S. A. HADJ, **R. CHERIF**, M. D. OULD ELHADJ. **2019**. Effet insecticide des extraits aqueux d'*Euphorbia guyoniana* (*Euphorbiaceae*) récoltée dans oued Sebseb (Sahara Algérien) sur le *Tibolium castaneum*. ***Lebanese Science Journal***, Vol. 20, No. 1, 16pages.
  **Site :** http://dx.doi.org/10.22453/LSJ-020.1.055-070 National Council for Scientific Research – Lebanon 2018© lsj.cnrs.edu.lb/vol-20-no-1-2019/

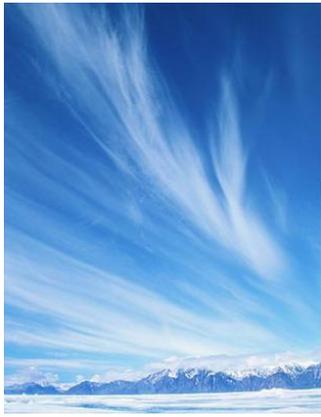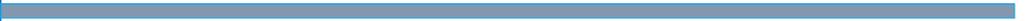

# *Publications nationales*

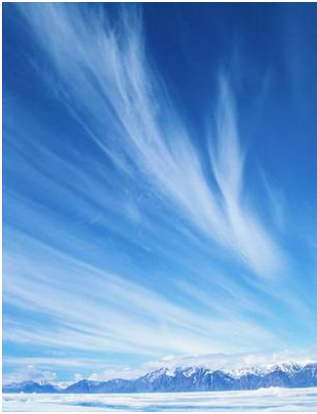

## Publications nationales

- KEMASSI ABDELLAH, DAREM SABRINE, **CHERIF REKIA**, BOUAL ZAKARIA, SADINE SALAH EDDINE, AGGOUNE MOHAMED SALAH, OULD EL HADJ-KHELIL AMINATA, et OULD ELHADJ MOHAMED DIDI., **2014.** Recherche et identification de quelques plantes médicinales à caractère hypoglycémiant de la pharmacopée traditionnelle des communautés de la vallée du M'Zab (Sahara septentrional Est Algérien). *Journal of Advanced Research in Science and Technology.* ISSN: 2352-9989. 1(1), 1-5.

  **Site :**

  https://www.researchgate.net/profile/Salah_Sadine/publication/260990891_Recherche_et_identification_de_quelques_plantes_medicinales_a_caractere_hypoglycemiant_de_la_pharmacopee_traditionnelle_des_communautes_de_la_vallee_du_M%27Zab_Sahara_septentrional_Est_Algerien/links/00463532f51de1d716000000/Recherche-et-identification-de-quelques-plantes-medicinales-a-caractere-hypoglycemiant-de-la-pharmacopee-traditionnelle-des-communautes-de-la-vallee-du-MZab-Sahara-septentrional-Est-Algerien.pdf

- KEMASSI A., BOUKHARI K., **CHERIF R.,** GHADA K., BENDAKEN N., BOUZIANE N.1, BOUAL Z., BOURAS N., OULD ELHADJ-KHELI. A. et OULD ELHADJ M.D., **2015.** Evaluation de l'effet larvicide de l'extrait aqueux d'*Euphorbia guyoniana* (Boiss. & Reut.)(*Euphorbiaceae*). ISSN : 1112 -7163.Vol.8 n°1 (2015) : 44 – 61Pages.

- **Site:**

  https://www.researchgate.net/publication/308054521_Evaluation_de_l'effet_larvicide_de_l'extrait_aqueux_d'Euphorbia_guyoniana_Boiss_Reut_Euphorbiaceae

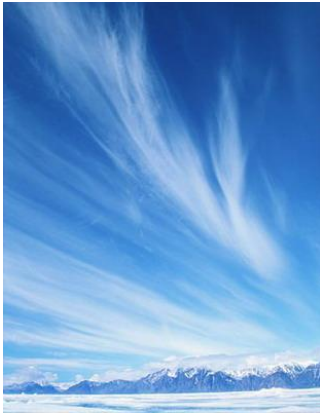

# Communications internationales

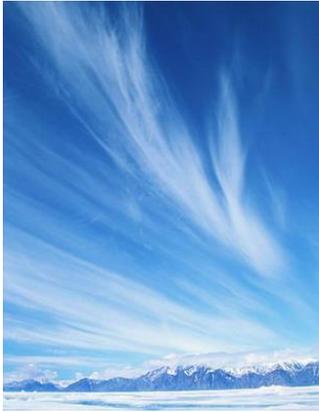

## Communications internationals

1. **CHERIF R**., A. KEMASSI, Z. BOUAL, N. BOUZIANE, F. BENBRAHIM, A. HADJSEYD, T. GHARIB1, A. OULD EL HADJ-KHELIL, M.L. SAKEUR et M.D. OULD EL HADJ**. 2014.** Activités biologiques des extraits aqueux de *Pergularia tomentosa* l. (Asclepiadaceae) in Congrès **International** sur le milieu aride « Ressources, Biodiversité, Environnement » (**CIMA**). Université de Ghardaïa (Algérie), les 09 et 10 décembre 2014.

2. **CHERIF Rekia**, KEMASSI Abdellah, BOUAL Zakaria, BOUZIANE Nawal, BENBRAHIM Fouzi, HADJSEYD Abdelkader, OULD EL HADJ-KHELIL Aminata, SAKEUR Mohamed LAKHDER and OULD EL HADJ Mohamed Didi, **2016**. Activités biologiques des extraits aqueux de *Pergularia tomentosa* l. (*Asclepiadaceae*). In 2$^{sd}$ Africa-**International** Allelopathy Congress in Sousse-Tunisia, on November, 16-19, 2016 *(AIAC-2016).*

3. KEMASSI A., DAREM S., **CHERIF R.,** BOUAL Z., SADINE S. E., AGGOUNE M. S., OULD EL HADJ-KHELIL A., & OULD ELHADJ M. D. **2017**. Recherche et identification de quelques plantes médicinales à caractère hypoglycémiant de la pharmacopée traditionnelle des communautés de la vallée du M'Zab (Sahara septentrional Est Algérien). **In** Séminaire **international** sur les plantes spontanées «Biodiversité, Préservation, Valorisation et Innovation» (**SNPS**).

4. *HEROUINI Amel. KEMASSI Abdellah, ABISMAIL LELA, KADICI Meriem1 AITAOUDIA Ahmed,; TAIBAOUI Zakaria, CHERIF Rekia* et *OULD EL HADJ Mohamed Didi.* **2019.** Evaluation du pouvoir larvicide des huiles de graines *Pergularia tomentosa.* in 30$^{ème}$ congres Internationale de l'Association Tunisienne biologiques (**ATSB**), les 25-28 Marsin Sousse. Tunisie.

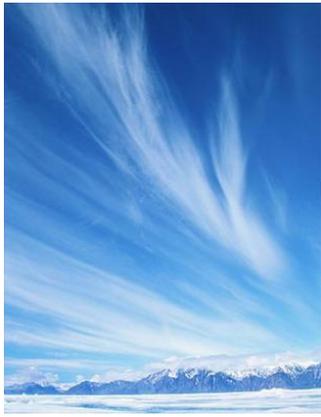
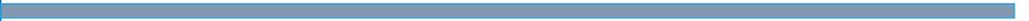

*Communications nationales*

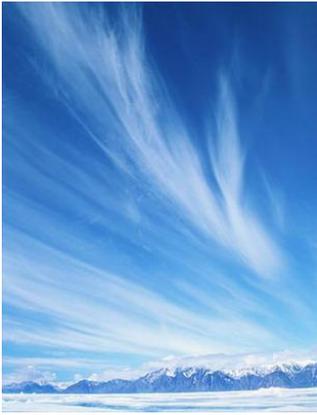

## Communications nationales

1. KEMASSI A., DAREM S., **CHERIF R.,** BOUAL Z., SADINE S. E., AGGOUNE M. S., OULD EL HADJ-KHELIL A., et OULD ELHADJ M. D., 2012. Inventaire des plantes spontanées à caractère hypoglécimantes utilisées dans la pharmacopée traditionnel dans la région de Ghardaïa, Université de Ghardaïa (Algérie), In Séminaire **National** sur les Plantes Spontanées du Sahara (**SNPS 2012**). Les 12 et 13 décembre 2012.

2. **CHERIF R**., A. KEMASSI, Z. BOUAL, N. BOUZIANE, F. BENBRAHIM, A. HADJSEYD, T. GHARIB1, A. OULD EL HADJ-KHELIL, M.L. SAKEUR et M.D. OULD EL HADJ. **2015.** Activités biologiques des extraits aqueux de *Pergularia tomentosa* l. (*Asclepiadaceae*) in Séminaire **International** sur l'agriculture en zones arides (**SNAZA**), les 17 et 18 Novembre 2015.

3. KEMASSI Abdellah, **CHERIF Rekia**, BOUAL Zakaria, BOUZIANE Nawal, BENBRAHIM Fouzi, HADJSEYD Abdelkader, OULD EL HADJ-KHELIL Aminata, SAKEUR Mohamed LAKHDER & OULDEL HADJ Mohamed Didi., **2015.** Inventaire de la flore messicole associées à la culture du blé dur *Triticum durum* L.(*Poaceae*) irriguée sous pivots dans la région d'Ouargla et recherche de l'activité allélopathique des extraits de quelques plantes spontanées du Sahara algérien. 'La céréaliculture dans les zones aride'- Ouargla le 10 mars 2015.

4. **CHERIF R**., DAREM S, HEROUINI A., et KEMASSI A., Inventaire des plantes spontanées à caractère hypoglécimantes utilisées dans la pharmacopée traditionnel dans la région de Ghardaïa, Université de Ghardaïa (Algérie). **2019.** In 1[er] journée d'études en biochimie sur le Stress Oxydatif et les Maladies Chroniques : Quelle Relation et Quelle Solution. Le 29 Janvier 2019.